\documentclass{aa}  

\usepackage{makecell}
\usepackage{booktabs}
\usepackage{multirow}
\usepackage{graphicx}
\usepackage[varg]{txfonts}
\usepackage{hyperref}
\usepackage{xcolor}
\usepackage{ulem}
\hypersetup{
    colorlinks=true,
    citecolor=blue,
    linkcolor=red,
    filecolor=magenta,
    urlcolor=cyan,
} 
\bibpunct{(}{)}{;}{a}{}{,}
\begin{document}   

\title{Featureless transmission spectra of 12 giant exoplanets observed by GTC/OSIRIS}

\author{
    C. Jiang\inst{1, 2}
    \and G. Chen\inst{1, 3}\thanks{\email{guochen@pmo.ac.cn}}
    \and E. Pall\'e\inst{4,5}
    \and F. Murgas\inst{4,5}
    \and H. Parviainen\inst{4,5}
    \and Y. Ma\inst{1}
    }

\institute{
    CAS Key Laboratory of planetary sciences, Purple Mountain Observatory, Chinese Academy of Sciences, 210023 Nanjing, China
    \and School of Astronomy and Space Science, University of Science and Technology of China, 230026 Hefei, China
    \and CAS Center for Excellence in Comparative Planetology, 230026 Hefei, China
    \and Instituto de Astrof\'isica de Canarias (IAC), E-38205 La Laguna, Tenerife, Spain
    \and Departamento de Astrof\'isica, Universidad de La Laguna (ULL), E-38206 La Laguna, Tenerife, Spain
}

\date{Received ...; accepted ...}

\abstract
{Exoplanet atmospheres are the key to understanding the nature of exoplanets. To this end, transit spectrophotometry provides us opportunities to investigate the physical properties and chemical compositions of exoplanet atmospheres.}
{We aim to detect potential atmospheric signatures in 12 gaseous giant exoplanets using transit spectrophotometry and we try to constrain their atmospheric properties.}
{The targets of interest were observed using transit spectrophotometry with the GTC OSIRIS instrument. We estimated the transit parameters and obtained the optical transmission spectra of the target planets using a Bayesian framework. We analyzed the spectral features in the transmission spectra based on atmospheric retrievals.}
{Most of the observed transmission spectra were found to be featureless, with only the spectrum of CoRoT-1b showing strong evidence for atmospheric features. However, in combination with the previously published near-infrared transmission spectrum, we found multiple interpretations for the atmosphere of CoRoT-1b due to the lack of decisive evidence for alkali metals or optical absorbers. }
{Featureless spectra are not necessarily indicative of cloudy atmospheres if they poorly constrain the altitudes of cloud decks. Precise constraints on the models of hazes and clouds strongly depend on the significance of the observed spectral features. Further investigations on these exoplanets, especially CoRoT-1b, are required to confirm the properties of their atmospheres.}

\keywords{planetary systems -- planets and satellites: individuals: CoRoT-1b, HAT-P-18b, HAT-P-57b, Qatar-1b, TrES-4b, WASP-2b, WASP-10b, WASP-32b, WASP-36b, WASP-39b, WASP-49b, WASP-156b -- planets and satellites: atmospheres -- techniques: spectroscopic -- Methods: data analysis}

\maketitle

\section{Introduction}

    Exoplanet atmospheres provide crucial insights into the nature of exoplanets. By analyzing their physical properties and chemical compositions, we can gain a deeper understanding of planet formation and evolution. One widely used technique for studying exoplanet atmospheres is transit spectroscopy \citep{2000ApJ...537..916S, 2001ApJ...553.1006B}. This method involves analyzing the transmission spectra of exoplanets to identify absorption and scattering signatures at the day-night terminators. From these spectra, we can derive parameters using simplified atmospheric radiative transfer models, estimate the abundances of various atomic and molecular species, and constrain the altitudes of clouds and hazes \citep{2019ARA&A..57..617M}. 
    
    For the large-aperture ground-based telescopes currently in operation, close-in transiting giant planets can be suitable targets for transit spectrophotometric observations. In addition to characterizing individual targets, it is also important to conduct population studies on large groups of exoplanets to identify statistical trends and suggest new directions for future research.  Many such studies have been performed on exoplanet transmission spectra using space-based observations from the instruments such as the Space Telescope Imaging Spectrograph (STIS) and Wide Field Camera 3 (WFC3) on the Hubble Space Telescope (HST) and the Infrared Array Camera (IRAC) on the Spitzer Space Telescope \citep{2016ApJ...823..109I, 2016Natur.529...59S, 2016ApJ...817L..16S, 2017ApJ...834...50B, 2017AJ....154..261C, 2017ApJ...847L..22F, 2018AJ....155..156T, 2018MNRAS.481.4698F, 2019MNRAS.482.1485P, 2019ApJ...887L..20W, 2020NatAs...4..951G, 2021A&A...648A.127B, 2021AJ....162...37R, 2022arXiv221100649E}. The STIS (G430L, G750L) spectra cover the optical wavebands and are used to study features such as Rayleigh scattering and alkali metal absorption (Na I and K I), while the WFC3 (G141) spectra mainly focus on the $\rm H_2O$ absorption features around 1.4~$\rm\mu m$. \cite{2016Natur.529...59S} compared the transmission spectra of ten hot Jupiters, from clear to cloudy, with theoretical models, and found that clouds and hazes, rather than primordial water depletion, were responsible for the weaker spectral signatures in the WFC3 wavebands. The retrieval analyses performed by \cite{2017ApJ...834...50B}, \cite{2018AJ....155..156T}, and \cite{2019MNRAS.482.1485P} constrained the $\rm H_2O$ abundances and the aerosol properties in exoplanet atmospheres, revealing subsolar $\rm H_2O$ abundances in some planets. \cite{2021AJ....162...37R} investigated the possibility of disequilibrium chemistry in exoplanet atmospheres using 62 transmission spectra observed by HST WFC3. They found that disequilibrium chemistry occurs in about half of the samples, and therefore plays an important role in interpreting exoplanet atmospheres. Most recently, \cite{2022arXiv221100649E} conducted a homogeneous analysis of the transmission spectra of 70 exoplanets observed by HST WFC3. Over half of the samples show strong evidence for atmospheric features. While the HST WFC3 G141 data alone were not sufficient to extract detailed trends from the whole population of objects, the researchers did identify general patterns of super-solar water abundances in most samples. 
    
    While many studies of exoplanet atmospheres rely on observations with distinct spectral features to effectively constrain the physical properties and chemical abundances, there are also plenty of observed targets with featureless spectra or nondetection results. This raises the question of whether we can learn anything about exoplanet atmospheres from inconclusive results. For instance, gas giant planets with cloudy atmospheres often have featureless or weakly characterized transmission spectra because high-altitude clouds attenuate the absorption signals of atomic and molecular gases across broad wavebands. Conversely, if a puffy gaseous planet is observed and its transmission spectrum is found to be flat, the nondetection of transmission signals may result from multiple factors such as the low signal-to-noise ratio (S/N) of the transmission spectrum, the depletion of the chemical composition, and the presence of clouds and/or hazes. By combining a large number of observed featureless transmission spectra, it may be possible to establish a relationship between atmospheric cloudiness and the physical properties of an exoplanet. This relationship could then be used to predict the cloudiness of other exoplanets, improving the efficiency of target selection for atmospheric characterization.

\begin{figure}[htbp]
        \centering
        \includegraphics[width=\linewidth]{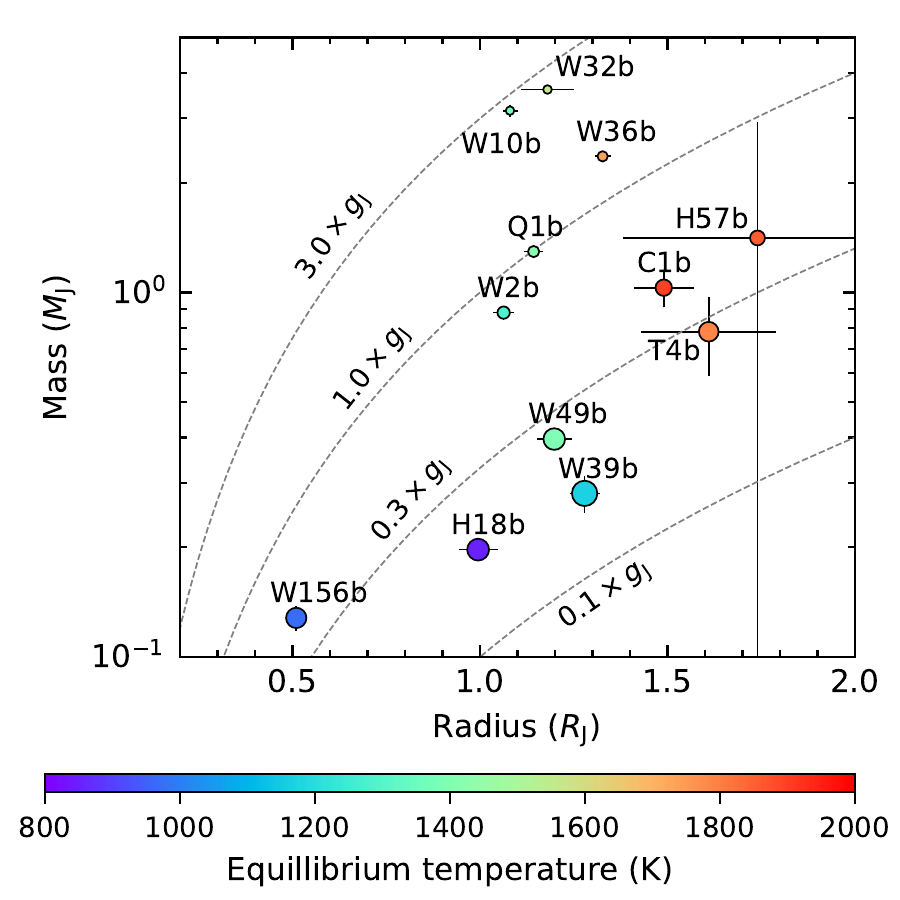}
        \caption{ Parameter distribution of the target planets. The letters ``C,'' ``H,'' ``Q,'' ``T,'' and ``W'' are short for ``CoRoT-,'' ``HAT-P-,'' ``Qatar-,'' ``TrES-,'' and ``WASP-,'' respectively. The colors indicate the equilibrium temperature of the planet. The marker size is linearly proportional to the atmospheric scale height relative to the planetary radius ($H/R_{\rm p} \propto R_{\rm p} T_{\rm eq} / M_{\rm p}$), where $H$ is the scale height, $R_{\rm p}$ is the planetary radius, $M_{\rm p}$ is the planetary mass, and $T_{\rm eq}$ is the equilibrium temperature. The adopted parameters are listed in Table~\ref{table:lc_priors}. The dashed lines are the isogravity lines, where $g_{\rm J}$ indicate Jupiter's gravity. }
        \label{fig:target_distribution}
    \end{figure}

    In this study, we aim to study the atmospheres of 12 gaseous giant exoplanets (CoRoT-1b, \citealt{2008A&A...482L..17B}; HAT-P-18b, \citealt{2011ApJ...726...52H}; HAT-P-57b, \citealt{2015AJ....150..197H}; Qatar-1b, \citealt{2011MNRAS.417..709A}; TrES-4b, \citealt{2007ApJ...667L.195M}; WASP-2b, \citealt{2007MNRAS.375..951C}; WASP-10b, \citealt{2009MNRAS.392.1585C}; WASP-32b, \citealt{2010PASP..122.1465M}; WASP-36b, \citealt{2012AJ....143...81S}; WASP-39b, \citealt{2011A&A...531A..40F}; WASP-49b, \citealt{2012A&A...544A..72L}; and WASP-156b, \citealt{2018A&A...610A..63D}) by detecting their spectral features using transit spectrophotometry. These planets span a wide range of parameters (Fig.~\ref{fig:target_distribution}). Some have large pressure scale heights that could produce significant atmospheric signatures if their atmospheres are clear (e.g., CoRoT-1b, HAT-P-18b, TrES-4b, WASP-39b, WASP-49b, and WASP-156b). Others have smaller scale heights and can be used to investigate under what circumstances the featureless spectra can be interpreted as cloudy atmospheres, or whether the activity of their host stars can affect the observed transmission spectra with weak features. By combining data from all targets, we aim to determine whether it is possible to predict the observability of atmospheric signatures based on certain planetary parameters. 
    
    This paper is organized as follows. In the next section, we summarize the observations and data reduction procedures. Section~\ref{sect:method_lightcurve} introduces our light-curve analysis methods, while Sect.~\ref{sect:method_retrieval} describes the details of our atmospheric retrievals. We further investigate several targets that have been studied in other literature and discussed the possible statistical trends in Sect.~\ref{sect:discussion}. Finally, we draw our conclusions in Sect.~\ref{sect:conclusions}. 

\section{GTC Observations and data reduction}
\label{sect:observations}

    In this work, a total of 12 transiting exoplanets were observed using the Optical System for Imaging and low-intermediate-Resolution Integrated Spectroscopy on the Gran Telescopio Canarias \citep[GTC OSIRIS;][]{2000SPIE.4008..623C}. We observed one transit for each target, except for WASP-36b, for which we observed two transits. The observations were carried out using the longslit spectroscopic mode of OSIRIS. The unvignetted field of view for OSIRIS is $7.8'\times7.8'$. The detector consists of a mosaic of two red-sensitive CCDs ($2048\times 4096$ pixels) with a $9.4''$ gap between them. The pixel scale is $0.254''$ after a $2\times2$ pixel binning in the spectroscopic mode. The readout speed is 200 kHz for the standard mode and 500 kHz for the fast readout mode. The adopted grisms were R1000B or R1000R, both with a resolution of $\sim$1000. The R1000B grism covers a wavelength range of 3630 -- 7880~\AA, while the R1000R grism covers a wavelength range of 5100 -- 10\,400~\AA. In each observation, a comparison star was simultaneously observed with the target star through a $12''$ or $40''$ slit. The calibration images (the flat, the bias, and the arc lines of Ne, Xe, and HgAr) were obtained on the same day as the science observations. Further details of the observations are provided in Table~\ref{table:obs_info}. 
        
    \begin{table*}[htbp]
    \caption{Observation summary.}  
    \label{table:obs_info} 
    \tiny
    \centering    
    \begin{tabular}{l l c c c c c c c c c}  
    \hline \hline
    Target star & Proposal ID 
    & Date (UT) & Grism & Slit
    & \makecell{Ref. star \\ (UCAC4)} 
    & \makecell{Readout \\ (kHz)} 
    & \makecell{Exp.\\time (s)} 
    & \makecell{Frame\\number}
    & \makecell{Airmass \\ (start -- min -- end)}  
    & \makecell{Seeing\tablefootmark{a} \\ (95\% interval)}\\  
    \hline 
    CoRoT-1 & GTC...17B\tablefootmark{b, c} & 2018-01-10 & R1000R & $40''$ & 435-022010 
            & 200 & 100 & 168 & 1.41 -- 1.18 -- 1.85 & $1.0'' - 1.5''$\\
    HAT-P-18 & GTC59-16A\tablefootmark{c} & 2016-04-28 & R1000R & $40''$ & 616-055015 
            & 500 & 7 & 840 & 1.64 -- 1.00 -- 1.02 & $0.9'' - 1.3''$\\ 
    HAT-P-57 & GTC16-19A\tablefootmark{c} & 2019-08-04 & R1000R & $12''$ & 503-079345 
        & 200 & 8 & 422 & 1.10 -- 1.05 -- 1.29 & $1.1'' - 2.7''$\\ 
    Qatar-1 & GTC9-14A\tablefootmark{d} & 2014-07-08 & R1000R & $40''$ & 776-040769 
        & 200 & 45 & 167 & 1.40 -- 1.24 -- 1.25 & $1.4'' - 2.4''$\\ 
    TrES-4 & GTC59-16A\tablefootmark{c} & 2016-07-01 & R1000R & $40''$ & 637-055982 
        & 200 & 13 & 608 & 1.07 -- 1.01 -- 1.96 & $0.8'' - 1.3''$\\ 
    WASP-2 & GTC54-14B\tablefootmark{d} & 2014-08-05 & R1000R & $40''$ & 482-122257 
        & 200 & 10 & 479 & 1.50 -- 1.08 -- 1.10 & $0.7'' - 1.2''$\\ 
    WASP-10 & GTC59-16A\tablefootmark{c} & 2016-08-16 & R1000B & $40''$ & 608-139057 
        & 500 & 30 & 459 & 1.29 -- 1.00 -- 1.21 & $1.0'' - 1.8''$\\ 
    WASP-32 & GTC50-13B\tablefootmark{d} & 2013-10-20 & R1000R & $40''$ & 457-000336 
        & 200 & 17 & 497 & 1.19 -- 1.13 -- 1.42 & $0.8'' - 1.4''$\\ 
    WASP-36 & GTC54-14B\tablefootmark{d} & 2015-01-17 & R1000R & $40''$ & 410-046712 
        & 200 & 65 & 157 & 1.64 -- 1.25 -- 1.28 & $1.5'' - 3.0''$\\ 
    WASP-36 & GTC54-14B\tablefootmark{d} & 2015-01-19 & R1000R & $40''$ & 410-046712 
        & 200 & 45 & 215 & 1.28 -- 1.25 -- 1.82 & $1.6'' - 4.4''$\\ 
    WASP-39 & GTC47-12B\tablefootmark{d} & 2013-02-24 & R1000B & $12''$ & 433-060837
        & 200 & 40 & 243 & 1.69 -- 1.18 -- 1.22 & $1.1'' - 2.0''$\\ 
    WASP-49 & GTC50-13B\tablefootmark{d} & 2013-11-15 & R1000R & $40''$ & 366-011140 
        & 200 & 30 & 384 & 1.88 -- 1.43 -- 1.48 & $2.2'' - 7.5''$\\ 
    WASP-156 & GTC...17B\tablefootmark{b, c} & 2017-10-03 & R1000R & $40''$ & 463-002875 
        & 200 & 10 & 516 & 2.20 -- 1.11 -- 1.14 & $1.0'' - 2.2''$\\ 
    \hline  
    \end{tabular}
    \tablefoot{
    \tablefoottext{a}{Seeing was measured by the full width at half maximum of the stellar spectrum along the spatial direction at the central wavelength.}
    \tablefoottext{b}{Full ID: GTCMULTIPLE2G-17B.}
    \tablefoottext{c}{PI: G. Chen.}
    \tablefoottext{d}{PI: E. Pallé.}
    } 

    \end{table*}

    We followed the data reduction procedures outlined in our previous work \citep{2022A&A...664A..50J} and used customized IDL and IRAF scripts to remove the bias, flat, sky, and cosmic rays. The optimal extraction algorithm \citep{1986PASP...98..609H} was used to produce 1D stellar spectra. Then, we aligned the spectra of the target and the reference stars based on the cross-correlation of telluric oxygen A- and B-bands in the wavelength solutions. The stellar flux counts were summed in narrow passbands and normalized by out-of-transit median values to produce spectroscopic light curves. We obtained white light curves in the same manner but used a broader passband that excluded the telluric oxygen A-band (755 -- 765 nm). When using the OSIRIS R1000R grism, we adopted uniform 10-nm passbands from 525 nm to 925nm and discarded spectra with wavelengths longer than 925 nm due to fringing patterns and second-order contamination. When using the R1000B grism, we adopted nonuniform passbands from 400 nm to 785 nm due to the very low response at the blue end. The specific passband setups are provided in Table~\ref{table:ts_R1000R1}, \ref{table:ts_R1000R2}, and \ref{table:ts_R1000B}.

\section{Transit light-curve analysis}
\label{sect:method_lightcurve}

    \subsection{Light-curve model}
    
    We adopted \texttt{PyTransit} \citep{2015MNRAS.450.3233P} to calculate the transit light-curve models. This Python package is implemented based on the transit models presented by \cite{2002ApJ...580L.171M} and \cite{2006A&A...450.1231G}. We assumed a quadratic limb-darkening law and estimated the prior constraints on the stellar limb-darkening coefficients $u_1$ and $u_2$ using the \texttt{LDTK} Python package \citep{2015MNRAS.453.3821P} with input stellar parameters ($T_{\rm eff}$, [M/H], and $\log g$) from the literature values (Table~\ref{table:stellar_parameters}). In broadband light-curve fitting, the free transit parameters were the planet-to-host-star radius ratio $R_{\rm p}/R_{\rm s}$, the central transit time $T_{\rm c}$, the semimajor axis relative to the host star radius $a/R_{\rm s}$, the inclination $i$, and the quadratic limb-darkening coefficients $u_1$ and $u_2$. We assumed uniform priors for $R_{\rm p}/R_{\rm s}$ and $T_{\rm c}$, Gaussian priors for $a/R_{\rm s}$ and $i$ based on the literature estimates, and Gaussian priors for $u_1$ and $u_2$ based on the \texttt{LDTK} estimates. The orbital period $P$ was fixed to the literature value, and circular orbits were assumed for all targets. Table~\ref{table:lc_priors} presents the input priors and values of the transit parameters for broadband light-curve fitting. In narrowband light-curve fitting, the free transit parameters were the wavelength-dependent transit parameters ($R_{\rm p}/R_{\rm s}$, $u_1$ and $u_2$), while the wavelength-independent parameters ($T_{\rm c}$, $a/R_{\rm s}$, and $i$) were fixed to the best-fit values derived from the broadband light-curve fitting. 

    \begin{table*}[htbp]
    \caption{ Input parameters and derived broadband limb darkening coefficients of the host stars.}  
    \label{table:stellar_parameters} 
    \small
    \centering    
    \begin{tabular}{l l l l l l l l c }  
    \hline\hline 
    Star & $T_{\rm eff}$ (K) & [M/H] & $M_{\rm s}$ ($M_\odot$) 
        & $R_{\rm s}$ ($R_\odot$) & $\log g_{\rm s}$ (cgs) \
        & $u_1$ \tablefootmark{a} & $u_2$ \tablefootmark{a} & Ref.\\  
    \hline 
    CoRoT-1     & $5950 \pm 150$ 
                & $-0.30\pm 0.25$ 
                & $0.95 \pm 0.15$ 
                & $1.11 \pm 0.05$ 
                & $4.25 \pm 0.30$ 
                & $0.4599\pm0.0030$
                & $0.1494\pm0.0048$
                & [1]\\
    HAT-P-18    & $4803 \pm 80$ 
                & $0.10 \pm 0.08$ 
                & $0.770 \pm 0.031$ 
                & $0.749 \pm 0.037$ 
                & $4.57 \pm 0.04$ 
                & $0.6094\pm0.0022$
                & $0.0994\pm0.0031$
                & [2]\\
    HAT-P-57    & $6330 \pm 124$ 
                & $-0.25 \pm 0.25$ 
                & $2.77 \pm 1.74$ 
                & $1.85 \pm 0.39$ 
                & $4.25 \pm 0.02$ 
                & $0.4183\pm0.0016$
                & $0.1587\pm0.0027$
                & [3]\\
    Qatar-1     & $5013 \pm 90$ 
                & $0.17 \pm 0.10$ 
                & $0.84 \pm 0.04$ 
                & $0.803 \pm 0.016$
                & $4.55 \pm 0.01$ 
                & $0.5412\pm0.0029$
                & $0.1195\pm0.0045$
                & [4]\\
    TrES-4      & $6200 \pm 75$ 
                & $0.14 \pm 0.09$ 
                & $1.08 \pm 0.38$ 
                & $1.66 \pm 0.19$ 
                & $4.06 \pm 0.02$ 
                & $0.4479\pm0.0013$
                & $0.1446\pm0.0019$
                & [3]\\
    WASP-2      & $5170 \pm 60$ 
                & $0.04 \pm 0.05$ 
                & $0.851 \pm 0.050$ 
                & $0.823 \pm 0.018$ 
                & $4.54 \pm 0.02$ 
                & $0.5119\pm0.0015$
                & $0.1339\pm0.0022$
                & [5] \\
    WASP-10     & $4675 \pm 100$ 
                & $0.03 \pm 0.20$ 
                & $0.75 \pm 0.04$ 
                & $0.698 \pm 0.012$ 
                & $4.63 \pm 0.01$ 
                & $0.6464\pm0.0033$
                & $0.0783\pm0.0044$
                & [6] \\
    WASP-32     & $6100 \pm 100$ 
                & $-0.13 \pm 0.10$ 
                & $1.10 \pm 0.03$ 
                & $1.11 \pm 0.05$ 
                & $4.39 \pm 0.03$ 
                & $0.4202\pm0.0015$
                & $0.1680\pm0.0023$
                & [7]\\
    WASP-36     & $5959 \pm 134$ 
                & $-0.26 \pm 0.10$ 
                & $1.081 \pm 0.034$ 
                & $0.985 \pm 0.014$ 
                & $4.49 \pm 0.01$ 
                & $0.4173\pm0.0019$
                & $0.1742\pm0.0030$
                & [8]\\
    WASP-39     & $5485 \pm 50$ 
                & $0.01 \pm 0.09$ 
                & $0.913 \pm 0.047$ 
                & $0.939 \pm 0.022$ 
                & $4.45 \pm 0.02$ 
                & $0.5527\pm0.0017$
                & $0.1279\pm0.0024$
                & [9]\\
    WASP-49     & $5600 \pm 160$ 
                & $-0.23 \pm 0.07$ 
                & $1.003 \pm 0.100$ 
                & $1.038 \pm 0.037$ 
                & $4.41 \pm 0.02$ 
                & $0.4527\pm0.0033$
                & $0.1584\pm0.0050$
                & [10] \\
    WASP-156    & $4910 \pm 61$ 
                & $0.24 \pm 0.12$ 
                & $0.842 \pm 0.052$ 
                & $0.76 \pm 0.03$ 
                & $4.60 \pm 0.06$ 
                & $0.5613\pm0.0021$
                & $0.1139\pm0.0032$
                & [11] \\    
    \hline  
    \end{tabular}

    \tablefoot{
        \tablefoottext{a}{Quadratic limb-darkening coefficients for broadband light-curve fitting.}
    }
    \tablebib{
        [1] \cite{2008A&A...482L..17B} [2] \cite{2011ApJ...726...52H} 
        [3] \cite{2017AJ....153..136S} [4] \cite{2017AJ....153...78C} 
        [5] \cite{2012MNRAS.426.1291S} [6] \cite{2009ApJ...692L.100J} 
        [7] \cite{2010PASP..122.1465M} [8] \cite{2016MNRAS.459.1393M} 
        [9] \cite{2018A&A...613A..41M} [10] \cite{2016A&A...587A..67L} 
        [11] \cite{2018A&A...610A..63D}.
    }
    \end{table*}
    
    \begin{table*}[htbp]
        \caption{ Input physical and orbital parameters of the planets.}  
        \label{table:lc_priors} 
        \small
        \centering    
        \begin{tabular}{l l l l l l l l c}  
        \hline\hline 
        Planets & $M_{\rm p}$ ($M_{\rm J}$) & $R_{\rm p}$ ($R_{\rm J}$) & $T_{\rm eq}$ (K) & $T_0$ ($\rm BJD_{TDB}$) \tablefootmark{a} & $P$ (day) & $a/R_{\rm s}$ & $i$ (deg) & Ref. \\  
        \hline 
        CoRoT-1b    & $1.03 \pm 0.12$
                    & $1.49 \pm 0.08$
                    & $1898 \pm 50$
                    & 4159.4532
                    & 1.5089557
                    & $4.92 \pm 0.08$ 
                    & $85.1 \pm 0.5$ 
                    & [1]\\
        HAT-P-18b   & $0.197 \pm 0.013$
                    & $0.995 \pm 0.052$
                    & $852 \pm 28$
                    & 4715.02174
                    & 5.508023
                    & $16.04 \pm 0.75$
                    & $88.8 \pm 0.3$
                    & [2] \\
        HAT-P-57b   & $1.41 \pm 1.52$
                    & $1.74 \pm 0.36$
                    & $1855 \pm 40$ \tablefootmark{b}
                    & 5113.48127
                    & 2.465300
                    & $5.82 \pm 0.09$
                    & $88.26 \pm 0.85$
                    & [3] \\
        Qatar-1b    & $1.294 \pm 0.050$
                    & $1.143 \pm 0.026$
                    & $1418 \pm 28$
                    & 6234.10322
                    & 1.42002420
                    & $6.25 \pm 0.07$
                    & $84.08 \pm 0.16$
                    & [4] \\
        TrES-4b     & $0.78 \pm 0.19$
                    & $1.61 \pm 0.18$
                    & $1785 \pm 40$ \tablefootmark{b}
                    & 4230.90530
                    & 3.553950 
                    & $6.04 \pm 0.23$
                    & $82.81 \pm 0.33$
                    & [3] \\ 
        WASP-2b     & $0.880 \pm 0.038$
                    & $1.063 \pm 0.028$
                    & $1286 \pm 17$
                    & 3991.51455
                    & 2.15222144
                    & $8.08 \pm 0.12$ 
                    & $84.81 \pm 0.17$ 
                    & [5] \\
        WASP-10b    & $3.15 \pm 0.12$
                    & $1.080 \pm 0.020$
                    & $1370 \pm 50$
                    & 4664.03091
                    & 3.0927616
                    & $11.65 \pm 0.11$
                    & $88.49 \pm 0.20$
                    & [6] \\
        WASP-32b    & $3.60 \pm 0.07$
                    & $1.18 \pm 0.07$
                    & $1560 \pm 50$
                    & 5151.0546
                    & 2.718659
                    & $7.8 \pm 0.3$ 
                    & $85.3 \pm 0.5$ 
                    & [7] \\ 
        WASP-36b    & $2.361 \pm 0.070$
                    & $1.327 \pm 0.021$
                    & $1733 \pm 19$
                    & 5569.83771
                    & 1.53736596
                    & $5.85 \pm 0.06$ 
                    & $83.15 \pm 0.13$ 
                    & [8] \\
        WASP-39b    & $0.281 \pm 0.032$
                    & $1.279 \pm 0.040$
                    & $1166 \pm 14$
                    & 5342.96913
                    & 4.0552941
                    & $11.07 \pm 0.17$
                    & $87.32 \pm 0.17$
                    & [9] \\
        WASP-49b    & $0.396 \pm 0.026$
                    & $1.198 \pm 0.046$
                    & $1399 \pm 41$
                    & 6267.68389
                    & 2.7817362
                    & $8.01 \pm 0.20$
                    & $84.48 \pm 0.13$
                    & [10] \\                                  
        WASP-156b   & $0.128 \pm 0.010$
                    & $0.51 \pm 0.02$
                    & $970 \pm 25$
                    & 4677.707
                    & 3.836169
                    & $12.8 \pm 0.5$
                    & $89.1\pm 0.8$
                    & [11] \\       
        \hline  
        \end{tabular}
        
        \tablefoot{
            \tablefoottext{a}{Initial transit epochs subtracting 2\,450\,000.}
            \tablefoottext{b}{Black-body equilibrium temperature calculated with the input stellar effective temperature and normalized semimajor axis ($T_{\rm eq} = T_{\rm eff} \left(R_{\rm s}/2a\right)^{1/2} \left[f(1-A_{\rm B})\right]^{1/4}$, \citealt{2005ApJ...626..523C}) assuming a Bond albedo of 0 and uniform heat redistribution ($f=1$).}
        }
        \tablebib{ 
            [1] \cite{2008A&A...482L..17B} [2] \cite{2011ApJ...726...52H} 
            [3] \cite{2017AJ....153..136S} [4] \cite{2017AJ....153...78C} 
            [5] \cite{2012MNRAS.426.1291S} [6] \cite{2009ApJ...692L.100J} 
            [7] \cite{2010PASP..122.1465M} [8] \cite{2016MNRAS.459.1393M} 
            [9] \cite{2018A&A...613A..41M} [10] \cite{2016A&A...587A..67L} 
            [11] \cite{2018A&A...610A..63D}.
        }
    \end{table*}
    
    Some of the target stars were reported to have companion or background stars (HAT-P-57, TrES-4, WASP-2, WASP-36b, and WASP-49b; see Appendix \ref{sect:companions} for more information). If a contamination star was included in the photometric aperture, the flux dilution effect would result in slight underestimations of transit depths. Consequently, a false slope might appear in the derived transmission spectrum if the chromatic flux ratio between the contamination star and the target star have a strong wavelength dependency. We followed several procedures to correct the effects of flux dilution. First, we estimated the stellar parameters of the companion or background stars based on their spectral types determined by previous research papers (TrES-4, \citealt{2015A&A...575A..23W}; WASP-2, \citealt{2020A&A...635A..73B}). We took values from the stellar color and effective temperature sequence provided by \cite{2013ApJS..208....9P} as approximation. Next, we generated model spectra of the target and the companion stars based on their stellar parameters ($T_{\rm eff}$, $\log g$, and [M/H]) via grid interpolation of the PHOENIX stellar atmosphere models \citep{2013A&A...553A...6H}. Then, we rescaled the stellar spectra in the OSIRIS passbands based on the measured flux ratio reported in previous literature (e.g., $\Delta i'~{\rm mag}=4.49\pm0.09$ for TrES-4 and its companion star; \citealt{2015A&A...575A..23W}). We also used Monte Carlo sampling to achieve error propagation from the uncertainties of stellar parameters to those of flux ratios. The obtained mean values and uncertainties were used for the Gaussian priors of flux ratios in the light-curve fitting for each passband. Finally, we corrected the flux dilution effect  with Eq. \ref{eq_dilution}:
    \begin{equation}\label{eq_dilution}
    f(t) = \frac{f^*(t) + \mathcal{F}}{1 + \mathcal{F}},
    \end{equation}
    where $\mathcal{F}$ is the companion-to-target flux ratio in the corresponding passband, and $f^*(t)$ is the transit light-curve model without flux dilution.
    
    We used Gaussian process (GP) to account for the noise components in the light-curve fitting, which was first introduced to retrieve robust transit models by \cite{2012MNRAS.419.2683G}. \cite{2022A&A...664A..50J} investigated the reliability and accuracy of GP regression in realistic spectroscopic light-curve analyses, and they considered GP regression in practice as a direct and effective way to characterize systematic noise. In this work, we follow the methods in \cite{2022A&A...664A..50J} to perform GP regression. We use the code \texttt{george} \citep{2015ITPAM..38..252A} to compute GP components, which allows multi-input GPs in light-curve modeling. For most of the targets, we adopted the vectors of time and seeing variation (characterized by the full width at half maximum, FWHM, of the stellar spectrum along the spatial direction at the central wavelength of each passband) as GP input vectors. We noticed that some of the light curves were largely affected by the rapid change of instrumental rotation angles. Therefore, for the light curves of HAT-P-18b, TrES-4b, WASP-10b, we also added the vector of instrumental rotation angles as GP inputs. The GP kernel functions were implemented by combining multiple 3/2-order Mat\'ern kernels \citep{2006gpml.book.....R}:
    \begin{equation} \label{eq:kernel_tw}
        k(\boldsymbol{r}) = \sigma_{\rm k}^2 \cdot k_{\rm M32}(r_t) 
        \cdot k_{\rm M32}(r_w),\\
        r_x = \frac{x_i - x_j}{\ell_x},
    \end{equation}
    or 
    \begin{equation} \label{eq:kernel_twr}
        k(\boldsymbol{r}) = \sigma_{\rm k}^2 \cdot k_{\rm M32}(r_t) 
        \cdot k_{\rm M32}(r_w) \cdot k_{\rm M32}(r_\alpha),
    \end{equation}
    where $\sigma_{\rm k}^2$ is the variance, $k_{\rm M32}(r)$ is the 3/2-order Mat\'ern function provided by \texttt{george}, $r_x$ is the normalized distance between two data points $x_i$ and $x_j$, $\ell_x$ is the length scale, the subscripts $t$, $w$, and $\alpha$ refer to the time, the seeing variation, and the rotation angle, respectively. A jitter term with a variance of $\sigma_{\rm n}^2$ was then added to the diagonal of the covariance matrix calculated by the kernel function to account for the underestimation of white noise. The GP parameters $\sigma_{\rm n}^2$, $\sigma_{\rm k}^2$, $\ell_t$, $\ell_w$, and $\ell_\alpha$ are free parameters in the light-curve fitting. We have also tried other possible combinations of GP kernels, although Eqs. \ref{eq:kernel_tw} and \ref{eq:kernel_twr} were considered as the best GP models according to the model comparison of Bayesian evidence.

    \subsection{Bayesian framework} \label{sect:BayesianFramework}
    
    We used the nested sampling algorithm \citep{2004AIPC..735..395S} to conduct Bayesian parameter estimation and model comparison, which is implemented by the code \texttt{PyMultiNest} \citep{2009MNRAS.398.1601F, 2014A&A...564A.125B}. The nested sampling algorithm is designed for efficient estimation of Bayesian evidence (marginalized likelihood):
    \begin{equation}
        \mathcal{Z(\boldsymbol{D}|H)} = \int \mathcal{L}(\boldsymbol{D}| \boldsymbol{\Theta},\mathcal{H})
        \pi (\boldsymbol{\Theta}|\mathcal{H}) \mathrm{d}^m\boldsymbol{\Theta},
    \end{equation}
    where $\boldsymbol{D}$ is the observed data, $\mathcal{H}$ is the model hypothesis, $\mathcal{L}$ is the likelihood, $\pi$ is the prior function, $\boldsymbol{\Theta}$ is the parameter vector, and $m$ is the model dimensionality. The posterior distributions of the parameters can be generated from the full sequence of nested sampling points. \texttt{PyMultiNest} computes the natural logarithmic Bayesian evidence $\ln \mathcal{Z}$. To compare two model hypotheses $\mathcal{H}_1$ and $\mathcal{H}_2$, we compute the difference of natural logarithmic evidence of the two hypotheses: $\Delta\ln \mathcal{Z}_{12}= \ln (\mathcal{Z}_1/\mathcal{Z}_2)$, which is equivalent to the logarithm of the Bayes factor. A decisive model preference is determined when $|\Delta\ln \mathcal{Z}| > 5$ according to the criteria proposed by \cite{1995JASA...90..773K}. In all fittings presented in this work, we used 1000 live points in the nested sampling to reach an evidence precision of $\sim$0.05. 
    
    \subsection{Results of light-curve fitting}

    \begin{figure*}[htbp]
        \centering
        \includegraphics[width=\linewidth]{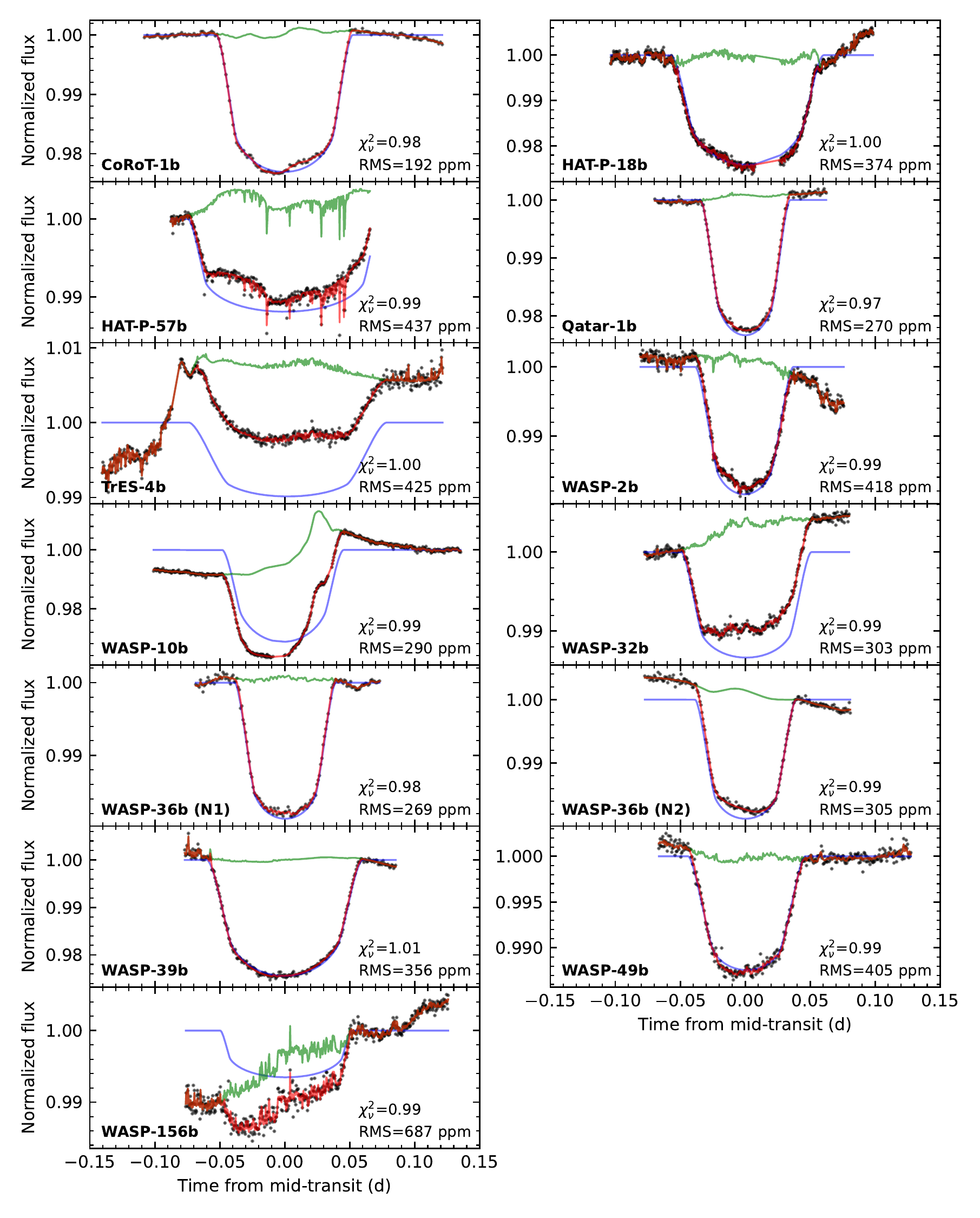}
        \caption{Best-fit white-light curves of all targets. The black dots are the observed data. The red lines are the best-fit curves. The green lines are the extracted GP systematics. The blue lines are the derived transit models.}
        \label{fig:lightcurves}
    \end{figure*}

    \begin{table*}[htbp]
        \caption{ Posterior estimates of transit parameters of all target planets in broadband light-curve fitting.}  
        \label{table:transit_posteriors} 
        \renewcommand\arraystretch{1.2}
        \centering    
        \begin{tabular}{l c c c c c c c}  
        \hline\hline 
        Planets     & $R_{\rm p}/R_{\rm s}$  & $a/R_{\rm s}$ & $i$ (deg) 
                    & $T_{\rm c}$ ($\rm BJD_{TDB}$) \tablefootmark{a} & $u_1$ & $u_2$ & $\mathcal{F}$ \tablefootmark{b} \\  
        \hline
        CoRoT-1b    & $0.1408^{+0.0017}_{-0.0013}$
                    & $4.96^{+0.04}_{-0.04}$
                    & $85.34^{+0.24}_{-0.24}$
                    & $8129.55008^{+0.00015}_{-0.00014}$
                    & $0.4595^{+0.0030}_{-0.0029}$
                    & $0.1488^{+0.0046}_{-0.0046}$ 
                    & - \\
        HAT-P-18b   & $0.1410^{+0.0038}_{-0.0042}$
                    & $16.05^{+0.25}_{-0.23}$
                    & $88.61^{+0.13}_{-0.12}$
                    & $7507.59488^{+0.00056}_{-0.00073}$
                    & $0.6094^{+0.0022}_{-0.0021}$
                    & $0.0994^{+0.0031}_{-0.0030}$ 
                    & - \\
        HAT-P-57b   & $0.1013^{+0.0055}_{-0.0049}$
                    & $5.83^{+0.05}_{-0.05}$
                    & $88.90^{+0.72}_{-0.64}$
                    & $8700.48541^{+0.00048}_{-0.00056}$
                    & $0.4183^{+0.0016}_{-0.0017}$
                    & $0.1586^{+0.0026}_{-0.0027}$ 
                    & $0.0289^{+0.0001}_{-0.0001}$\\
        Qatar-1b    & $0.1474^{+0.0017}_{-0.0016}$
                    & $6.30^{+0.04}_{-0.04}$
                    & $84.13^{+0.08}_{-0.09}$
                    & $6847.55351^{+0.00009}_{-0.00009}$
                    & $0.5409^{+0.0028}_{-0.0028}$
                    & $0.1191^{+0.0044}_{-0.0045}$
                    & - \\
        TrES-4b     & $0.1004^{+0.0056}_{-0.0055}$
                    & $5.99^{+0.16}_{-0.17}$
                    & $82.70^{0.21}_{-0.21}$
                    & $7571.59992^{+0.00277}_{-0.00490}$
                    & $0.4479^{+0.0013}_{-0.0013}$
                    & $0.1446^{+0.0018}_{-0.0018}$  
                    & $0.0136^{+0.0011}_{-0.0011}$\\ 
        WASP-2b     & $0.1366^{+0.0029}_{-0.0029}$
                    & $8.02^{+0.06}_{-0.06}$
                    & $84.81^{+0.08}_{-0.08}$
                    & $6875.49209^{+0.00020}_{-0.00018}$
                    & $0.5119^{+0.0015}_{-0.0014}$
                    & $0.1339^{+0.0022}_{-0.0021}$  
                    & $0.0250^{+0.0020}_{-0.0020}$\\
        WASP-10b    & $0.1588^{+0.0039}_{-0.0040}$
                    & $11.73^{+0.09}_{-0.10}$
                    & $88.36^{+0.13}_{-0.13}$
                    & $7617.59323^{+0.00024}_{-0.00024}$
                    & $0.6465^{+0.0033}_{-0.0032}$
                    & $0.0786^{+0.0042}_{-0.0044}$ 
                    & - \\
        WASP-32b    & $0.1119^{+0.0023}_{-0.0023}$
                    & $7.88^{+0.08}_{-0.09}$
                    & $85.29^{+0.12}_{-0.12}$
                    & $6586.51016^{+0.00017}_{-0.00018}$
                    & $0.4202^{+0.0014}_{-0.0015}$
                    & $0.1679^{+0.0022}_{-0.0023}$ 
                    & - \\
        WASP-36b (N1)    & $0.1341^{+0.0019}_{-0.0015}$
                    & $5.84^{+0.03}_{-0.03}$
                    & $83.41^{+0.07}_{-0.07}$
                    & $7039.55971^{+0.00016}_{-0.00015}$
                    & $0.4170^{+0.0018}_{-0.0018}$
                    & $0.1736^{+0.0029}_{-0.0028}$  
                    & $0.0102^{+0.0005}_{-0.0005}$\\
        WASP-36b (N2)    & -
                    & -
                    & -
                    & $7042.63429^{+0.00015}_{-0.00014}$
                    & -
                    & - 
                    & - \\
        WASP-39b    & $0.1444^{+0.0020}_{-0.0026}$
                    & $11.31^{+0.08}_{-0.09}$
                    & $87.64^{+0.08}_{-0.10}$
                    & $6348.67915^{+0.00016}_{-0.00014}$
                    & $0.5526^{+0.0017}_{-0.0017}$
                    & $0.1277^{+0.0024}_{-0.0023}$ 
                    & - \\
        WASP-49b    & $0.1128^{+0.0016}_{-0.0017}$
                    & $7.97^{+0.07}_{-0.07}$
                    & $84.45^{+0.08}_{-0.08}$
                    & $6612.62024^{+0.00019}_{-0.00019}$
                    & $0.4529^{+0.0032}_{-0.0033}$
                    & $0.1587^{+0.0049}_{-0.0049}$  
                    & $0.0080^{+0.0008}_{-0.0008}$\\
        WASP-156b   & $0.0730^{+0.0037}_{-0.0037}$
                    & $12.45^{+0.41}_{-0.41}$
                    & $88.56^{+0.66}_{-0.46}$
                    & $8030.52097^{+0.00036}_{-0.00037}$
                    & $0.5613^{+0.0022}_{-0.0023}$
                    & $0.1139^{+0.0031}_{-0.0032}$ 
                    & - \\
        \hline  
        \end{tabular}
        
        \tablefoot{
            \tablefoottext{a}{Central transit time subtracting 2\,450\,000.}
            \tablefoottext{b}{Companion-to-host star flux ratio.}
        } 
    \end{table*}  
    
    The best-fit white-light curves are shown in Fig.~\ref{fig:lightcurves}. The derived transit parameters of all target planets are listed in Table~\ref{table:transit_posteriors}. We show the posterior distribution of the transit parameters for CoRoT-1b in Fig.~\ref{fig:corners_white_c1b}, while those of the other planets are available at {\tt ScienceDB} \footnote{https://cstr.cn/31253.11.sciencedb.08305}. The posterior estimates of the transit parameters are consistent with literature values. As shown in Fig.~\ref{fig:lightcurves}, the light curves of HAT-P-18b, HAT-P-57b, TrES-4b, WASP-10b, and WASP-156b were considerably affected by large amplitudes of systematic noise, where those of HAT-P-18b, TrES-4b, and WASP-10b could be well characterized by adding a GP component of instrumental rotation angles. The light curves of HAT-P-57b and WASP-156b were mostly affected by seeing variation and their extracted systematics are highly correlated with the vector of FWHM, while the multi-input GP models considering time and FWHM could still extract the transit signals and the systematic noise correctly.
    
    We did not remove the common-mode patterns for narrowband light curves because typical common-mode reduction methods such as the ``divide-white'' method might underestimate the uncertainties of chromatic transit depths \citep{2022A&A...664A..50J}. The resulting chromatic transit depths of all target planets are listed in Table~\ref{table:ts_R1000R1} (CoRoT-1b, HAT-P-18b, HAT-P-57b, Qatar-1b, and TrES-4b), \ref{table:ts_R1000R2} (WASP-2b, WASP-32b, WASP-36b, WASP-49b, and WASP-156b), and \ref{table:ts_R1000B} (WASP-10b and WASP-39b). 

    To ensure the consistency of our results when analyzing data for WASP-36b, we jointly fitted the light curves from two nights of observations. We simultaneously fitted the two light curves in the same waveband with the same transit parameters (except for $T_{\rm c}$) but different GP parameters. The total likelihood function was calculated as the sum of the log-likelihoods of the two fits. In addition, we also verified whether the obtained transmission spectra were consistent when fitting the light curves separately for each night, and we found that the obtained spectra agreed well in the bluer part, while there were 1- to 2-$\sigma$ differences near the red end. Given that WASP-36 is a metal-poor G2 dwarf with low stellar activity \citep{2012AJ....143...81S}, it is less likely that the host star activity would significantly affect the transmission spectra of WASP-36b and cause variations at the red end. Therefore, it is reasonable to fit both data sets of WASP-36b together to obtain better constraints.        

\section{Atmospheric retrieval}
\label{sect:method_retrieval}

    \subsection{Model setups}
    
    We use the Python package \texttt{petitRADTRANS} (hereafter \texttt{pRT}; \citealt{2019A&A...627A..67M}) to model the transmission spectra of exoplanet atmospheres. We assumed one-dimensional atmospheres with isothermal atmospheric profiles. The opacities of different atmospheric layers are contributed by gas absorption, collision-induced absorption, and Rayleigh-like scattering. We used the equilibrium chemistry model and the free chemistry model to separately determine the corresponding chemical abundances of gas absorption. When assuming chemical equilibrium, \texttt{pRT} uses 4D linear interpolation to obtain the mass fractions of 20 species ($\rm H_2$, $\rm H$, $\rm He$, $\rm H_2O$, $\rm H_2S$, $\rm HCN$, $\rm CH_4$, $\rm C_2H_2$, $\rm CO$, $\rm CO_2$, $\rm NH_3$, $\rm PH_3$, $\rm SiO$, $\rm TiO$, $\rm VO$, $\rm Na$, $\rm K$, $\rm FeH$, $\rm e^-$, $\rm H^-$; see Table~\ref{table:line_lists} for reference) over the model grids precalculated with \texttt{easyCHEM} \citep{2017ApJ...850..150B, 2017A&A...600A..10M}. The grid dimensions are temperature ($T\in[60, 4000]$~K with 100 equidistant points), pressure ($P\in[10^{-8}, 10^3]$~bar with 100 equidistant points in log space), metallicity ($\rm [M/H]\in[-2, 3]$ with 40 equidistant points) and carbon-to-oxygen ratio ($\rm C/O\in[0.1, 1.6]$ with 20 equidistant points). The collision-induced absorption consisted of the continuum opacities of $\rm H_2$--$\rm H_2$ and $\rm H_2$--$\rm He$ pairs. The scattering features were characterized by Rayleigh scattering opacities of $\rm H_2$ and $\rm He$ scaled with a free parameter ($f_{\rm haze}$). The model also considered a uniform and opaque cloud deck with a cloud-top pressure of $P_{\rm c}$, under which the atmosphere is completely obscured. The transmission spectra were first computed at a spectral resolution of $\lambda/\Delta\lambda=1000$ and then rebinned to the same passbands as the observed data.
    
    We used the same Bayesian framework illustrated in Sect.~\ref{sect:BayesianFramework} to fit the atmospheric models and retrieve the posterior parameters and model evidence. To evaluate the absorption signatures in each transmission spectrum, we performed model comparison with the following four hypotheses: 
    $\mathcal{H}_0$, a null model with no atmosphere (a flat spectrum); 
    $\mathcal{H}_1$, an atmosphere with only cloud and scattering features; 
    $\mathcal{H}_2$, an atmosphere with cloud, scattering, and gas absorption features assuming equilibrium chemical abundances; 
    $\mathcal{H}_3$, an atmosphere with cloud, scattering, and gas absorption features assuming free chemical abundances.
    The null hypothesis ($\mathcal{H}_0$) has only one free parameter: the planet radius ($R_p$). The scattering-only hypothesis ($\mathcal{H}_1$) has six free parameters: the reference pressure ($P_0$), the planet radius ($R_{\rm p}$) at $P_0$, the planet mass ($M_{\rm p}$), the atmospheric temperature ($T$), the cloud-top pressure ($P_{\rm c}$), and the scaling factor of Rayleigh scattering ($f_{\rm haze}$). The equilibrium chemistry hypothesis ($\mathcal{H}_2$) includes all the parameters in $\mathcal{H}_1$, plus two parameters, the metallicity ([M/H]) and the carbon-to-oxygen ratio (C/O), for determining the chemical abundances. The free chemistry hypothesis ($\mathcal{H}_3$) includes all the parameters in $\mathcal{H}_1$, plus the mass fractions of the major species ($\rm Na$, $\rm K$, and $\rm H_2O$), keeping a total mass fraction of unity and a $\rm He/H_2$ mass ratio of 0.3. For the planets with higher equilibrium temperatures (CoRoT-1b, HAT-P-57b, TrES-4b, WASP-32b, WASP-36b, and WASP-49b), the optical absorbers TiO and VO were also considered in their free chemistry models. Uninformative priors were adopted for all the free parameters except for $R_{\rm p}$ and $M_{\rm p}$, which were constrained by the Gaussian priors based on the literature values in Table~\ref{table:lc_priors}. We estimated the Bayesian evidence of $\mathcal{H}_0$, $\mathcal{H}_1$, $\mathcal{H}_2$, and $\mathcal{H}_3$ for each transmission spectrum using the nested sampling algorithm. We then compared the Bayesian evidence of $\mathcal{H}_1$, $\mathcal{H}_2$, and $\mathcal{H}_3$ with that of $\mathcal{H}_0$ to derive the model inferences using the criteria illustrated in Sect.~\ref{sect:BayesianFramework}.
    
    \subsection{Retrieval results}
    \label{sect:retrieval_results}
           
\begin{figure*}[htbp]
        \centering
        \includegraphics[width=\linewidth]{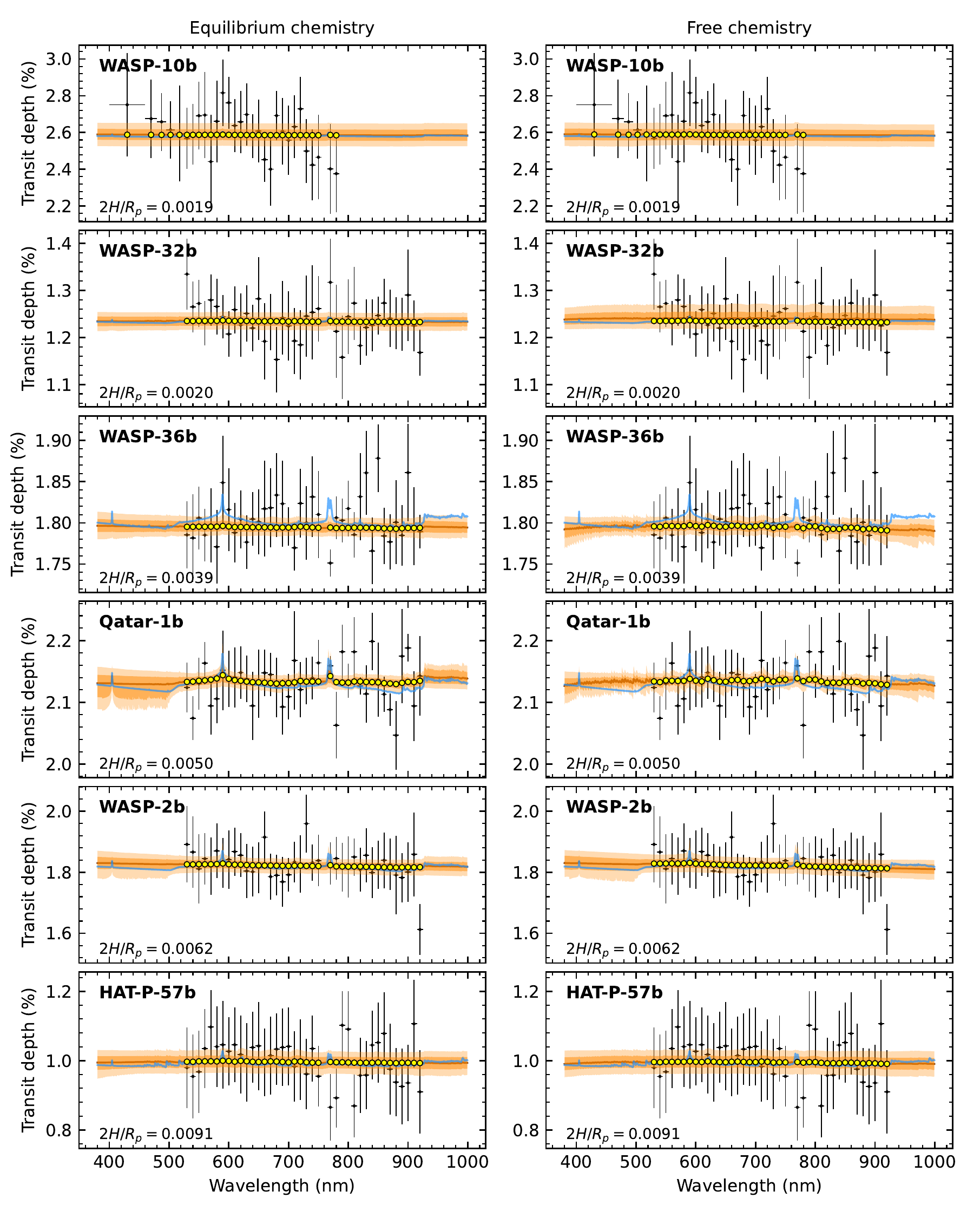}
        \caption{ Retrieval results of the transmission spectra of HAT-P-57b, Qatar-1b, WASP-2b, WASP-10b, and WASP-32b, assuming equilibrium chemistry (left column) and free chemistry (right column). The black points with error bars are the observed data, where the points in the oxygen A-band have been excluded (755 -- 765 nm). The orange lines and the shaded areas are the median, $1\sigma$-, and $2\sigma$-intervals of the retrieved spectra. The blue lines are the fiducial models assuming clear atmospheres with equilibrium temperature, $1\times$ solar metallicity, and a solar C/O of 0.53, and the planetary radii have been adjusted to match the observed data. The panels are sorted by the normalized scale heights ($2H/R_{\rm p}$) from top to bottom.}
        \label{fig:ts_grids_A}
    \end{figure*}

\begin{figure*}[htbp]
        \centering
        \includegraphics[width=\linewidth]{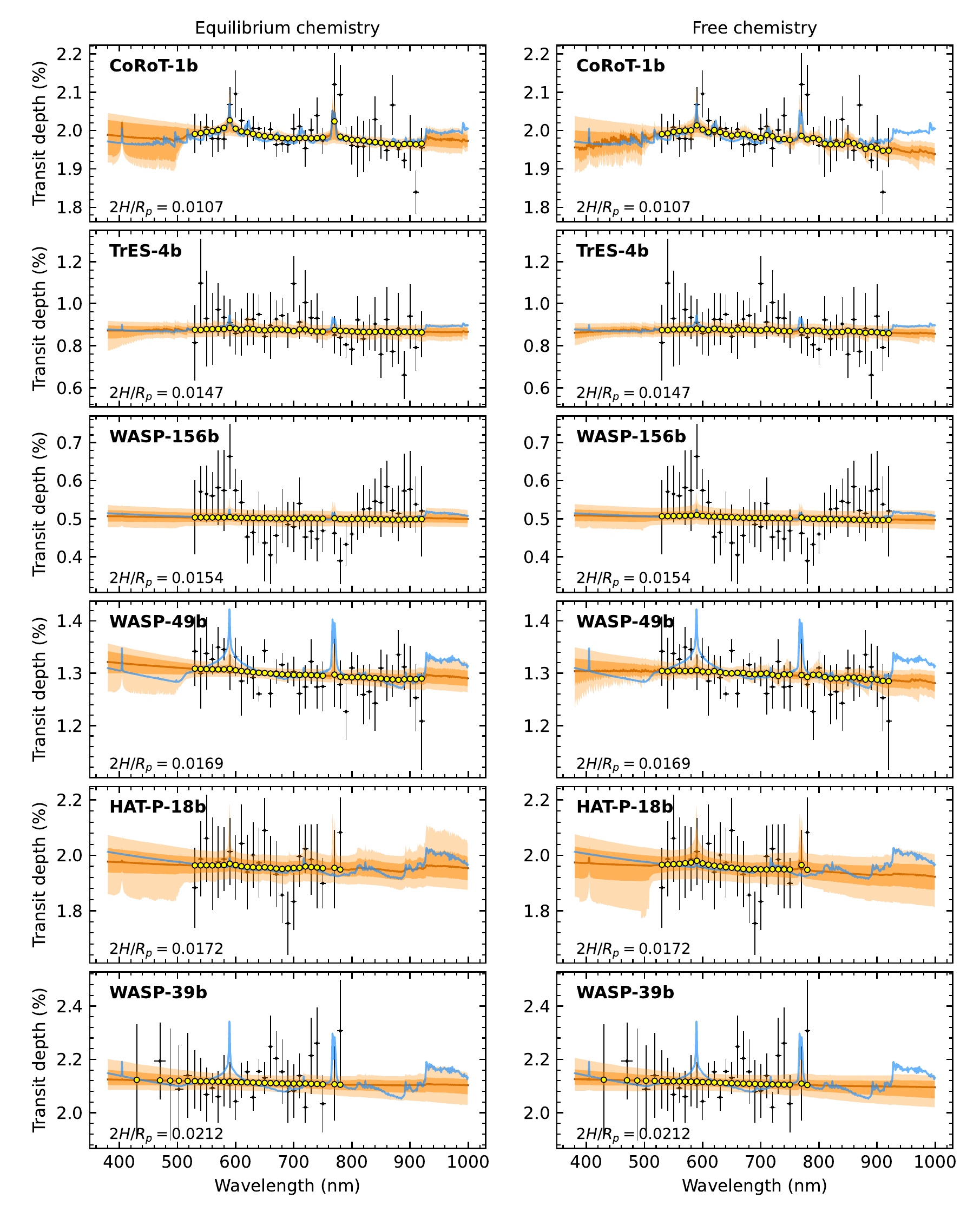}
        \caption{ Retrieval results of the transmission spectra of CoRoT-1b, HAT-P-18b, TrES-4b, WASP-39b, WASP-49b, and WASP-156b, assuming equilibrium chemistry (left column) and free chemistry (right column). The black points with error bars are the observed data, where the points in the oxygen A-band have been excluded (755 -- 765 nm). The orange lines and the shaded areas are the median, $1\sigma$-, and $2\sigma$-intervals of the retrieved spectra. The blue lines are the fiducial models assuming clear atmospheres with equilibrium temperature, $1\times$ solar metallicity, and a solar C/O of 0.53, and the planetary radii have been adjusted to match the observed data. The panels are sorted by the normalized scale heights ($2H/R_{\rm p}$) from top to bottom.}
        \label{fig:ts_grids_B}
    \end{figure*}

    We show the transmission spectra retrieved with the equilibrium chemistry models ($\mathcal{H}_2$) and the free chemistry models ($\mathcal{H}_3$) in Figs.~\ref{fig:ts_grids_A} and \ref{fig:ts_grids_B}, where we also present a fiducial transmission spectrum for each target assuming clear atmospheres with equilibrium temperature, $1\times$ solar metallicity, and a solar C/O of 0.53 for comparison. The posterior estimates of atmospheric retrievals are listed in Table~\ref{table:atm_posteriors_equilibrium} (equilibrium chemistry) and Table~\ref{table:atm_posteriors_free} (free chemistry). The joint distributions of the retrieved parameters for CoRoT-1b are shown in Figs.~\ref{fig:corners_retrievals_c1b_ec} and \ref{fig:corners_retrievals_c1b_fc}, while those for the other planets are available at {\tt ScienceDB} \footnote{https://cstr.cn/31253.11.sciencedb.08305}. According to Bayesian model comparison (Table~\ref{table:evidence}), most of the transmission spectra were found to be featureless, and there was no significant difference between the results from the equilibrium chemistry models and the free chemistry models, implying that the strong constraint of the chemical equilibrium assumption is not the primary cause of null inferences. The exoplanet CoRoT-1b was the only target with significant scattering features and possible absorption signatures of atomic Na and K. 
    
    \begin{table}[htbp]
    \caption{ Bayesian evidence of the atmospheric retrievals.}  
    \label{table:evidence} 
    \tiny
    \centering
    \begin{tabular}{l c c c c}  
    \hline\hline 
    Planets
    & $\ln\mathcal{Z}\left(\frac{\mathcal{H}_1}{\mathcal{H}_0}\right)$\tablefootmark{a}
    & $\ln\mathcal{Z}\left(\frac{\mathcal{H}_2}{\mathcal{H}_0}\right)$\tablefootmark{b}
    & $\ln\mathcal{Z}\left(\frac{\mathcal{H}_3}{\mathcal{H}_0}\right)$\tablefootmark{c}
    & Inferences \\
    \hline 
    CoRoT-1b  
        & $ 5.67 $ & $ 5.88 $ & $5.84$ &
        Strong features \\ 
    HAT-P-18b 
        & $ -0.25 $ & $ 0.04 $ & $-0.25$ &
        Featureless \\
    HAT-P-57b  
        & $ 0.11 $ & $ 0.12 $ & $0.11$ &
        Featureless \\
    Qatar-1b  
        & $ -0.29 $ & $ 0.54 $ & $0.33$ &
        Featureless \\
    TrES-4b 
        & $ 0.14 $ & $ 0.27 $ & $0.03$ &
        Featureless \\
    WASP-2b 
        & $ 0.75 $ & $ 0.67 $ & $0.79$ &
        Featureless \\
    WASP-10b  
        & $ 0.05 $ & $ -0.07 $ &  $-0.03$ &
        Featureless \\
    WASP-32b  
        & $ 0.15 $ & $ 0.18 $ & $0.11$ &
        Featureless \\
    WASP-36b 
        & $ -0.44 $ & $ -0.56 $ & $-0.82$ &
        Featureless \\
    WASP-39b 
        & $ 0.48 $ & $ 0.46 $ & $0.20$ &
        Featureless \\
    WASP-49b  
        & $ 0.52 $ & $ 0.42 $ & $-0.60$ &
        Featureless \\                                   
    WASP-156b  
        & $ 0.25 $ & $ 0.34 $ & $0.29$ &
        Featureless \\
    \hline  
    \end{tabular}
    
    \tablefoot{
        \tablefoottext{a}{Log-evidence of the scattering model against that of the null model.}
        \tablefoottext{b}{Log-evidence of the equilibrium chemistry model against that of the null model.}
        \tablefoottext{c}{Log-evidence of the free chemistry model against that of the null model.}
    }
    
    \end{table}
    
    We found that some of the retrieval results have considerable differences from the fiducial models. For instance, the fiducial model of CoRoT-1b exhibited the absorption signatures from both alkali metals and TiO at solar abundances due to its high equilibrium temperature ($T_{\rm eq}=1898\pm50~{\rm K}$, \citealt{2008A&A...482L..17B}). Although its retrieval results also indicate a relatively clear atmosphere with a cloud-top pressure of $0.67\pm1.70~\log_{10}{\rm bar}$ from equilibrium chemistry and $0.25\pm1.81~\log_{10}{\rm bar}$ from free chemistry, the equilibrium chemistry model prefers an atmosphere with a high C/O ($\sim$1.5), thus lower abundances and weaker features for TiO and VO. On the other hand, the free chemistry model showed bimodal mass fractions of TiO, suggesting that the far wings of alkali lines can also be fitted with the TiO absorption features. There are two ways to resolve this divergence. One is to verify the line core and wings of the sodium D-lines at higher resolution. On the other hand, we can use a broader wavelength coverage of the transmission spectra to better constrain the continuum spectrum of TiO and VO. We would further discuss these two tests for CoRoT-1b in Sect.~\ref{sect:individuals}. 

    The exoplanets WASP-36b, WASP-39b, and WASP-49b also showed different features between their fiducial transmission spectra and the retrieved spectra. All of their fiducial models indicate very strong absorption signatures of Na and K, but none were detected in the observed spectra. The cloud and haze parameters for WASP-36b remained weakly constrained, with the small scale height of its atmosphere and the low signal-to-noise ratio (S/N) of the observed spectra likely responsible for the nondetection. The exoplanet WASP-39b has a very large scale height and was therefore expected to show distinct alkali lines. Previous research has also reported the possible presence of atomic Na and/or K in its atmosphere \citep[e.g.,][]{2016ApJ...827...19F, 2016ApJ...832..191N, 2019AJ....158..144K, 2023Natur.614..670F, 2023Natur.614..659R}. However, our observation found no evidence for the alkali metals, but a very high-altitude cloud cover ($\log_{10} P_{\rm c} = -3.37 \pm2.45~\log_{10}{\rm bar}$). In the case of WASP-49b, the retrieval results showed a $\sim$$270\times$ enhanced Rayleigh-like scattering, which greatly obscured the alkali absorption lines. Although a hazy atmosphere was able to explain the muted atmospheric features of WASP-49b, the corresponding Bayesian evidence was not strong enough to obtain a decisive inference for a hazy atmosphere (Table~\ref{table:evidence}).  

    For WASP-156b, both the fiducial model and the retrieval results suggested weak atmospheric features. However, the observed spectrum showed very strange and unexplained bumps at both the blue and the red ends (520 -- 600 nm; 800 -- 900 nm). These two features could not be interpreted by any species provided by the \texttt{pRT} and were likely to originate from light-curve systematic noise (Fig.~\ref{fig:lightcurves}). 

    In our study of HAT-P-57b, we attempted to improve the constraint on its mass through atmospheric retrievals. The atmospheric scale height, which determines the amplitudes of spectral features in the planetary transmission spectra, is inversely proportional to the planetary mass. If the mass is small enough, the corresponding atmospheric features should be detectable unless obscured by high-altitude clouds. As such, the observed transmission spectra can help constrain the planetary mass to some extent, particularly when strong atmospheric features are present \citep[e.g., ][]{2013Sci...342.1473D, 2020ApJ...896..107C}. However, in the case of HAT-P-57b, we did not detect any atmospheric signals in its transmission spectrum. Due to model degeneracy between the planetary mass and radius, the mass of HAT-P-57b remains poorly constrained and its posterior estimate is close to the prior estimate. Further constraints on the atmospheric scale height and the mass of HAT-P-57b depend on more precise measurements of its atmospheric signatures. 

\section{Discussion}
\label{sect:discussion}

\subsection{Further investigation on individual targets} 
\label{sect:individuals}

    \subsubsection{Dubious spectral features of CoRoT-1b}

    \begin{figure*}[htbp]
        \centering
        \includegraphics[width=\linewidth]{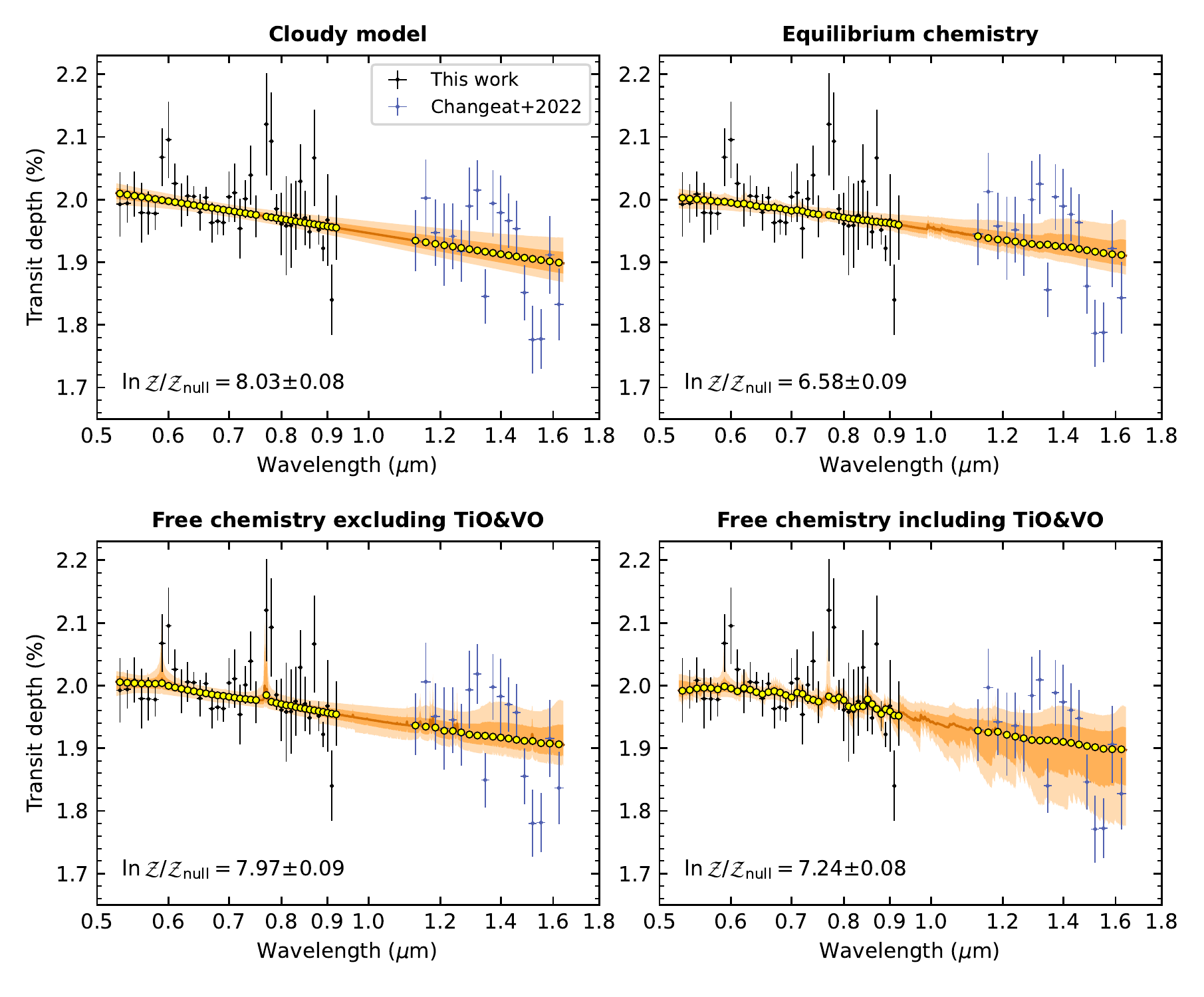}
        \caption{Retrieval results of the joint transmission spectra of CoRoT-1b using different model assumptions. The orange lines are the posteriors of the retrieved models. The yellow dots are models downsampled to the data passbands. The Bayesian evidence for each hypothesis is denoted as $\ln\mathcal{Z}$. The median offsets of the NIR spectra in the retrievals have been added to the error bars in each panel, which are $-92$ ppm (upper left), $10$ ppm (upper right), $-54$ ppm (lower left), and $-146$ ppm (lower right).}
        \label{fig:joint_retrieval_22}
    \end{figure*}  

    \cite{2014ApJ...785..148R} first studied the transits of CoRoT-1b and TrES-4b observed with the HST WFC3 instrument (G141 grism). They presented featureless NIR spectra and detected no clear source of opacities. The resulting spectra had a low S/R due to the staring observation mode. \cite{2022ApJS..260....3C}, \cite{2022AJ....164...19G}, and \cite{2022arXiv221100649E} reanalyzed the same NIR dataset and confirmed a NIR transmission spectrum with weak spectral features. Here we combine the NIR spectra published by \cite{2022ApJS..260....3C} with the OSIRIS optical spectrum to further constrain the atmospheric models of CoRoT-1b. The basic model setups of \texttt{pRT} were same as those in Sect.~\ref{sect:method_retrieval}. Additionally, a free transit depth offset $\delta$ following a uniform prior of $\mathcal{U}(-0.002, 0.002)$ was added to the WFC3 spectrum to compensate for the potential systematic biases caused by different light-curve analysis methods and systematic noise. We compared the following model hypotheses in the joint retrievals: 
    $\mathcal{H}_0$, a null model with no atmosphere (a flat spectrum); 
    $\mathcal{H}_1$, a cloudy and/or hazy atmosphere without absorption signatures; 
    $\mathcal{H}_2$, an atmosphere assuming equilibrium chemical abundances; 
    $\mathcal{H}_3$, an atmosphere assuming free chemical abundances with six species: Na, K, $\rm H_2O$, $\rm CH_4$, CO, and $\rm CO_2$; 
    $\mathcal{H}_4$, an atmosphere assuming free chemical abundances including all the species in $\mathcal{H}_3$, plus TiO and VO. 

    As shown in Fig.~\ref{fig:joint_retrieval_22}, the results of the joint retrievals were partially consistent with our previous analysis using only the OSIRIS spectrum in Fig.~\ref{fig:ts_grids_B}. Unlike the retrieval results of the equilibrium chemistry model from the optical transmission spectra, the joint spectra did not show significant alkali metal absorption signals in the retrieved models, but only strong scattering features. The spectrum retrieved with the equilibrium chemistry model was similar to that retrieved with a hazy model without gas absorption. In contrast, the free chemistry model emphasized the tentative spectral features in the joint data. However, different results were obtained when the optical absorbers (TiO and VO) were included or excluded in the model. When TiO and VO were excluded, the results showed weak absorption features of the alkali atoms Na and K. The corresponding volume mixing ratio of Na was $-5.02^{+2.41}_{-3.50}$ dex, approximately one order of magnitude lower than that obtained using the optical data alone, while the volume mixing ratio of K was $-5.49^{+2.12}_{-3.08}$ dex, slightly higher than that obtained using the optical data alone. When TiO and VO were included in the model, the spectral features of TiO and VO overwhelmed those of Na and K, resulting in unconstrained abundances of Na and K. However, both free chemistry models with or without TiO and VO explained the observed data equally well. The small differences in their Bayesian evidence suggest that all four scenarios can explain the observed data from a Bayesian model comparison perspective. The absence of water absorption features in the near-infrared band is an important reason for undistinguished chemistry models. 

\begin{figure}[htbp]
        \centering
        \includegraphics[width=\linewidth]{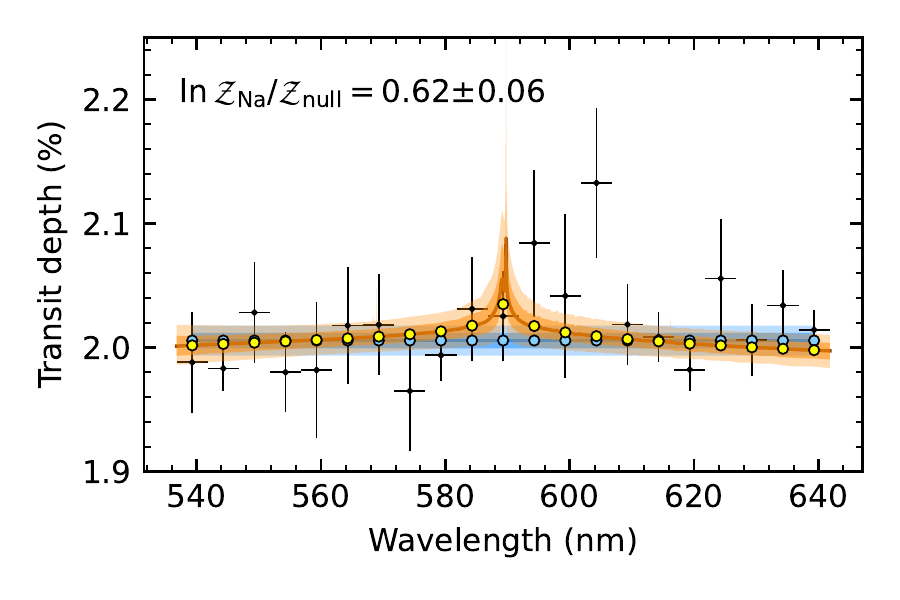}
        \caption{Transmission spectrum of CoRoT-1b zoomed into 536.8 -- 641.8 nm and centered at the sodium D-lines. The black points with error bars are the observed data. The orange lines are the posteriors of free chemistry models with presence of atomic Na, while the blue lines are those of null models. The dots correspond to the models after a wavelength binning of 5 nm.}
        \label{fig:verify_sodium}
    \end{figure}

    We further examined the narrower passbands from 536.8 nm to 641.8 nm with a wavelength binning of 5 nm to verify the sodium D-lines centered at $\sim$589.3 nm (Fig.~\ref{fig:verify_sodium}). We performed a free chemistry retrieval in the presence of atomic Na using a higher resolution line list with $\lambda/\Delta\lambda=10^4$ (downsampled from $\lambda/\Delta\lambda=10^6$). According to Bayesian model comparison, there was no strong evidence for the sodium D-lines ($\Delta \ln\mathcal{Z}/\mathcal{Z}_{\rm null} = 0.62 \pm 0.06$), indicating nondetection of sodium when using narrower passbands. The potassium D-lines were difficult to detect reliably due to their proximity to the oxygen A-band (758 -- 770 nm), resulting in lower S/N and stronger systematics in spectroscopic light curves. We found that the observed peaks had offsets of 5 -- 10 nm toward the red end from the line cores of Na and K, which could not be explained by any physical process from the planet or host star. The wavelength calibration in data reduction showed no anomalies that could cause such a large offset (Fig.~\ref{fig:stellar_spectra_c1b}). Therefore, the suspicious peaks in the OSIRIS spectrum of CoRoT-1b may be outliers or false signals. This suggests that when probing line features based on low-dispersion transmission spectroscopy, it is necessary to carefully analyze the core and wing parts to verify that the detected signal appears correctly at the line core and is not influenced by the outliers from the line wings \citep[e.g.,][]{2022AJ....164..173C}. 

    In summary, the absorption features in the atmosphere of CoRoT-1b still lack effective constraints. The equilibrium chemistry model prefers an atmosphere with weak features, but this interpretation has a flaw in the equilibrium chemistry itself, that is, the planetary atmosphere may not be in a state of chemical equilibrium. Limited by the precalculated chemistry grids, a proper characterization of the data may not be obtained even if the parameter space is fully explored. On the other hand, the free chemistry models provided stronger spectral signatures. However, according to the narrowband analysis, we could not exclude the possibility that the detected absorption lines of alkali metals were spurious signals. The model with the presence of TiO and VO seems reasonable, as \cite{2022ApJS..260....3C} have presented strong evidence for the presence of VO in the terminator region based on the NIR transmission spectrum of CoRoT-1b, and decisive evidence for $\rm H_2O$, VO, and $\rm H^-$ in the dayside atmosphere based on the thermal emission of CoRoT-1b observed by the HST and the Spitzer Space Telescopes. However, it is still possible that those detected features might be fluctuations in the systematic noise of the transmission spectra, especially in the optical part. A high-confidence detection of TiO and VO could be obtained from high-resolution observations. Overall, we did not have a clear detection in the atmosphere of CoRoT-1b based on the current data. Future data with higher S/N and broader wavelength coverage are needed to further investigate its atmosphere. For example, JWST observations with continuous coverage in the near-infrared wavelengths can be used to validate the detection results of CoRoT-1b.  

    \subsubsection{Featureless spectra of HAT-P-18b, Qatar-1b, TrES-4b, WASP-39b, and WASP-49b}

\begin{figure}[htbp]
        \centering
        \includegraphics[width=\linewidth]{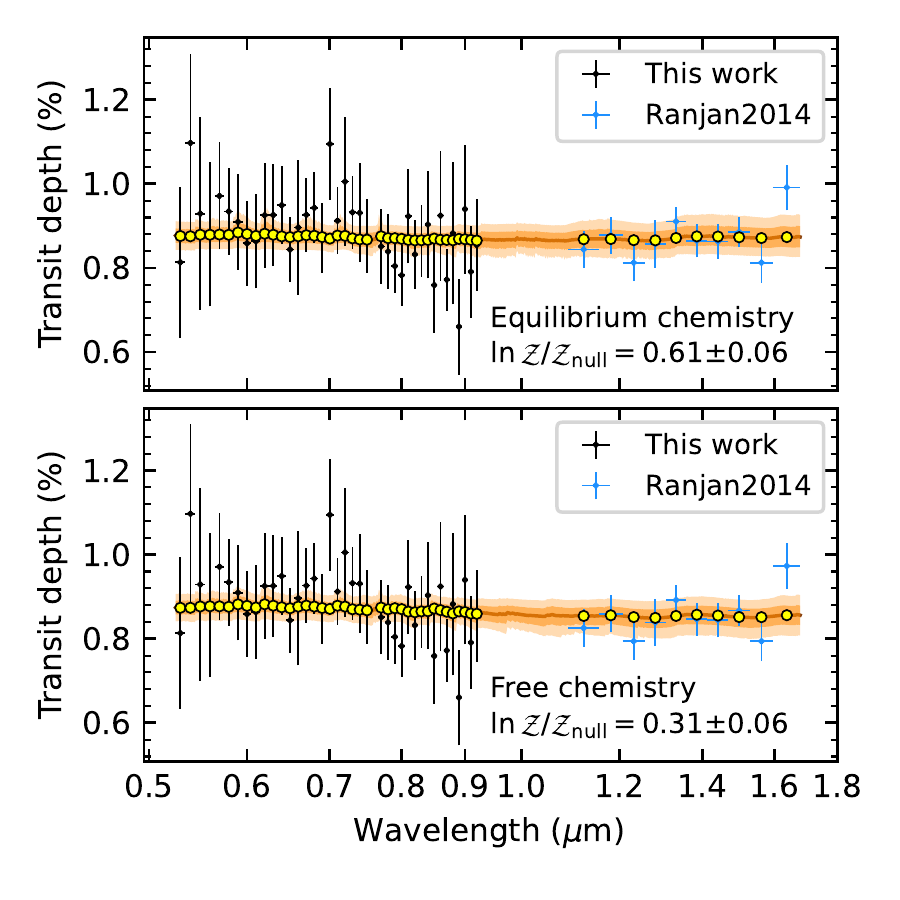}
        \caption{Retrieval results of the optical-to-NIR spectra of TrES-4b. The results from the equilibrium chemistry model and the free chemistry model are shown in the upper and the lower panels, respectively. The orange line and the shaded areas indicate the posteriors distribution of the retrieved spectra. The yellow dots correspond to the models rebinned to the data passbands. The median offsets of the NIR spectra are $-246$~ppm in the upper panel and $-430$~ppm in the lower panel.}
        \label{fig:retrieval_tres4b}
    \end{figure}

    The exoplanets TrES-4b and WASP-39b have the largest pressure scale heights among the targets studied. However, their atmospheric transmission spectra did not exhibit any significant spectral features. For TrES-4b, we performed additional atmospheric retrievals joint with the WFC3 transmission spectrum \citep{2014ApJ...785..148R} considering three scenarios: a null hypothesis, an equilibrium chemistry hypothesis, and a free chemistry hypothesis. The results showed a flat and featureless optical-to-NIR transmission spectrum of TrES-4b with no strong Bayesian evidence for any atmospheric features (Fig.~\ref{fig:retrieval_tres4b}). One explanation for such results is cloud obscuration. If most absorption and scattering features were obscured by opaque clouds, the cloud top should be constrained to a very high altitude. However, the obtained posterior estimates of the cloud-top pressure were weakly constrained ($-0.80\pm2.26~\log_{10}{\rm bar}$ for equilibrium chemistry and $-1.15\pm2.33~\log_{10}{\rm bar}$ for free chemistry). Thus, while high-altitude cloud cover is a plausible explanation for the featureless spectrum of TrES-4b, conclusive evidence is still lacking.
    
\begin{figure}[htbp]
        \centering
        \includegraphics[width=\linewidth]{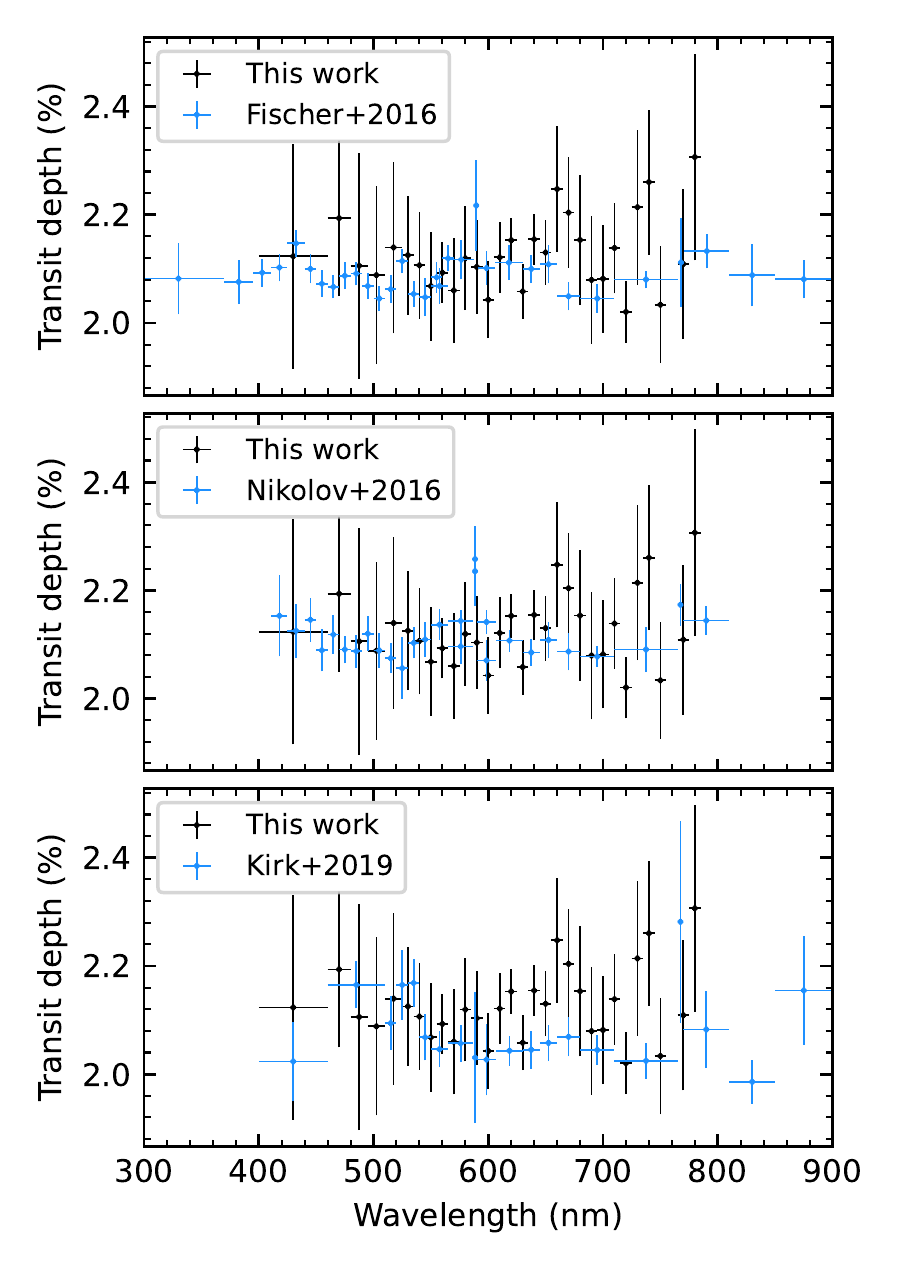}
        \caption{Comparison between the transmission spectra of WASP-39b obtained in this work and the results found in the literature}.
        \label{fig:w39_comparison}
    \end{figure}

    Previous observations of WASP-39b have mostly focused on the optical band using the HST \citep{2016ApJ...827...19F}, the Very Large Telescope \citep{2016ApJ...832..191N} and the William Herschel Telescope \citep{2019AJ....158..144K}, while more recently there have also been  observations with the James Webb Space Telescope (JWST) with full coverage from optical to near-infrared \citep{2023Natur.614..653A, 2023Natur.614..664A, 2023Natur.614..670F, 2023Natur.614..659R}. \cite{2016ApJ...827...19F} and \cite{2016ApJ...832..191N} reported the detection of atomic Na and K, while \cite{2019AJ....158..144K} presented only a tentative detection of K. \cite{2018AJ....155...29W} reported a precise constraint on the $\rm H_2O$ abundance after combining their HST WFC3 data with previously published data. According to these observations, the atmosphere of WASP-39b has significant absorption features and is considered to be clear. We compared the OSIRIS spectrum with the previously published optical spectra and show that our transmission spectrum is still in agreement with the literature results (Fig.~\ref{fig:w39_comparison}). However, the relatively large observational errors of our results prevented the detection of any atmospheric signatures, including the previously reported alkali D-lines. 
    
    In addition to WASP-39b, the published transmission spectra of HAT-P-18b, Qatar-1b, and WASP-49b also showed no significant absorption signatures in low resolution observations. \cite{2017MNRAS.468.3907K} presented an optical transmission spectrum of HAT-P-18b with Rayleigh scattering features. High dispersion spectroscopy revealed an extended atmosphere for HAT-P-18b with the detection of metastable helium \citep{2021ApJ...909L..10P}. Thus, it is possible that the scattering of the high-altitude haze obscured the alkali features in its atmosphere. \cite{2016A&A...587A..67L} and \cite{2017A&A...603A..20V} reported the featureless optical transmission spectra of WASP-49b and Qatar-1b, in agreement with our results. The transmission signals of WASP-49b and Qatar-1b are $\sim$221~ppm and $\sim$109~ppm per unit scale height. If they have clear atmospheres, the atmospheric features should be revealed with the ideal precision of the ground-based instruments including the GTC OSIRIS, the FORS2 on the Very Large Telescope (VLT), or the GMOS on the Gemini telescopes. Thus, the nondetection of WASP-49b and Qatar-1b could be attributed to the presence of high-altitude hazes, which should be confirmed by further observations in bluer wavebands. A more conservative explanation is that the systematic noise in transit light curves prevented the observed transmission spectra from reaching the theoretically expected precision, making the otherwise clear signals indistinguishable. 

    \subsubsection{Effect of stellar activities in the cases of WASP-10b and WASP-32b}

    According to \cite{2009MNRAS.398.1827S} and \cite{2014MNRAS.440.3392B}, the host stars WASP-10 (K5V) and WASP-32 (G0V) are both active stars with photometric modulations induced by starspots. \cite{2009MNRAS.398.1827S} reported that WASP-10 has a brightness modulation with a period of $\sim$11.95 days and amplitudes of 6.3 -- 10.1 mmag. \cite{2014MNRAS.440.3392B} reported a rotation period of $11.6 \pm 1.0$ days for WASP-32. In addition, \cite{2011MNRAS.411.1204M, 2011A&A...535A...7M} reported possible periodic TTVs in the transit light curves of WASP-10b, while \cite{2013MNRAS.430.3032B} suggested that the observed TTVs of WASP-10b should be attributed to starspot occultation rather than a companion planet. These studies suggest that the effect of stellar activity should be considered when analyzing the transmission spectra of WASP-10b and WASP-32b. These planets have very small atmospheric scale heights due to their large masses, making the transmission signals originated from their atmospheres too weak to be detected with current observational precision. However, this provides a unique opportunity to study the effects of stellar activity on planetary transmission spectra. \citet{2018ApJ...853..122R,2019AJ....157...96R} discussed possible spectral features introduced by unocculted starspots and faculae during the transit, hereafter called the stellar contamination effect. Here we adopted their simplified parametric model to characterize such effects. Based on Bayesian model evidence, we analyzed whether the unocculted activity regions would affect the observed transmission spectra of WASP-10b and WASP-32b. 
    
    Following \cite{2018ApJ...853..122R}, the transmission spectrum modified by stellar contamination becomes
    \begin{equation}
        D_{\lambda, \rm{obs}} = \frac{D_{\lambda, \rm{clear}}}
            {1-f\left(1 - \frac{F_\lambda(T_{\rm het})}{F_\lambda(T_{\rm phot})}\right)},
    \end{equation}
    where $D_{\lambda, \rm{obs}}$ is the observed chromatic transit depth at the wavelength $\lambda$, $D_{\lambda, \rm{clear}}$ is the calculated chromatic transit depth for a clear stellar photosphere, $f$ is the fraction of unocculted heterogeneous regions, $F_\lambda$ is the model stellar spectrum, $T_{\rm het}$ and $T_{\rm phot}$ are the temperatures of the heterogeneous regions and the quiescent photosphere, respectively. The temperatures $T_{\rm het}$ and $T_{\rm phot}$ were estimated with Stefan-Boltzmann law:
    \begin{equation}
        \sigma T_{\rm eff}^4 \approx f\sigma T_{\rm het}^4 + (1-f)\sigma T_{\rm phot}^4.
    \end{equation}
    The stellar contamination model is therefore controlled by three free parameters: $T_{\rm eff}$, $T_{\rm het}$, and $f$, where $T_{\rm eff}$ follows a Gaussian prior from the literature values listed in Table~\ref{table:stellar_parameters}, $T_{\rm het}$ follows a uniform prior of $\mathcal{U}(2100, 9000)$ K, and $f$ follows a uniform prior of $\mathcal{U}(0, 0.5)$. 
    
\begin{figure}[htbp]
        \centering
        \includegraphics[width=\linewidth]{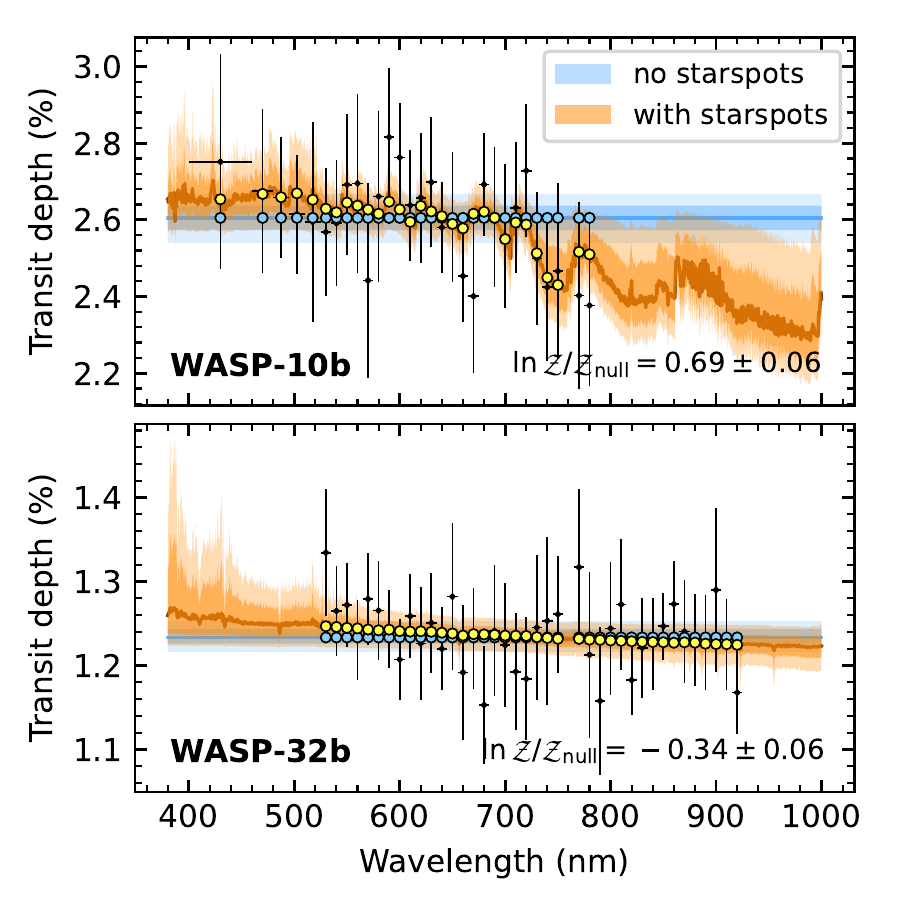}
        \caption{Transmission spectra of WASP-10b (upper panel) and WASP-32b (lower panel) interpreted with stellar contamination models. The black points with error bars are the observed data. The blue lines and shaded areas are the retrieved null models, while the orange ones are the retrieved models with unocculted starspots. The circles are the best-fit spectra after wavelength binning.}
        \label{fig:stellar_activity}
    \end{figure}

    Figure~\ref{fig:stellar_activity} presents the results of the atmospheric retrievals taking int account the effect of stellar contamination. The tentative slopes observed in the transmission spectra of WASP-10b and WASP-32b could not be attributed to Rayleigh scattering features due to their small atmospheric scale heights, but could be explained by the stellar contamination induced by unocculted starspots to some extent. The retrieved parameters of the stellar contamination model yielded a starspot temperature of $3490^{+278}_{-484}$~K and a starspot fraction of $0.39^{+0.04}_{-0.04}$ for WASP-10b, while a starspot temperature of $4781^{+950}_{-1589}$~K and a starspot fraction of $0.13^{+0.12}_{-0.08}$ for WASP-32b. The temperature contrasts between the photosphere and the starspots were therefore $\sim$1200 K for WASP-10 and $\sim$1300 K for WASP-32, which were basically consistent with the statistical trends presented in \cite{2005LRSP....2....8B}. Although we successfully retrieved the parameters from the stellar contamination models, the corresponding Bayesian evidence was very weak for either WASP-10b ($\Delta \ln \mathcal{Z} = 0.69$) or WASP-32b ($\Delta \ln \mathcal{Z} = -0.34$). The host star WASP-10 has a K5 spectral type with rich contamination features at wavelengths greater than than 0.7~$\mu$m. In contrast, WASP-32 is a G0 star with contamination features are mainly concentrated at the bluer end ($\sim$0.4~$\mu$m). However, as shown in Fig.~\ref{fig:stellar_activity}, the transmission spectrum of WASP-10b was observed with the OSIRIS R1000B grism, covering only a wavelength range of 0.40 -- 0.78~$\mu$m and lacking important constraints from the 0.8 to 1~$\mu$m range. On the other hand, the spectrum of WASP-32b was observed with the R1000R grism, missing the bluer wavelength range that best constrains the stellar contamination models, which makes it difficult to obtain strong model evidence. Therefore, although it is theoretically feasible to constrain the models of stellar spots and/or faculae based on the stellar contamination effect on exoplanet transmission spectra, a decisive inference for the presence of unocculted starspots requires significant differences between the quiescent and the active regions of the stellar photosphere and precise measurements of the transmission spectra.
    
\subsection{Possible causes of the featureless spectra}

    In Sect.~\ref{sect:retrieval_results}, we showed that most of the target planets exhibited featureless transmission spectra across a range of atmospheric scale heights. Here we explored two possible reasons for these featureless spectra: large transit depth errors relative to small atmospheric scale heights and coverage by high altitude clouds and/or hazes. These factors often act in concert to produce featureless spectra. Determining which factor is dominant is crucial for accurately interpreting transmission spectra. To this end, we conducted a population study of the observed targets.

    \begin{figure*}[htbp]
        \centering
        \includegraphics[width=\linewidth]{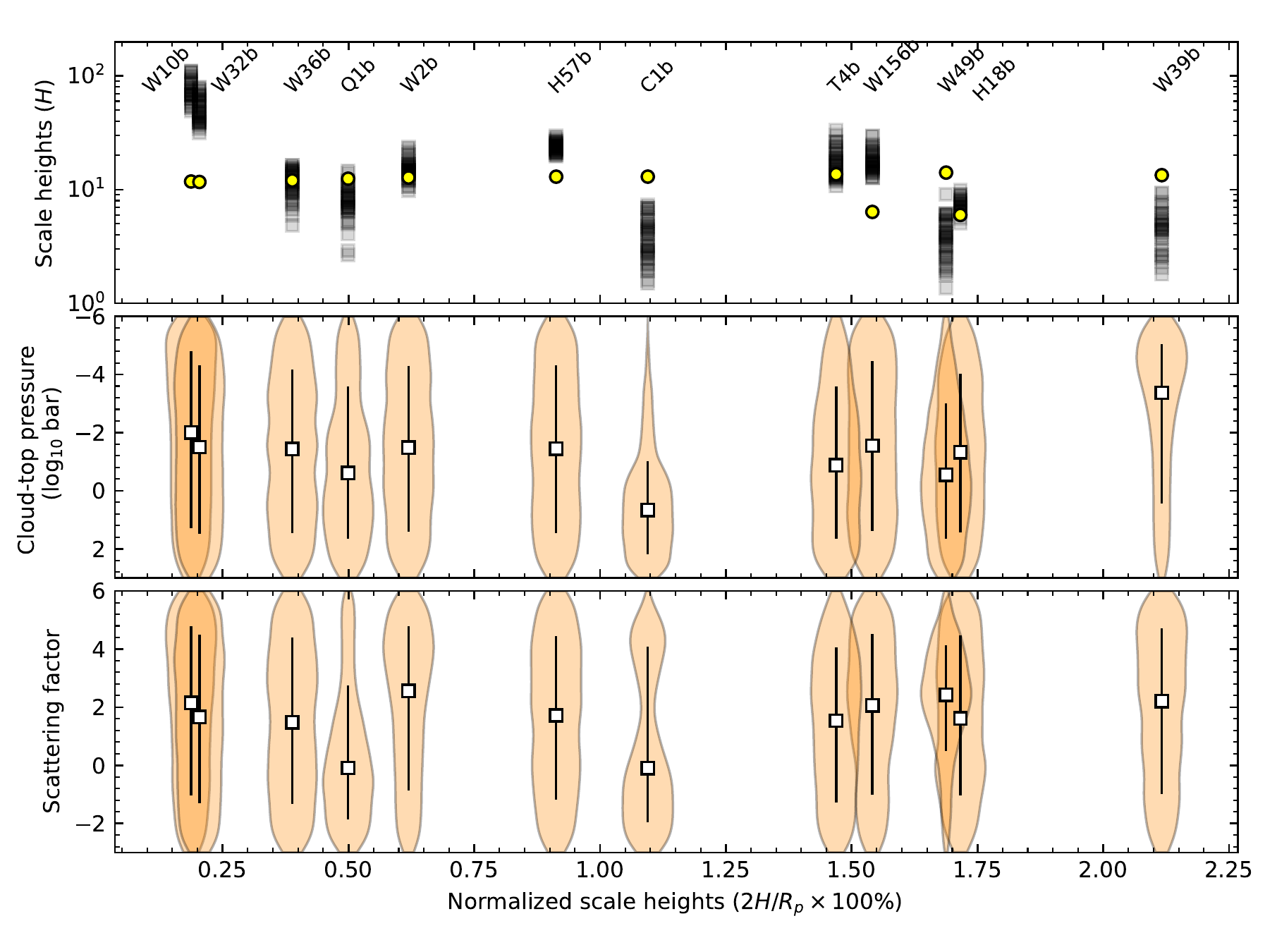}
        \caption{Population analysis on the observed targets. {\it Upper panel:} Observational errors of the chromatic transit depths in unit of scale heights (gray squares) compared with the nominal amplitudes of the sodium D-lines assuming clear atmospheres (yellow dots). {\it Middle panel:} Posterior distributions of the cloud-top pressures from the retrieved atmospheric models assuming chemical equilibrium. {\it Lower panel:} Posterior distributions of the scattering factors from the retrieved atmospheric models assuming chemical equilibrium. The error bars indicate the $1\sigma$ credible intervals. }
        \label{fig:population}
    \end{figure*}   

    We first compared the observational errors of the transmission spectra with the planetary atmospheric scale heights to see whether the observational precision allows the detection of major atmospheric signatures if the planets have clear atmospheres. The OSIRIS transmission spectra mainly cover the spectral features at optical wavelengths, where the signatures of alkali atoms could be dominant in the absence of other optical absorbers such as TiO and VO. Therefore, we adopted the absorption amplitude of the sodium D-lines centered at 589.3~nm under the assumption of a clear atmosphere and solar abundances to characterize the strength of transmission signals at visible wavelengths. The sodium absorption depths were rebinned in the passband of 585 -- 595 nm, consistent with that of the observed spectroscopic light curves. The amplitudes of Na absorption and the transit depth errors are normalized by the planetary scale heights for convenience:
    \begin{equation}
        \begin{aligned}
        \Tilde{D}_{\rm Na} = \frac{D_{\rm Na} - D_{\rm disk}} {D_{\rm disk} \cdot (2H/R_{\rm p})}, \\
        \Tilde{\varepsilon}(D_\lambda) = \frac{\varepsilon(D_\lambda)}{D_{\rm disk} \cdot (2H/R_{\rm p})},
        \end{aligned}
    \end{equation}
    where $D_{\rm Na}$ is the calculated transit depth in the passband of 585 -- 595 nm, $D_{\rm disk}=R_{\rm p}^2/R_{\rm s}^2$ is the transit depth of the opaque planetary disk (approximated by our calculated broadband transit depths), $\varepsilon(D_\lambda)$ is the observational error of the chromatic transit depth, and $2H/R_{\rm p}$ is the previously defined normalized scale height. The quantities $\Tilde{D}_{\rm Na}$ and $\Tilde{\varepsilon}(D_\lambda)$ are the numbers of scale heights that the amplitude of Na absorption and the observational error are equivalent to. Apparently, in order to detect the spectral features in a clear atmosphere at high confidence levels, the observational errors, $\Tilde{\varepsilon}(D_\lambda)$, must be much smaller than $\Tilde{D}_{\rm Na}$ in all optical passbands. Figure~\ref{fig:population} shows that $\Tilde{D}_{\rm Na}$ is typically distributed between 5 -- 15 times scale heights, while most of the observational errors of the transmission spectra are larger than $\Tilde{D}_{\rm Na}$, suggesting that the observational precision does not allowed us to detect any atmospheric signatures in a single transit observation even if they posses clear atmospheres. Only three targets, CoRoT-1b, WASP-39b, and WASP-49b, had relatively better precision of the transmission spectra, so their transmission spectra reached the required precision for detecting the sodium D-lines. In this case, the nondetection of sodium is either due to the absence or depletion of sodium in their atmospheres, or due to the presence of high-altitude clouds, hazes, and/or optical absorbers. 
    
    Next, we illustrate whether high-altitude clouds and hazes could be the dominant factor for the nondetection in the observed transmission spectra. As introduced in Sect.~\ref{sect:method_retrieval}, we assumed a uniform opaque cloud deck in the planetary atmospheric model, simply characterized by a cloud-top pressure. Figure~\ref{fig:population} shows the posterior estimates of the cloud-top pressure from the atmospheric retrieval (using only the optical data). We found that for most of the targets with large uncertainties in the transmission spectra, the posterior estimates of the cloud-top pressure remained unconstrained and were close to the uniform prior of $\mathcal{U}(-6, 3)$ ($\rm \log_{10}~bar$). CoRoT-1b was the only target with a relatively low cloud top ($\log_{10}~P_{\rm c} = 0.67 ^{+1.51}_{-1.79}$), although its scattering factor showed a bimodal distribution due to the indistinct absorption features. The nondetection of WASP-39b and WASP49b resulted in high-altitude clouds for the former and enhanced Rayleigh-like scattering for the latter. However, for other planets with large scale heights (e.g., HAT-P-18b, TrES-4b, and WASP-156b), the cloud-top pressures were still weakly constrained from their flat transmission spectra due to the very large transit depth uncertainties. We performed additional test retrievals for all targets assuming a fixed solar C/O and a fixed solar metallicity, but even such a strong prior forcing spectral features did not significantly improve the fitted cloud and haze parameters. This suggests that featureless transmission spectra of exoplanets with large scale heights do not necessarily indicate the presence of high-altitude clouds. \cite{2016ApJ...820...78L} discussed the degeneracy between the atmospheric scale heights and clouds. Smaller scale heights due to either a low temperature, a high gravity, and/or a high mean molecular weight would make it much more difficult to constrain the cloud model. Precise constraints on the cloud model heavily depend on the pressure broadening features of the absorption signatures. Therefore, it is quite difficult to constrain the cloud-top pressure in the absence of spectral features. High altitude clouds should not be used as a universal explanation for featureless transmission spectra.     

\begin{figure*}[htbp]
        \centering
        \includegraphics[width=\linewidth]{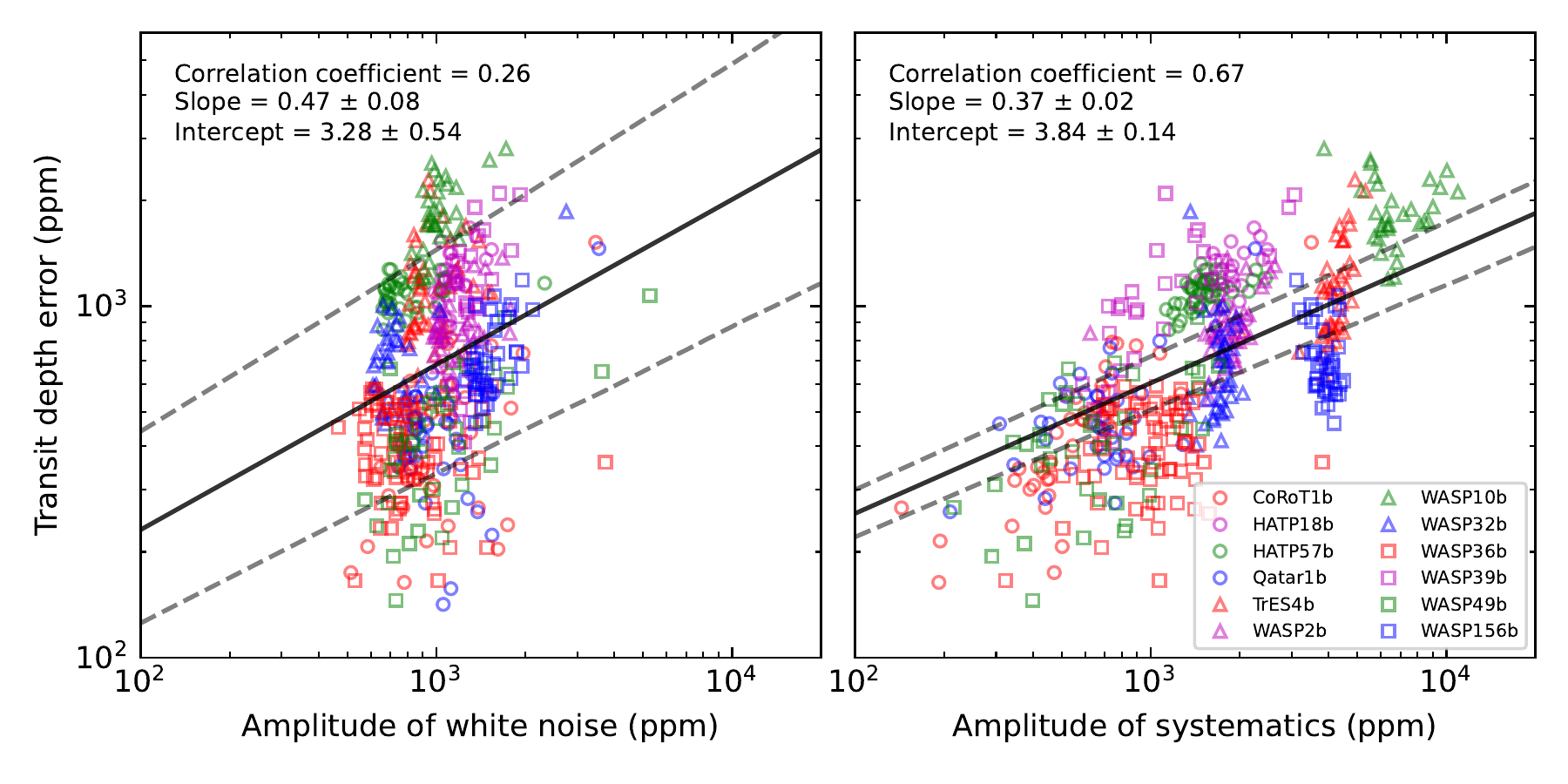}
        \caption{Correlations between the chromatic transit depth errors and light-curve noise. {\it Left panel:} transit depth errors and light-curve white noise. {\it Right panel:} transit depth errors and light-curve systematic noise. The solid lines are the best-fit linear regression results, and the dashed lines indicate the $1\sigma$ credible intervals.}
        \label{fig:error_noise}
    \end{figure*}

    In conclusion, the large uncertainties of the transmission spectra are considered to be the dominant reason for the nondetection of transmission signals in this work. The uncertainties of a transmission spectrum largely depend on the noise levels of the transit light curves. In Fig.~\ref{fig:error_noise}, we analyzed the possible relationship between the chromatic transit depth errors and the light-curve noise and found some correlation between the transit depth errors and the light-curve systematics. This suggests that the systematic noise is the dominant component affecting the light-curve analyses and resulting in large uncertainties of the transmission spectra in ground-based observations. While specific light-curve analysis methods can be used to suppress common-mode noise and reduce overall uncertainty (e.g., the \texttt{divide-white} method; \citealt{2013MNRAS.436.2974G, 2014AJ....147..161S}), these methods may reduce accuracy or introduce spurious artificial features \citep{2022A&A...664A..50J}. As such, these methods were not applied in our light-curve analyses. The resulting observational errors are acceptable for single transit observations and allow for robust estimation of atmospheric models. However, more transit observations are necessary to improve the observational precision and confirm the atmospheric conditions of those targets with large scale heights, such as WASP-39b.

\section{Conclusions}
\label{sect:conclusions}

    In this work, we analyzed the atmospheres of 12 transiting gaseous giant planets using transit spectrophotometry. The transit observations were conducted with the GTC OSIRIS instrument in the optical wavebands. We used a robust GP framework for transit light-curve analysis and obtained a low-resolution transmission spectrum for each target. We then performed one-dimensional atmospheric retrievals on the observed transmission spectra. Through Bayesian model comparison, the transmission spectra for most of the target planets are found to be flat and featureless. Following is the summary on the individual targets:
    \begin{description}
        \item[CoRoT-1b:] Among the observed spectra in this work, only that of CoRoT-1b exhibited strong atmospheric signatures. However, we could not find a decisive model inference when combined with the published WFC3 data of CoRoT-1b. When assuming chemical equilibrium, strong scattering features are preferred in the joint spectrum. When assuming free chemical abundances, the joint spectrum could be explained by either the absorption features of Na and K or those of TiO and VO, where additional evidence for VO has been presented in \cite{2022ApJS..260....3C}. We suggested that the atmosphere of CoRoT-1b should be relatively clear but has a depleted water abundance. Further observations are necessary for verify its atmospheric features. 
        
        \item[TrES-4b:] We did not find any spectral features in the OSIRIS or the published WFC3 spectra. Thus its atmospheric model is still poorly constrained, though it has a relatively large scale height. 

        \item[WASP-10b and WASP-32b:] The atmospheric features in their atmospheres are too weak to detect because of the small scale heights. The effects of stellar contamination could be used to explain the tentative slopes in their transmission spectra, although no strong evidence was found for the contamination model because the wavelength range covered did not capture the most important spectral features for either target. 
        
        \item[HAT-P-57b:] Previous measurements provided only a weak constraint on the mass of this planet ($1.41 \pm 1.52~M_{\rm J}$) due to its A-type host star. We attempted to improve the mass constraint by fitting its atmospheric model with the observed transmission spectrum, but were unsuccessful due to the absence of significant atmospheric signatures.   

        \item[The others:] No atmospheric features were detected in HAT-P-18b, Qatar-1b, WASP-2b, WASP-36b, WASP-39b, WASP-49b, and WASP-156b, mainly because the systematics in the OSIRIS transit light curves substantially reduced the S/N of their transmission spectra. 
    \end{description}

    We also performed a population study on the obtained transmission spectra, which showed that large observational errors, rather than high-altitude clouds, are the main reason for the nondetection of spectral features for most targets. As such, the possibility of clear atmospheres for these exoplanets cannot be ruled out. Further observations are needed to improve the S/N of the transmission spectra for planets with large pressure scale heights, including HAT-P-18b, TrES-4b, WASP-39b, WASP-49b, and WASP-156b, so that the potential spectral features can be identified. It should be noted that featureless transmission spectra should not be simply interpreted as cloudy atmospheres. With current ground-based instruments, it is difficult to break the degeneracy between clouds and metallicity. Therefore, we can only make relatively conservative estimates and inferences based on atmospheric retrievals and model comparisons. A broader wavelength coverage for transit spectroscopy is critical for the characterization of exoplanet atmospheres since many molecular absorption features are not detectable in the optical wavebands. The high-precision instruments on JWST in the near- to mid-infrared wavebands, such as NIRSpec, NIRISS, NIRCam, and MIRI, will greatly improve the constraints on the exoplanet atmospheric models \citep[e.g., ][]{2014PASP..126.1134B, 2016ApJ...817...17G, 2016PASP..128i4401S, 2017A&A...600A..10M, 2018PASP..130k4402B}, and thus our understanding of exoplanets.

\begin{acknowledgements}
    We thank the anonymous referee for their valuable comments and suggestions. G. Chen acknowledges the support by the National Natural Science Foundation of China (grant Nos. 12122308, 42075122), the B-type Strategic Priority Program of the Chinese Academy of Sciences (grant No. XDB41000000), Youth Innovation Promotion Association CAS (2021315), and the Minor Planet Foundation of the Purple Mountain Observatory. This work is based on the observations made with the Gran Telescopio Canarias installed at the Spanish Observatorio del Roque de los Muchachos of the Instituto de Astrofísica de Canarias in the island of La Palma.
\end{acknowledgements}

\bibliographystyle{aa}
\bibliography{references}

\begin{appendix}

\section{Targets with background star contamination}
\label{sect:companions}

\begin{figure}[htbp]
        \centering
        \includegraphics[width=\linewidth]{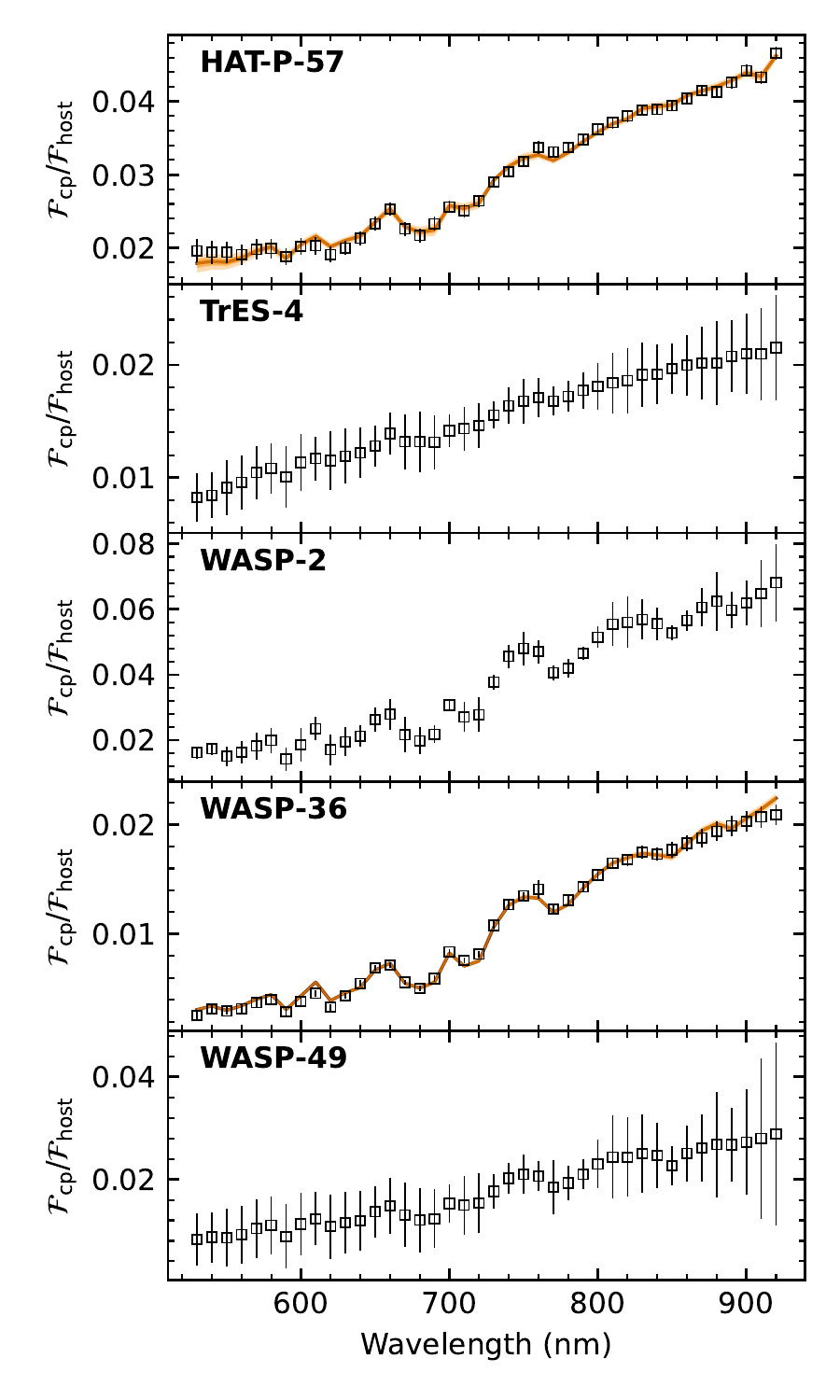}
        \caption{Companion-to-host flux ratios in each narrow passband. The orange lines in the panels of HAT-P-57 and WASP-36 are the best-fit models for companion-star parameter estimation, while the uncertainties are too small to be shown.}
        \label{fig:fluxratios}
    \end{figure}

    Five of the targets have nearby stars that would cause flux dilution in the transit light curves. Of these, only the contamination stars of HAT-P-57 and WASP-36 could be spatially resolved in the OSIRIS spectroscopic observations. For the contamination stars of TrES-4, WASP-2, and WASP-49, we inferred their stellar parameters based on previous research papers (Table~\ref{table:contamination_star_parameters}). We then used Monte Carlo sampling to generate stellar spectra with the PHOENIX model and calculate the flux ratios and corresponding errors in each passband (Fig.~\ref{fig:fluxratios}). For those of HAT-P-57 and WASP-36 that could be spatially resolved in our observations, we fit the stellar PSFs along the spatial direction with Gaussian functions to directly measure their chromatic flux ratios. The chromatic flux ratios were then rebinned to the narrow passbands of the transmission spectra to correct for the flux dilution in spectroscopic light-curve fitting. As shown in Fig.~\ref{fig:fluxratios}, the narrowband flux ratios range from 0.005 to 0.05, resulting in nonnegligible flux dilutions. We note that all the flux ratios have a positive slope, indicating that the contamination stars are redder than their corresponding target stars. 
   
    We also attempted to retrieve the stellar parameters of the contamination stars for HAT-P-57 and WASP-36 from the measured flux ratios. The forward modeling of the flux ratios was done by interpolating the parameter grids of the PHOENIX model spectra, where the free parameters were the effective temperature $T_{\rm eff}$, the gravity $\log g$, and the metallicity [M/H] of the contamination star, as well as a scaling factor, $(R_2/R_1)^2 / (D_2/D_1)^2$, since we do not know the relative distance from the two stars to the observer ($D_2/D_1$). The retrieval algorithm was achieved by \texttt{PyMultiNest}, which is the same as our light-curve fitting and atmospheric retrievals. The mass $M$ and radius $R$ of the contamination star could then be calculated using the retrieved posteriors of $T_{\rm eff}$, $\log g$, and [M/H]. However, due to the limitations in the parameter ranges ($T_{\rm eff}$: 2300 -- 11\,800 K; $\log g$: 3.0 -- 6.0; [M/H]: $-$1.0 -- 1.0) and grid steps (100 -- 200 K for $T_{\rm eff}$; 0.5 for $\log g$ and [M/H]) of the PHOENIX model, we could only obtain very rough estimates and classification for these companions based on the relative flux measurements, and the estimated parameters may be inaccurate if the companions are not main-sequence stars.

    \begin{table*}[htbp]
        \caption{Retrieved or presumed stellar parameters of the contamination stars.}  
        \label{table:contamination_star_parameters} 
        \tiny
        \centering    
        \begin{tabular}{l l l l l l l l l l}  
        \hline\hline 
        Target & Status\tablefootmark{a} & $\Delta{\rm mag}$\tablefootmark{b}
        & $T_{\rm eff}$ (K) & $M$ ($M_\odot$)\tablefootmark{c} & $R$ ($R_\odot$)\tablefootmark{c}
        & $\log g$ (cgs) & [M/H] & Spectral type & Comment\\  
        \hline
        HAT-P-57B
            & C [1]
            & $3.85 \pm 0.04$
            & $3659^{+130}_{-114}$
            & $0.44^{+0.05}_{-0.04}$
            & $0.82^{+0.80}_{-0.39}$
            & $4.30^{+0.53}_{-0.56}$
            & $-0.41^{+0.29}_{-0.26}$
            & K8 -- M3
            & Retrieved \\
        HAT-P-57C
            & C [1]
            & $3.85 \pm 0.04$
            & $10265^{+979}_{-1194}$
            & $3.26^{+0.64}_{-0.67}$
            & $1.06^{+1.59}_{-0.55}$
            & $4.90^{+0.65}_{-0.75}$
            & $-0.06^{+0.63}_{-0.59}$
            & B8 -- A6
            & Retrieved \\
        WASP-36     
            & A [2]
            & $5.32 \pm 0.07$
            & $3156^{+37}_{-27}$
            & $0.38^{+0.02}_{-0.01}$
            & $2.19^{+0.72}_{-0.61}$
            & $3.37^{+0.25}_{-0.22}$
            & $-0.90^{+0.11}_{-0.07}$
            & M3 -- M5 
            & Retrieved \\
        TrES-4      
            & C [3]
            & $4.56 \pm 0.17$ 
            & $\mathcal{U}(3560, 5100)$
            & -
            & -
            & $\mathcal{U}(4.56, 4.78)$
            & $\mathcal{U}(-1.0, 1.0)$ 
            & K2 – M2 [3]
            & Presumed \\
        WASP-2      
            & C [4]
            & $3.72 \pm 0.15$ 
            & $\mathcal{N}(3523, 24)$ [4]
            & -
            & -
            & $\mathcal{U}(4.56, 4.89)$
            & $\mathcal{U}(-1.0, 1.0)$  
            & K2 -- M3 [3] 
            & Presumed \\
        WASP-49  
            & A [5, 6]
            & $4.40 \pm 0.39$ 
            & $\mathcal{U}(2380, 5270)$
            & -
            & -
            & $\mathcal{U}(4.56, 5.32)$
            & $\mathcal{U}(-1.0, 1.0)$ 
            & K0 -- M9 
            & Presumed \\ 
        \hline  
        \end{tabular}

        \tablefoot{
            \tablefoottext{a}{Status of the contamination star: Ambiguous (A), Background (B), or Companion (C).}
            \tablefoottext{b}{The difference of magnitudes between the contamination star(s) and the target star converted from the estimated flux ratio in the broadband (525 -- 925 nm). In the case of HAT-P-57, $\Delta {\rm mag}$ consists of the two binary components.}
            \tablefoottext{c}{The masses and radii were inferred from the PHOENIX model grids after the retrievals, which were not free parameters.}
        }
        \tablebib{[1] \cite{2015AJ....150..197H} [2] \cite{2012AJ....143...81S} 
        [3] \cite{2015A&A...575A..23W} [4] \cite{2020A&A...635A..73B} 
        [5] \cite{2016A&A...587A..67L} [6] \cite{2016A&A...589A..58E}.} 
    \end{table*}    

    \subsection{HAT-P-57}

    According to \cite{2015AJ....150..197H}, there is a binary object $\sim$$2.667''$ separated from HAT-P-57, which is likely to be physically associated with our target star. Its two components are separated by $\sim$$0.225''$. Therefore in the GTC OSIRIS spectroscopic observation, we could only resolve the binaries as a whole by fitting the PSFs of the stellar spectra along the spatial direction, but could not resolve the two components in that system. \cite{2015AJ....150..197H} measured the relative magnitudes between the binary components and HAT-P-57 in the $H$- and $L'$-band, which are $\Delta H_{\rm B} = 2.82 \pm 0.10$ and $\Delta L'_{\rm B} = 2.72 \pm 0.09$ for one of the components, while $\Delta H_{\rm C} = 3.83 \pm 0.11$ and $\Delta L'_{\rm C} = 3.16 \pm 0.10$ for the other one. They also reported that the binary components have masses of $0.61 \pm 0.10~M_\odot$ and $0.53 \pm 0.08~M_\odot$. 

    We conducted a retrieval calculation on the stellar parameters of the binary objects. We considered the sum of the two individual components as the total spectrum of the binary object. Thus, the free parameters in this retrieval were the effective temperature $T_{\rm eff,B}$, the logarithmic gravity $\log g_{\rm B}$, and the metallicity $\rm [M/H]_{B}$ for the component B, and $T_{\rm eff,C}$, $\log g_{\rm C}$, and $\rm [M/H]_{C}$ for the component C. We adopted Gaussian priors on the same parameters for HAT-P-57 (Table~\ref{table:stellar_parameters}) to achieve error propagation. The fitting results are shown in Fig.~\ref{fig:fluxratios} and the retrieved parameters are listed Table~\ref{table:contamination_star_parameters}. We found that the two objects in the binary system have very different temperatures, which is $\sim$3659~K for one star and $\sim$10\,265~K for the other. The one with the lower temperature is well consistent with the mass constraint from \cite{2015AJ....150..197H}, but the other one seems not a K-type or M-type dwarf. We note that the observed flux ratios for HAT-P-57 (the top panel in Fig.~\ref{fig:fluxratios}) is relatively flat near the blue end ($<600$~nm). This indicated an object with a temperature comparable to or even higher than that of HAT-P-57. However, considering that these three stars are physically bounded, their distances to the observer should be quite the same. If the object of the higher temperature is an A-type dwarf with a retrieved mass of $3.26^{+0.64}_{-0.67}$~$M_\odot$ and a radius of $1.06^{+1.59}_{-0.55}$~$R_\odot$, it should be as much bright as the host star, which is not the case. Therefore, we speculate that this object might be a white dwarf, and thus the PHOENIX model grids were not suitable for it. Further investigations are required to confirm this white dwarf candidate.
    
    \subsection{TrES-4}

    A companion star has been reported by \cite{2013MNRAS.428..182B},  \cite{2013MNRAS.433.2097F}, \cite{2015A&A...575A..23W}, and \cite{2015A&A...579A.129W}.  According to \cite{2015A&A...575A..23W}, there is a companion star to TrES-4 at a separation of $1.596''$ and a magnitude contrast of  $\Delta i'=4.49 \pm 0.08~{\rm mag}$, and its spectral type is estimated to be K2 -- M2. Based on their results, we further estimate the effective temperature, and the surface gravity of the companion star using the empirical spectral type-color sequence from \cite{2013ApJS..208....9P}. We lack information on the metallicity of the contamination star. Therefore, we assumed an uninformative prior constraint of $\mathcal{U}(-1, 1)$ on its metallicity, which is the maximum range supported by the PHOENIX model grids.

    \subsection{WASP-2}
    
    A nearby star of WASP-2 has been reported by \cite{2007MNRAS.375..951C}, \cite{2009A&A...498..567D}, \cite{2013MNRAS.428..182B}, \cite{2013AJ....146....9A}, \cite{2015ApJ...800..138N}, \cite{2015A&A...575A..23W}, \cite{2016A&A...589A..58E}, and \cite{2020A&A...635A..73B}. According to \cite{2015A&A...575A..23W}, it is potentially a companion star at a separation of $\sim$$0.71''$ from WASP-2 with a magnitude contrast of $3.51 \pm 0.04$ in the $i'$ band and $3.26 \pm 0.06$ in the $z'$ band. Based on the $i'-z'$ color, they determined a spectral type of K2 -- M3 for the companion star. \cite{2020A&A...635A..73B} reported a magnitude contrast of $2.55 \pm 0.07$ in the $K$ band, and they further constrained its mass and effective temperature to be $0.40 \pm 0.02$ $M_\odot$ and $3523 \pm 24$ K, respectively. 

    \subsection{WASP-36}
    
    According to \cite{2012AJ....143...81S}, there are four fainter background stars that are close to WASP-36. They are separated from WASP-36 by approximately $4''$, $7''$, $13''$, and $17''$. These four stars are roughly in a straight line with WASP-36 in the sky plane, and are perpendicular to the dispersion direction in the GTC OSIRIS observations in this work. Therefore, although these background stars were simultaneously included in the slit, with an optimized aperture size of $4.3''$, only the nearest one had a noticeable effect on the spectrum of WASP-36. And because the two stars could be spatially resolved by the GTC, we could solve their flux ratios by fitting their PSF profiles in each passband and then correct the flux dilution effect in the light-curve fitting. The calculated flux ratio in the broadband (525 -- 925 nm) is $0.0102 \pm 0.0005$. The chromatic flux ratios increase from $0.0026 \pm 0.0004$ at 530 nm to $0.0209 \pm 0.0009$ at 920 nm. If the dilution effect was not corrected, an tentative scattering-like slope would appear in the obtained transmission spectrum. 
    
    The retrieval calculation of stellar parameters showed that the nearby contamination star should be an M-type star with an effective temperature of $3156^{+37}_{-27}$ K. The PHOENIX model spectra suggested that the contaminated star has a low gravity ($\log g = 3.37^{+0.25}_{-0.22}$ cgs) and therefore might be a red giant. Further observations are required to confirm its luminosity class.
    
    \subsection{WASP-49}

    There is a nearby star at a separation of $2.2''$ from WASP-49, and its flux dilution effect is nonnegligible. \cite{2016A&A...587A..67L} reported a magnitude contrast of $\Delta z=4.30 \pm 0.12$, and \cite{2016A&A...589A..58E} reported a magnitude contrast of $\Delta r_{\rm TCI}=4.979 \pm 0.018$, which indicates that the nearby star is much fainter and redder than WASP-49. It is not clear whether this nearby star is physically bounded to WASP-49, and its spectral type and stellar parameters have not been determined yet. Considering that WASP-49 is a G6V star, we could only assume that the contamination star has a spectral type of K0 -- M9 so as to obtain rough estimation on its effective temperature, radius, and mass. 

\section{Additional tables and figures}

\begin{figure*}[htbp]
    \centering
    \includegraphics[width=\linewidth]{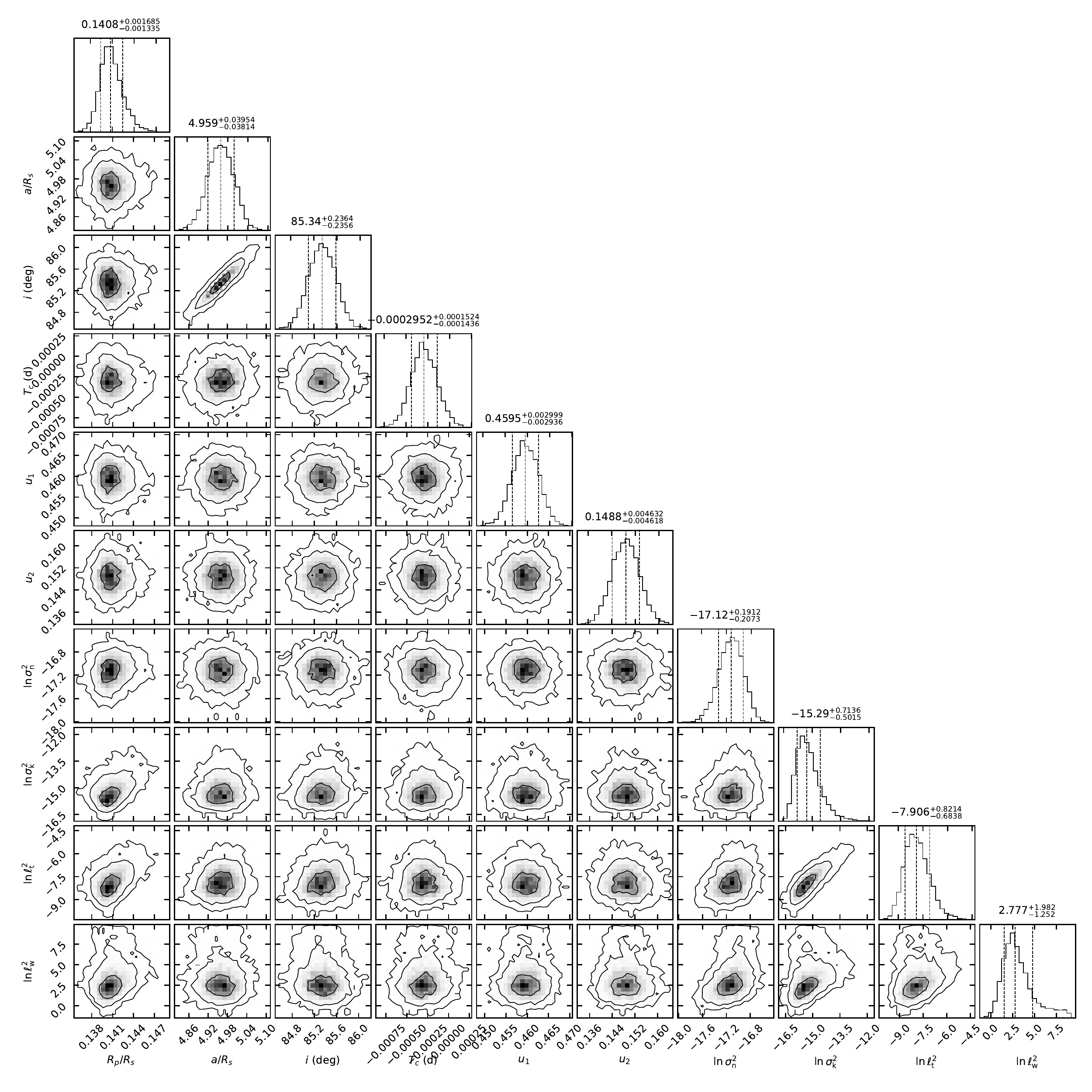}
    \caption{ Posterior distribution of the parameters from the white-light-curve fitting of CoRoT-1b. The contours indicate 1- to 3-$\sigma$ credible intervals. The parameters from left to right are $R_{\rm p}/R_{\rm s}$: planet-to-star radius ratio; $a/R_{\rm s}$: orbital semimajor axis relative to star radius; $i$: orbital inclination, $T_{\rm c}$: observed central transit time subtracting the predicted central transit time ($\rm BJD_{TDB}$); $u_1$ and $u_2$: quadratic limb-darkening coefficients; $\sigma_{\rm n}^2$: variance of the GP jitter term; $\sigma_{\rm k}^2$: variance of the GP kernel function; $\ell_{\rm t}$ and $\ell_{\rm w}$: length scales of the input variables (the time vector and the FWHM of the PSF).}
    \label{fig:corners_white_c1b}
\end{figure*}

\begin{table*}[htbp]
    \caption{Transmission spectra of CoRoT-1b, HAT-P-18b, HAT-P-57b, Qatar-1b, and TrES-4b.}
    \label{table:ts_R1000R1} 
    \centering   
    \begin{tabular}{l c c c c c}  
    \hline\hline 
    Passbands (nm) 
    & \multicolumn{5}{c}{Planet-to-star radius ratio ($R_{\rm p}/R_{\rm s}$)} \\
     & CoRoT-1b & HAT-P-18b & HAT-P-57b & Qatar-1b & TrES-4b\\
    \hline 
525 -- 535 & $0.1412 \pm 0.0018$ & $0.1372 \pm 0.0055$ & $0.0990 \pm 0.0059$ & $0.1457 \pm 0.0014$ & $0.0902 \pm 0.0103$ \\ 
535 -- 545 & $0.1412 \pm 0.0011$ & $0.1410 \pm 0.0048$ & $0.0977 \pm 0.0063$ & $0.1440 \pm 0.0012$ & $0.1047 \pm 0.0106$ \\
545 -- 555 & $0.1417 \pm 0.0013$ & $0.1436 \pm 0.0055$ & $0.0984 \pm 0.0061$ & $0.1459 \pm 0.0014$ & $0.0964 \pm 0.0125$ \\
555 -- 565 & $0.1407 \pm 0.0017$ & $0.1403 \pm 0.0060$ & $0.1017 \pm 0.0056$ & $0.1471 \pm 0.0012$ & $0.0938 \pm 0.0093$ \\
565 -- 575 & $0.1407 \pm 0.0012$ & $0.1406 \pm 0.0047$ & $0.1047 \pm 0.0051$ & $0.1447 \pm 0.0016$ & $0.0985 \pm 0.0064$ \\
575 -- 585 & $0.1406 \pm 0.0010$ & $0.1409 \pm 0.0041$ & $0.1020 \pm 0.0057$ & $0.1451 \pm 0.0014$ & $0.0967 \pm 0.0053$ \\
585 -- 595 & $0.1438 \pm 0.0016$ & $0.1419 \pm 0.0043$ & $0.1023 \pm 0.0062$ & $0.1467 \pm 0.0022$ & $0.0954 \pm 0.0060$ \\
595 -- 605 & $0.1448 \pm 0.0021$ & $0.1399 \pm 0.0041$ & $0.1013 \pm 0.0049$ & $0.1462 \pm 0.0016$ & $0.0927 \pm 0.0055$ \\
605 -- 615 & $0.1423 \pm 0.0011$ & $0.1429 \pm 0.0049$ & $0.1023 \pm 0.0053$ & $0.1463 \pm 0.0017$ & $0.0929 \pm 0.0060$ \\
615 -- 625 & $0.1411 \pm 0.0012$ & $0.1393 \pm 0.0049$ & $0.1009 \pm 0.0055$ & $0.1466 \pm 0.0014$ & $0.0962 \pm 0.0065$ \\
625 -- 635 & $0.1416 \pm 0.0011$ & $0.1415 \pm 0.0042$ & $0.0997 \pm 0.0057$ & $0.1462 \pm 0.0013$ & $0.0962 \pm 0.0064$ \\
635 -- 645 & $0.1416 \pm 0.0006$ & $0.1405 \pm 0.0045$ & $0.1019 \pm 0.0050$ & $0.1447 \pm 0.0019$ & $0.0974 \pm 0.0047$ \\
645 -- 655 & $0.1407 \pm 0.0010$ & $0.1446 \pm 0.0041$ & $0.1021 \pm 0.0062$ & $0.1459 \pm 0.0019$ & $0.0919 \pm 0.0043$ \\
655 -- 665 & $0.1415 \pm 0.0006$ & $0.1400 \pm 0.0039$ & $0.0998 \pm 0.0049$ & $0.1466 \pm 0.0013$ & $0.0947 \pm 0.0085$ \\
665 -- 675 & $0.1401 \pm 0.0011$ & $0.1390 \pm 0.0043$ & $0.1007 \pm 0.0056$ & $0.1464 \pm 0.0012$ & $0.0963 \pm 0.0045$ \\
675 -- 685 & $0.1402 \pm 0.0008$ & $0.1363 \pm 0.0037$ & $0.1016 \pm 0.0051$ & $0.1457 \pm 0.0016$ & $0.0971 \pm 0.0043$ \\
685 -- 695 & $0.1401 \pm 0.0007$ & $0.1325 \pm 0.0043$ & $0.1019 \pm 0.0055$ & $0.1447 \pm 0.0015$ & $0.0933 \pm 0.0044$ \\
695 -- 705 & $0.1416 \pm 0.0014$ & $0.1354 \pm 0.0038$ & $0.1020 \pm 0.0055$ & $0.1452 \pm 0.0013$ & $0.1046 \pm 0.0064$ \\
705 -- 715 & $0.1418 \pm 0.0016$ & $0.1413 \pm 0.0039$ & $0.0991 \pm 0.0051$ & $0.1472 \pm 0.0027$ & $0.0955 \pm 0.0041$ \\
715 -- 725 & $0.1398 \pm 0.0017$ & $0.1423 \pm 0.0031$ & $0.1002 \pm 0.0058$ & $0.1456 \pm 0.0015$ & $0.1003 \pm 0.0076$ \\
725 -- 735 & $0.1415 \pm 0.0011$ & $0.1409 \pm 0.0045$ & $0.0980 \pm 0.0052$ & $0.1461 \pm 0.0013$ & $0.0965 \pm 0.0044$ \\
735 -- 745 & $0.1428 \pm 0.0017$ & $0.1400 \pm 0.0055$ & $0.1017 \pm 0.0047$ & $0.1463 \pm 0.0012$ & $0.0965 \pm 0.0060$ \\
745 -- 755 & $0.1405 \pm 0.0012$ & $0.1378 \pm 0.0033$ & $0.0977 \pm 0.0045$ & $0.1471 \pm 0.0013$ & $0.0936 \pm 0.0046$ \\
755 -- 765 & $0.1509 \pm 0.0051$ & $0.1357 \pm 0.0042$ & $0.0849 \pm 0.0072$ & $0.1406 \pm 0.0051$ & $0.0950 \pm 0.0047$ \\
765 -- 775 & $0.1456 \pm 0.0028$ & $0.1397 \pm 0.0051$ & $0.0930 \pm 0.0051$ & $0.1469 \pm 0.0005$ & $0.0923 \pm 0.0046$ \\
775 -- 785 & $0.1447 \pm 0.0027$ & $0.1443 \pm 0.0044$ & $0.0944 \pm 0.0045$ & $0.1436 \pm 0.0019$ & $0.0916 \pm 0.0048$ \\
785 -- 795 & $0.1409 \pm 0.0008$ & ...                 & $0.1050 \pm 0.0047$ & $0.1477 \pm 0.0015$ & $0.0897 \pm 0.0035$ \\ 
795 -- 805 & $0.1400 \pm 0.0018$ & ...                 & $0.1044 \pm 0.0053$ & $0.1461 \pm 0.0009$ & $0.0885 \pm 0.0042$ \\ 
805 -- 815 & $0.1399 \pm 0.0028$ & ...                 & $0.0932 \pm 0.0049$ & $0.1477 \pm 0.0019$ & $0.0961 \pm 0.0058$ \\ 
815 -- 825 & $0.1400 \pm 0.0024$ & ...                 & $0.0978 \pm 0.0055$ & $0.1458 \pm 0.0005$ & $0.0912 \pm 0.0045$ \\ 
825 -- 835 & $0.1405 \pm 0.0019$ & ...                 & $0.0979 \pm 0.0051$ & $0.1454 \pm 0.0016$ & $0.0929 \pm 0.0045$ \\ 
835 -- 845 & $0.1424 \pm 0.0021$ & ...                 & $0.1022 \pm 0.0049$ & $0.1483 \pm 0.0016$ & $0.0950 \pm 0.0070$ \\ 
845 -- 855 & $0.1404 \pm 0.0016$ & ...                 & $0.1026 \pm 0.0057$ & $0.1462 \pm 0.0021$ & $0.0871 \pm 0.0067$ \\ 
855 -- 865 & $0.1396 \pm 0.0010$ & ...                 & $0.1038 \pm 0.0057$ & $0.1454 \pm 0.0010$ & $0.0961 \pm 0.0079$ \\ 
865 -- 875 & $0.1438 \pm 0.0027$ & ...                 & $0.0988 \pm 0.0067$ & $0.1445 \pm 0.0009$ & $0.0879 \pm 0.0042$ \\ 
875 -- 885 & $0.1397 \pm 0.0009$ & ...                 & $0.0968 \pm 0.0051$ & $0.1431 \pm 0.0019$ & $0.0939 \pm 0.0096$ \\ 
885 -- 895 & $0.1386 \pm 0.0007$ & ...                 & $0.0962 \pm 0.0057$ & $0.1475 \pm 0.0026$ & $0.0813 \pm 0.0073$ \\ 
895 -- 905 & $0.1403 \pm 0.0026$ & ...                 & $0.0967 \pm 0.0057$ & $0.1479 \pm 0.0008$ & $0.0969 \pm 0.0081$ \\ 
905 -- 915 & $0.1356 \pm 0.0020$ & ...                 & $0.1052 \pm 0.0060$ & $0.1447 \pm 0.0019$ & $0.0889 \pm 0.0062$ \\ 
915 -- 925 & $0.1398 \pm 0.0018$ & ...                 & $0.0954 \pm 0.0063$ & $0.1464 \pm 0.0022$ & $0.0924 \pm 0.0059$ \\
    \hline  
    \end{tabular}
    
    \tablefoot{
    The spectral images of the comparison star for HAT-P-18 overlapped with a column of bad pixels, for which the spectra with wavelengths larger than 785 nm were discarded.
    }
\end{table*}

\begin{table*}[htbp]
    \caption{Transmission spectra of WASP-2b, WASP-32b, WASP-36b, WASP-49b, and WASP-156b.} 
    \label{table:ts_R1000R2}
    \centering
    \begin{tabular}{l c c c c c}  
    \hline\hline 
    Passbands (nm) 
    & \multicolumn{5}{c}{Planet-to-star radius ratio ($R_{\rm p}/R_{\rm s}$)} \\
     & WASP-2b & WASP-32b & WASP-36b &  WASP-49b & WASP-156b\\
    \hline 
525 -- 535 & $0.1375 \pm 0.0045$ & $0.1155 \pm 0.0033$ & $0.1336 \pm 0.0015$ & $0.1158 \pm 0.0029$ & $0.0710 \pm 0.0069$ \\ 
535 -- 545 & $0.1366 \pm 0.0043$ & $0.1125 \pm 0.0024$ & $0.1335 \pm 0.0020$ & $0.1140 \pm 0.0029$ & $0.0756 \pm 0.0045$ \\
545 -- 555 & $0.1346 \pm 0.0042$ & $0.1128 \pm 0.0022$ & $0.1344 \pm 0.0014$ & $0.1157 \pm 0.0030$ & $0.0752 \pm 0.0049$ \\
555 -- 565 & $0.1358 \pm 0.0031$ & $0.1109 \pm 0.0021$ & $0.1336 \pm 0.0012$ & $0.1152 \pm 0.0013$ & $0.0748 \pm 0.0041$ \\
565 -- 575 & $0.1349 \pm 0.0035$ & $0.1131 \pm 0.0024$ & $0.1341 \pm 0.0017$ & $0.1162 \pm 0.0016$ & $0.0763 \pm 0.0063$ \\
575 -- 585 & $0.1368 \pm 0.0033$ & $0.1125 \pm 0.0026$ & $0.1331 \pm 0.0017$ & $0.1160 \pm 0.0009$ & $0.0758 \pm 0.0071$ \\
585 -- 595 & $0.1353 \pm 0.0030$ & $0.1115 \pm 0.0021$ & $0.1360 \pm 0.0021$ & $0.1141 \pm 0.0015$ & $0.0815 \pm 0.0052$ \\
595 -- 605 & $0.1357 \pm 0.0035$ & $0.1099 \pm 0.0022$ & $0.1348 \pm 0.0019$ & $0.1154 \pm 0.0016$ & $0.0758 \pm 0.0037$ \\
605 -- 615 & $0.1367 \pm 0.0029$ & $0.1122 \pm 0.0022$ & $0.1337 \pm 0.0013$ & $0.1134 \pm 0.0029$ & $0.0737 \pm 0.0039$ \\
615 -- 625 & $0.1362 \pm 0.0026$ & $0.1107 \pm 0.0031$ & $0.1343 \pm 0.0013$ & $0.1143 \pm 0.0012$ & $0.0673 \pm 0.0052$ \\
625 -- 635 & $0.1343 \pm 0.0033$ & $0.1118 \pm 0.0027$ & $0.1333 \pm 0.0012$ & $0.1137 \pm 0.0018$ & $0.0682 \pm 0.0045$ \\
635 -- 645 & $0.1342 \pm 0.0031$ & $0.1104 \pm 0.0022$ & $0.1343 \pm 0.0012$ & $0.1123 \pm 0.0006$ & $0.0710 \pm 0.0047$ \\
645 -- 655 & $0.1347 \pm 0.0031$ & $0.1132 \pm 0.0039$ & $0.1342 \pm 0.0015$ & $0.1159 \pm 0.0009$ & $0.0661 \pm 0.0080$ \\
655 -- 665 & $0.1384 \pm 0.0030$ & $0.1092 \pm 0.0037$ & $0.1348 \pm 0.0022$ & $0.1123 \pm 0.0012$ & $0.0636 \pm 0.0062$ \\
665 -- 675 & $0.1336 \pm 0.0028$ & $0.1110 \pm 0.0027$ & $0.1348 \pm 0.0018$ & $0.1142 \pm 0.0010$ & $0.0675 \pm 0.0054$ \\
675 -- 685 & $0.1338 \pm 0.0025$ & $0.1074 \pm 0.0033$ & $0.1354 \pm 0.0019$ & $0.1147 \pm 0.0010$ & $0.0712 \pm 0.0050$ \\
685 -- 695 & $0.1330 \pm 0.0031$ & $0.1114 \pm 0.0035$ & $0.1350 \pm 0.0019$ & $0.1140 \pm 0.0008$ & $0.0696 \pm 0.0044$ \\
695 -- 705 & $0.1339 \pm 0.0020$ & $0.1106 \pm 0.0034$ & $0.1340 \pm 0.0009$ & $0.1142 \pm 0.0012$ & $0.0692 \pm 0.0048$ \\
705 -- 715 & $0.1354 \pm 0.0025$ & $0.1092 \pm 0.0032$ & $0.1330 \pm 0.0010$ & $0.1123 \pm 0.0018$ & $0.0735 \pm 0.0046$ \\
715 -- 725 & $0.1352 \pm 0.0034$ & $0.1088 \pm 0.0034$ & $0.1351 \pm 0.0015$ & $0.1129 \pm 0.0018$ & $0.0672 \pm 0.0046$ \\
725 -- 735 & $0.1400 \pm 0.0034$ & $0.1116 \pm 0.0039$ & $0.1342 \pm 0.0017$ & $0.1150 \pm 0.0019$ & $0.0684 \pm 0.0034$ \\
735 -- 745 & $0.1353 \pm 0.0022$ & $0.1119 \pm 0.0045$ & $0.1353 \pm 0.0018$ & $0.1129 \pm 0.0018$ & $0.0669 \pm 0.0045$ \\
745 -- 755 & $0.1356 \pm 0.0031$ & $0.1123 \pm 0.0031$ & $0.1345 \pm 0.0019$ & $0.1129 \pm 0.0021$ & $0.0684 \pm 0.0041$ \\
755 -- 765 & $0.1373 \pm 0.0047$ & $0.1109 \pm 0.0083$ & $0.1341 \pm 0.0013$ & $0.1107 \pm 0.0048$ & $0.0784 \pm 0.0059$ \\
765 -- 775 & $0.1353 \pm 0.0026$ & $0.1148 \pm 0.0040$ & $0.1323 \pm 0.0006$ & $0.1140 \pm 0.0029$ & $0.0680 \pm 0.0041$ \\
775 -- 785 & $0.1358 \pm 0.0026$ & $0.1101 \pm 0.0045$ & $0.1344 \pm 0.0009$ & $0.1131 \pm 0.0020$ & $0.0624 \pm 0.0049$ \\
785 -- 795 & $0.1348 \pm 0.0030$ & $0.1076 \pm 0.0042$ & $0.1343 \pm 0.0010$ & $0.1108 \pm 0.0024$ & $0.0658 \pm 0.0042$ \\ 
795 -- 805 & $0.1348 \pm 0.0037$ & $0.1115 \pm 0.0036$ & $0.1348 \pm 0.0013$ & $0.1145 \pm 0.0013$ & $0.0678 \pm 0.0039$ \\ 
805 -- 815 & $0.1360 \pm 0.0022$ & $0.1128 \pm 0.0034$ & $0.1336 \pm 0.0010$ & $0.1138 \pm 0.0018$ & $0.0712 \pm 0.0043$ \\ 
815 -- 825 & $0.1345 \pm 0.0027$ & $0.1087 \pm 0.0019$ & $0.1353 \pm 0.0020$ & $0.1122 \pm 0.0025$ & $0.0725 \pm 0.0043$ \\ 
825 -- 835 & $0.1362 \pm 0.0033$ & $0.1105 \pm 0.0027$ & $0.1364 \pm 0.0019$ & $0.1124 \pm 0.0024$ & $0.0726 \pm 0.0035$ \\ 
835 -- 845 & $0.1341 \pm 0.0031$ & $0.1107 \pm 0.0025$ & $0.1329 \pm 0.0015$ & $0.1115 \pm 0.0024$ & $0.0739 \pm 0.0040$ \\ 
845 -- 855 & $0.1351 \pm 0.0029$ & $0.1116 \pm 0.0018$ & $0.1371 \pm 0.0015$ & $0.1144 \pm 0.0015$ & $0.0736 \pm 0.0061$ \\ 
855 -- 865 & $0.1356 \pm 0.0034$ & $0.1128 \pm 0.0023$ & $0.1336 \pm 0.0013$ & $0.1136 \pm 0.0020$ & $0.0765 \pm 0.0044$ \\ 
865 -- 875 & $0.1351 \pm 0.0031$ & $0.1114 \pm 0.0028$ & $0.1333 \pm 0.0013$ & $0.1134 \pm 0.0014$ & $0.0722 \pm 0.0041$ \\ 
875 -- 885 & $0.1338 \pm 0.0048$ & $0.1109 \pm 0.0025$ & $0.1342 \pm 0.0019$ & $0.1156 \pm 0.0021$ & $0.0717 \pm 0.0051$ \\ 
885 -- 895 & $0.1335 \pm 0.0030$ & $0.1108 \pm 0.0025$ & $0.1336 \pm 0.0014$ & $0.1145 \pm 0.0020$ & $0.0757 \pm 0.0064$ \\ 
895 -- 905 & $0.1342 \pm 0.0036$ & $0.1136 \pm 0.0043$ & $0.1364 \pm 0.0022$ & $0.1138 \pm 0.0026$ & $0.0760 \pm 0.0067$ \\ 
905 -- 915 & $0.1363 \pm 0.0050$ & $0.1107 \pm 0.0024$ & $0.1340 \pm 0.0018$ & $0.1119 \pm 0.0029$ & $0.0734 \pm 0.0050$ \\ 
915 -- 925 & $0.1270 \pm 0.0033$ & $0.1081 \pm 0.0023$ & $0.1338 \pm 0.0008$ & $0.1099 \pm 0.0043$ & $0.0721 \pm 0.0081$ \\
    \hline  
    \end{tabular}
\end{table*}

\begin{table}[htbp]
    \caption{Transmission spectra of WASP-10b and WASP-39b.}  
    \centering
    \label{table:ts_R1000B} 
    \begin{tabular}{l c c}  
    \hline\hline 
    Passbands (nm) 
    & \multicolumn{2}{c}{Planet-to-star radius ratio ($R_{\rm p}/R_{\rm s}$)} \\
     & WASP-10b & WASP-39b \\
    \hline 
    400 -- 460 & $0.1659 \pm 0.0085$ & $0.1457 \pm 0.0074$ \\ 
    460 -- 480 & $0.1636 \pm 0.0065$ & $0.1481 \pm 0.0050$ \\ 
    480 -- 495 & $0.1630 \pm 0.0048$ & $0.1451 \pm 0.0073$ \\ 
    495 -- 510 & $0.1617 \pm 0.0048$ & $0.1445 \pm 0.0057$ \\ 
    510 -- 525 & $0.1611 \pm 0.0082$ & $0.1463 \pm 0.0055$ \\  
    525 -- 535 & $0.1603 \pm 0.0052$ & $0.1458 \pm 0.0038$ \\ 
    535 -- 545 & $0.1610 \pm 0.0051$ & $0.1451 \pm 0.0034$ \\ 
    545 -- 555 & $0.1641 \pm 0.0056$ & $0.1438 \pm 0.0035$ \\ 
    555 -- 565 & $0.1642 \pm 0.0072$ & $0.1447 \pm 0.0019$ \\ 
    565 -- 575 & $0.1563 \pm 0.0082$ & $0.1435 \pm 0.0034$ \\ 
    575 -- 585 & $0.1631 \pm 0.0069$ & $0.1456 \pm 0.0033$ \\ 
    585 -- 595 & $0.1678 \pm 0.0054$ & $0.1450 \pm 0.0030$ \\ 
    595 -- 605 & $0.1662 \pm 0.0043$ & $0.1429 \pm 0.0025$ \\ 
    605 -- 615 & $0.1624 \pm 0.0045$ & $0.1456 \pm 0.0023$ \\ 
    615 -- 625 & $0.1630 \pm 0.0052$ & $0.1467 \pm 0.0014$ \\ 
    625 -- 635 & $0.1643 \pm 0.0052$ & $0.1435 \pm 0.0018$ \\ 
    635 -- 645 & $0.1606 \pm 0.0037$ & $0.1468 \pm 0.0016$ \\ 
    645 -- 655 & $0.1615 \pm 0.0053$ & $0.1459 \pm 0.0021$ \\ 
    655 -- 665 & $0.1566 \pm 0.0039$ & $0.1499 \pm 0.0038$ \\ 
    665 -- 675 & $0.1550 \pm 0.0065$ & $0.1485 \pm 0.0034$ \\ 
    675 -- 685 & $0.1641 \pm 0.0040$ & $0.1467 \pm 0.0041$ \\ 
    685 -- 695 & $0.1615 \pm 0.0057$ & $0.1442 \pm 0.0041$ \\ 
    695 -- 705 & $0.1600 \pm 0.0060$ & $0.1443 \pm 0.0034$ \\ 
    705 -- 715 & $0.1622 \pm 0.0053$ & $0.1462 \pm 0.0028$ \\ 
    715 -- 725 & $0.1652 \pm 0.0053$ & $0.1421 \pm 0.0020$ \\ 
    725 -- 735 & $0.1581 \pm 0.0055$ & $0.1488 \pm 0.0048$ \\ 
    735 -- 745 & $0.1557 \pm 0.0062$ & $0.1504 \pm 0.0044$ \\ 
    745 -- 755 & $0.1570 \pm 0.0073$ & $0.1426 \pm 0.0038$ \\ 
    755 -- 765 & $0.1585 \pm 0.0069$ & $0.1391 \pm 0.0043$ \\ 
    765 -- 775 & $0.1550 \pm 0.0079$ & $0.1452 \pm 0.0048$ \\ 
    775 -- 785 & $0.1542 \pm 0.0069$ & $0.1519 \pm 0.0063$ \\ 
    \hline  
    \end{tabular}
    
\end{table}

\begin{table}[htbp]
    \caption{References of the adopted opacity line lists.}  
    \label{table:line_lists} 
    \centering
    \begin{tabular}{l l l}  
    \hline\hline 
    Species & Database & References \\
    \hline 
    \multicolumn{3}{c}{Gas absorption}\\ 
    Na \tablefootmark{a,b} & VALD & {\cite{2019A&A...628A.120A}} \\ 
    K \tablefootmark{a,b} & VALD &  {\cite{2016A&A...589A..21A}} \\
    $\rm H_2O$ \tablefootmark{a,b} & ExoMolOP & \cite{2018MNRAS.480.2597P} \\ 
    $\rm H_2S$ \tablefootmark{a} & ExoMolOP & \cite{2016MNRAS.460.4063A}\\ 
    HCN \tablefootmark{a} & ExoMolOP & \cite{2014MNRAS.437.1828B} \\ 
    $\rm CH_4$ \tablefootmark{a,b} & ExoMolOP & {\cite{2017A&A...605A..95Y}}\\ 
    $\rm C_2H_2$ \tablefootmark{a} & ExoMolOP & \cite{2020MNRAS.493.1531C}\\ 
    CO \tablefootmark{a,b} & ExoMolOP & \cite{2015ApJS..216...15L}\\ 
    $\rm CO_2$ \tablefootmark{a,b} & ExoMolOP & \cite{2020MNRAS.496.5282Y}\\ 
    $\rm NH_3$ \tablefootmark{a} & ExoMolOP & \cite{2019MNRAS.490.4638C}\\ 
    $\rm PH_3$ \tablefootmark{a} & ExoMolOP & \cite{2015MNRAS.446.2337S}\\ 
    SiO \tablefootmark{a} & ExoMolOP & \cite{2013MNRAS.434.1469B}\\ 
    TiO \tablefootmark{a,b} & ExoMolOP & \cite{2019MNRAS.488.2836M}\\ 
    VO \tablefootmark{a,b} & ExoMolOP & \cite{2016MNRAS.463..771M}\\ 
    FeH \tablefootmark{a} & ExoMolOP & {\cite{2010A&A...523A..58W}}\\
    \hline 
    \multicolumn{3}{c}{Collision-induced absorption}\\
    $\rm H_2$--$\rm H_2$    & - & [1, 2, 3]\\
    $\rm H_2$--He           & - & [1, 2, 3]\\
    \hline
    \multicolumn{3}{c}{Rayleigh scattering}\\ \\
    $\rm H_2$ & - & \cite{1962ApJ...136..690D}\\
    He & - & \cite{1965PPS....85..227C}\\
    \hline  
    \end{tabular}
    
    \tablefoot{
        \tablefoottext{a}{Species used in the equilibrium chemistry model.}
        \tablefoottext{b}{Species used in the free chemistry model, where $\rm CH_4$, CO and $\rm CO_2$ are only used in the optical-to-NIR joint retrievals for CoRoT-1b and TrES-4b.}
    }    
    \tablebib{
        [1] \cite{1988ApJ...326..509B} [2] \cite{1989ApJ...341..549B} [3] \cite{2012JQSRT.113.1276R}.
    }
\end{table}

\begin{table*}[htbp]
    \caption{Posterior estimates of the atmospheric retrievals using the OSIRIS data alone (equilibrium chemistry hypothesis).}  
    \label{table:atm_posteriors_equilibrium} 
    \renewcommand\arraystretch{1.2}
    \centering
    \tiny
    \begin{tabular}{l lc lc lc lc}  
    \hline\hline 
    & \multicolumn{2}{c}{CoRoT-1b} 
    & \multicolumn{2}{c}{HAT-P-18b} 
    & \multicolumn{2}{c}{HAT-P-57b} 
    & \multicolumn{2}{c}{Qatar-1b} \\

    Parameters & Priors & Posteriors 
                & Priors & Posteriors 
                & Priors & Posteriors 
                & Priors & Posteriors \\
    \hline 
    $\log_{10}P_0$ (bar) 
        & $\mathcal{U}(-6, 3)$ & $-0.52^{+2.23}_{-2.94}$ 
            & $\mathcal{U}(-6, 3)$ & $-1.19^{+2.41}_{-2.45}$ 
                & $\mathcal{U}(-6, 3)$ & $-1.78^{+3.11}_{-2.77}$ 
                    & $\mathcal{U}(-6, 3)$ & $-0.83^{+2.27}_{-2.68}$ \\    
    $R_p~(R_{\rm J})$ 
        & $\mathcal{N}(1.49, 0.08)$ & $1.524^{+0.040}_{-0.035}$ 
            & $\mathcal{N}(0.995, 0.052)$ & $1.012^{+0.029}_{-0.038}$ 
                & $\mathcal{N}(1.74, 0.36)$ & $1.824^{+0.023}_{-0.032}$ 
                    & $\mathcal{N}(1.143, 0.026)$ & $1.160^{+0.011}_{-0.014}$ \\
    $M_{\rm p}~(M_{\rm J})$ 
        & $\mathcal{N}(1.03, 0.12)$ & $1.00^{+0.12}_{-0.11}$ 
            & $\mathcal{N}(0.197, 0.013)$ & $0.20^{+0.01}_{-0.01}$ 
                & $\mathcal{N}(1.41, 1.52)$ & $3.53^{+1.57}_{-1.83}$ 
                    & $\mathcal{N}(1.294, 0.050)$ & $1.29^{+0.05}_{-0.04}$ \\
    $T~(\rm K)$ 
        & $\mathcal{U}(500, 2500)$ & $1551^{+500}_{-258}$ 
            & $\mathcal{U}(500, 1500)$ & $934^{+343}_{-288}$ 
                & $\mathcal{U}(500, 3000)$ & $1825^{+737}_{-896}$ 
                    & $\mathcal{U}(500, 2000)$ & $1342^{+398}_{-468}$ \\
    $\log_{10} P_{\rm c}$ (bar)
        & $\mathcal{U}(-6, 3)$ & $0.67^{+1.51}_{-1.79}$ 
            & $\mathcal{U}(-6, 3)$ & $-1.32^{+2.74}_{-2.79}$ 
                & $\mathcal{U}(-6, 3)$ & $-1.45^{+2.90}_{-2.96}$ 
                    & $\mathcal{U}(-6, 3)$ & $-0.61^{+2.26}_{-3.11}$ \\
    $f_{\rm haze}$ (dex) 
        & $\mathcal{U}(-3, 6)$ & $-0.10^{+4.18}_{-1.92}$ 
            & $\mathcal{U}(-3, 6)$ & $1.61^{+2.85}_{-2.74}$ 
                & $\mathcal{U}(-3, 6)$ & $1.72^{+2.72}_{-2.96}$ 
                    & $\mathcal{U}(-3, 6)$ & $-0.09^{+2.83}_{-1.83}$ \\
    C/O
        & $\mathcal{U}(0.1, 1.6)$ & $1.13^{+0.32}_{-0.49}$ 
            & $\mathcal{U}(0.1, 1.6)$ & $0.83^{+0.49}_{-0.45}$ 
                & $\mathcal{U}(0.1, 1.6)$ & $0.80^{+0.51}_{-0.46}$ 
                    & $\mathcal{U}(0.1, 1.6)$ & $0.77^{+0.52}_{-0.45}$ \\
    {}[M/H]
        & $\mathcal{U}(-1, 3)$ & $-0.14^{+0.85}_{-0.53}$ 
            & $\mathcal{U}(-1, 3)$ & $1.32^{+1.13}_{-1.44}$ 
                & $\mathcal{U}(-1, 3)$ & $0.95^{+1.27}_{-1.24}$ 
                    & $\mathcal{U}(-1, 3)$ & $1.20^{+0.99}_{-1.13}$ \\
    \hline 

    & \multicolumn{2}{c}{TrES-4b} 
    & \multicolumn{2}{c}{WASP-2b} 
    & \multicolumn{2}{c}{WASP-10b} 
    & \multicolumn{2}{c}{WASP-32b}\\
    
    Parameters & Priors & Posteriors 
                & Priors & Posteriors 
                & Priors & Posteriors 
                & Priors & Posteriors \\
    \hline 
    $\log_{10}P_0$ (bar) 
        & $\mathcal{U}(-6, 3)$ & $-2.34^{+3.14}_{-2.45}$
            & $\mathcal{U}(-6, 3)$ & $-0.29^{+2.03}_{-2.57}$
                & $\mathcal{U}(-6, 3)$ & $0.01^{+2.04}_{-3.00}$
                    & $\mathcal{U}(-6, 3)$ & $-1.45^{+2.95}_{-2.93}$ \\    
    $R_p~(R_{\rm J})$ 
        & $\mathcal{N}(1.61, 0.18)$ & $1.522^{+0.042}_{-0.056}$
            & $\mathcal{N}(1.063, 0.028)$ & $1.085^{+0.016}_{-0.021}$
                & $\mathcal{N}(1.080, 0.020)$ & $1.110^{+0.008}_{-0.008}$
                    & $\mathcal{N}(1.18, 0.07)$ & $1.223^{+0.008}_{-0.010}$ \\
    $M_{\rm p}~(M_{\rm J})$ 
        & $\mathcal{N}(0.78, 0.19)$ & $0.79^{+0.17}_{-0.17}$
            & $\mathcal{N}(0.880, 0.038)$ & $0.88^{+0.04}_{-0.04}$
                & $\mathcal{N}(3.15, 0.12)$ & $3.14^{+0.11}_{-0.11}$
                    & $\mathcal{N}(3.60, 0.07)$ & $3.60^{+0.07}_{-0.07}$ \\
    $T~(\rm K)$ 
        & $\mathcal{U}(500, 2500)$ & $1710^{+522}_{-783}$
            & $\mathcal{U}(500, 2000)$ & $1335^{+412}_{-502}$
                & $\mathcal{U}(500, 2000)$ & $1364^{+426}_{-510}$
                    & $\mathcal{U}(500, 2500)$ & $1503^{+646}_{-651}$ \\
    $\log_{10} P_{\rm c}$ (bar)
        & $\mathcal{U}(-6, 3)$ & $-0.88^{+2.53}_{-2.84}$
            & $\mathcal{U}(-6, 3)$ & $-1.48^{+2.88}_{-2.92}$
                & $\mathcal{U}(-6, 3)$ & $-2.01^{+3.29}_{-2.86}$
                    & $\mathcal{U}(-6, 3)$ & $-1.50^{+2.97}_{-2.91}$ \\
    $f_{\rm haze}$ (dex) 
        & $\mathcal{U}(-3, 6)$ & $1.54^{+2.51}_{-2.92}$
            & $\mathcal{U}(-3, 6)$ & $2.57^{+2.22}_{-3.53}$
                & $\mathcal{U}(-3, 6)$ & $2.15^{+2.63}_{-3.33}$
                    & $\mathcal{U}(-3, 6)$ & $1.67^{+2.84}_{-3.05}$ \\
    C/O
        & $\mathcal{U}(0.1, 1.6)$ & $0.85^{+0.48}_{-0.47}$
            & $\mathcal{U}(0.1, 1.6)$ & $0.90^{+0.44}_{-0.51}$
                & $\mathcal{U}(0.1, 1.6)$ & $0.89^{+0.47}_{-0.52}$
                    & $\mathcal{U}(0.1, 1.6)$ & $0.87^{+0.48}_{-0.48}$ \\
    {}[M/H]
        & $\mathcal{U}(-1, 3)$ & $0.75^{+1.40}_{-1.12}$
            & $\mathcal{U}(-1, 3)$ & $0.80^{+1.19}_{-1.16}$
                & $\mathcal{U}(-1, 3)$ & $0.80^{+1.25}_{-1.21}$
                    & $\mathcal{U}(-1, 3)$ & $0.88^{+1.35}_{-1.26}$ \\
    \hline 

    & \multicolumn{2}{c}{WASP-36b} 
    & \multicolumn{2}{c}{WASP-39b} 
    & \multicolumn{2}{c}{WASP-49b} 
    & \multicolumn{2}{c}{WASP-156b} \\ 
    
    Parameters & Priors & Posteriors 
                & Priors & Posteriors 
                & Priors & Posteriors 
                & Priors & Posteriors \\
    \midrule 
    $\log_{10}P_0$ (bar) 
        & $\mathcal{U}(-6, 3)$ & $-2.64^{+3.12}_{-2.15}$
            & $\mathcal{U}(-6, 3)$ & $-1.29^{+2.37}_{-2.05}$
                & $\mathcal{U}(-6, 3)$ & $-3.67^{+2.33}_{-1.45}$
                    & $\mathcal{U}(-6, 3)$ & $-0.83^{+2.33}_{-2.67}$\\    
    $R_{\rm p}~(R_{\rm J})$ 
        & $\mathcal{N}(1.327, 0.021)$ & $1.311^{+0.006}_{-0.007}$
            & $\mathcal{N}(1.279, 0.040)$ &  $1.295^{+0.031}_{-0.034}$
                & $\mathcal{N}(1.198, 0.046)$ & $1.176^{+0.027}_{-0.024}$
                    & $\mathcal{N}(0.51, 0.02)$ & $0.519^{+0.013}_{-0.016}$\\
    $M_{\rm p}~(M_{\rm J})$ 
        & $\mathcal{N}(2.361, 0.070)$ & $2.36^{+0.07}_{-0.07}$
            & $\mathcal{N}(0.281, 0.032)$ & $0.28^{+0.03}_{-0.03}$
                & $\mathcal{N}(0.396, 0.026)$ & $0.40^{+0.02}_{-0.02}$
                    & $\mathcal{N}(0.128, 0.010)$ & $0.13^{+0.01}_{-0.01}$ \\
    $T~(\rm K)$ 
        & $\mathcal{U}(500, 2500)$ &  $1050^{+792}_{-352}$
            & $\mathcal{U}(500, 2000)$ & $988^{+549}_{-334}$
                & $\mathcal{U}(500, 2000)$ &  $1197^{+407}_{-442}$
                    & $\mathcal{U}(500, 1500)$ & $964^{+323}_{-305}$\\
    $\log_{10} P_{\rm c}$ (bar)
        & $\mathcal{U}(-6, 3)$ &  $-1.44^{+2.90}_{-2.81}$
            & $\mathcal{U}(-6, 3)$ &  $-3.37^{+3.80}_{-1.73}$
                & $\mathcal{U}(-6, 3)$ & $-0.55^{+2.19}_{-2.56}$
                    & $\mathcal{U}(-6, 3)$ & $-1.55^{+2.94}_{-2.97}$\\
    $f_{\rm haze}$ (dex) 
        & $\mathcal{U}(-3, 6)$ & $1.48^{+2.92}_{-2.92}$
            & $\mathcal{U}(-3, 6)$ &  $2.21^{+2.51}_{-3.27}$
                & $\mathcal{U}(-3, 6)$ &  $2.43^{+1.70}_{-2.05}$
                    & $\mathcal{U}(-3, 6)$ & $2.06^{+2.47}_{-3.20}$\\
    C/O
        & $\mathcal{U}(0.1, 1.6)$ &  $0.82^{+0.50}_{-0.49}$
            & $\mathcal{U}(0.1, 1.6)$ & $0.84^{+0.45}_{-0.45}$
                & $\mathcal{U}(0.1, 1.6)$ & $0.88^{+0.47}_{-0.48}$
                    & $\mathcal{U}(0.1, 1.6)$ & $0.84^{+0.47}_{-0.49}$\\
    {}[M/H]
        & $\mathcal{U}(-1, 3)$ & $1.47^{+1.09}_{-1.58}$
            & $\mathcal{U}(-1, 3)$ &  $0.90^{+1.40}_{-1.24}$
                & $\mathcal{U}(-1, 3)$ &  $0.82^{+1.40}_{-1.18}$
                    & $\mathcal{U}(-1, 3)$ & $1.04^{+1.26}_{-1.26}$\\
    \hline
    \end{tabular}
    
    \tablefoot{
        $P_0$: reference pressure; $R_{\rm p}$: planet radius at the reference pressure; $M_{\rm p}$: planet mass; $T$: temperature of the isothermal atmosphere; $P_{\rm c}$: cloud-top pressure; $f_{\rm haze}$: Rayleigh-like scattering factor; C/O: carbon-to-oxygen ratio of the atmosphere; [M/H]: metallicity of the atmosphere.
    } 
    
\end{table*}

\begin{table*}[htbp]
    \caption{Posterior estimates of the atmospheric retrievals using the OSIRIS data alone (free chemistry hypothesis).}  
    \label{table:atm_posteriors_free} 
    \centering
    \tiny
    \begin{tabular}{l lc lc lc lc}  
    \hline\hline 
    & \multicolumn{2}{c}{CoRoT-1b} 
    & \multicolumn{2}{c}{HAT-P-18b} 
    & \multicolumn{2}{c}{HAT-P-57b} 
    & \multicolumn{2}{c}{Qatar-1b} \\
    
    Parameters & Priors & Posteriors 
                & Priors & Posteriors 
                & Priors & Posteriors 
                & Priors & Posteriors \\
    \hline 
$\log_{10}P_0$ (bar) 
& $\mathcal{U}(-6, 3)$ & $-1.07^{+2.46}_{-2.74}$
    & $\mathcal{U}(-6, 3)$ &  $-1.99^{+2.45}_{-2.22}$
        & $\mathcal{U}(-6, 3)$ & $-2.01^{+2.90}_{-2.57}$
            & $\mathcal{U}(-6, 3)$ & $-1.91^{+2.77}_{-2.32}$\\    
$R_p~(R_{\rm J})$ 
& $\mathcal{N}(1.49, 0.08)$  & $1.522^{+0.028}_{-0.033}$
    & $\mathcal{N}(0.995, 0.052)$ & $1.007^{+0.035}_{-0.040}$
        & $\mathcal{N}(1.74, 0.36)$  & $1.816^{+0.023}_{-0.034}$
            & $\mathcal{N}(1.143, 0.026)$ & $1.156^{+0.009}_{-0.012}$\\
$M_{\rm p}~(M_{\rm J})$ 
& $\mathcal{N}(1.03, 0.12)$  &  $1.04^{+0.10}_{-0.10}$
    & $\mathcal{N}(0.197, 0.013)$ &  $0.20^{+0.01}_{-0.01}$
        & $\mathcal{N}(1.41, 1.52)$  &  $3.39^{+1.60}_{-1.65}$
            & $\mathcal{N}(1.294, 0.050)$ & $1.30^{+0.04}_{-0.04}$\\
$T~(\rm K)$ 
& $\mathcal{U}(500, 2500)$ & $1257^{+671}_{-423}$
    & $\mathcal{U}(500, 1500)$ & $812^{+331}_{-213}$
        & $\mathcal{U}(500, 3000)$ & $1677^{+831}_{-772}$
            & $\mathcal{U}(500, 2000)$ & $998^{+514}_{-316}$\\
$\log_{10} P_{\rm c}$ (bar)
& $\mathcal{U}(-6, 3)$ &  $0.25^{+1.74}_{-2.07}$
    & $\mathcal{U}(-6, 3)$ &  $-1.75^{+2.94}_{-2.55}$
        & $\mathcal{U}(-6, 3)$ &  $-1.33^{+2.75}_{-2.88}$
            & $\mathcal{U}(-6, 3)$ & $-1.38^{+2.69}_{-2.95}$\\
$f_{\rm haze}$ (dex) 
& $\mathcal{U}(-3, 6)$ & $0.35^{+2.66}_{-2.04}$
    & $\mathcal{U}(-3, 6)$ & $1.85^{+2.50}_{-3.01}$
        & $\mathcal{U}(-3, 6)$ & $1.56^{+2.66}_{-2.72}$
            & $\mathcal{U}(-3, 6)$ & $0.67^{+2.86}_{-2.29}$\\
$\rm MMR_{Na}$ (dex)
& $\mathcal{U}(-10, 0)$ &  $-2.75^{+1.60}_{-3.86}$
    & $\mathcal{U}(-10, 0)$ &  $-4.60^{+2.94}_{-3.21}$
        & $\mathcal{U}(-10, 0)$ &  $-5.07^{+3.07}_{-3.26}$
            & $\mathcal{U}(-10, 0)$ & $-5.50^{+3.06}_{-2.83}$\\
$\rm MMR_{K}$ (dex)
& $\mathcal{U}(-10, 0)$ &  $-5.32^{+2.55}_{-2.75}$
    & $\mathcal{U}(-10, 0)$ &  $-5.64^{+3.32}_{-2.69}$
        & $\mathcal{U}(-10, 0)$ &  $-5.10^{+3.00}_{-2.97}$
            & $\mathcal{U}(-10, 0)$ & $-3.44^{+2.31}_{-3.56}$\\
$\rm MMR_{H_2O}$  (dex)
& $\mathcal{U}(-10, 0)$ &  $-5.57^{+2.78}_{-2.64}$
    & $\mathcal{U}(-10, 0)$ &  $-4.94^{+3.30}_{-3.21}$
        & $\mathcal{U}(-10, 0)$ &  $-5.32^{+3.18}_{-2.92}$
            & $\mathcal{U}(-10, 0)$ & $-4.75^{+3.05}_{-3.16}$\\
$\rm MMR_{TiO}$  (dex)
& $\mathcal{U}(-10, 0)$ &  $-7.16^{+3.88}_{-1.98}$
    & n/a & n/a
        & $\mathcal{U}(-10, 0)$ &  $-5.05^{+2.92}_{-3.06}$
            & $\mathcal{U}(-10, 0)$ & $-6.12^{+3.95}_{-2.23}$\\
$\rm MMR_{VO}$ (dex)
& $\mathcal{U}(-10, 0)$ &  $-7.73^{+1.84}_{-1.37}$
    & n/a & n/a
        & $\mathcal{U}(-10, 0)$ &  $-5.16^{+3.06}_{-2.89}$
            & $\mathcal{U}(-10, 0)$ & $-5.56^{+3.83}_{-2.92}$\\
    
    \hline\hline

    & \multicolumn{2}{c}{TrES-4b} 
    & \multicolumn{2}{c}{WASP-2b} 
    & \multicolumn{2}{c}{WASP-10b} 
    & \multicolumn{2}{c}{WASP-32b}\\
    
    Parameters & Priors & Posteriors 
                & Priors & Posteriors 
                & Priors & Posteriors 
                & Priors & Posteriors \\
    \hline 
$\log_{10}P_0$ (bar) 
& $\mathcal{U}(-6, 3)$ & $-2.78^{+3.14}_{-2.17}$
    & $\mathcal{U}(-6, 3)$ & $-0.02^{+1.88}_{-2.53}$
        & $\mathcal{U}(-6, 3)$ &  $-0.20^{+2.27}_{-2.99}$
            & $\mathcal{U}(-6, 3)$ & $-1.39^{+2.87}_{-2.83}$ \\    
$R_p~(R_{\rm J})$ 
& $\mathcal{N}(1.61, 0.18)$  & $1.508^{+0.038}_{-0.052}$
    & $\mathcal{N}(1.063, 0.028)$ &  $1.079^{+0.017}_{-0.019}$
        & $\mathcal{N}(1.080, 0.020)$ & $1.109^{+0.008}_{-0.009}$
            & $\mathcal{N}(1.18, 0.07)$ &  $1.221^{+0.009}_{-0.010}$\\
$M_{\rm p}~(M_{\rm J})$ 
& $\mathcal{N}(0.78, 0.19)$  &  $0.79^{+0.17}_{-0.17}$
    & $\mathcal{N}(0.880, 0.038)$ &  $0.88^{+0.04}_{-0.03}$
        & $\mathcal{N}(3.15, 0.12)$ &  $3.14^{+0.12}_{-0.10}$
            & $\mathcal{N}(3.60, 0.07)$ & $3.60^{+0.06}_{-0.06}$\\
$T~(\rm K)$ 
& $\mathcal{U}(500, 2500)$ &  $1414^{+651}_{-611}$
    & $\mathcal{U}(500, 2000)$ &  $1256^{+452}_{-464}$
        & $\mathcal{U}(500, 2000)$ &  $1361^{+426}_{-546}$
            & $\mathcal{U}(500, 2500)$ & $1515^{+619}_{-637}$\\
$\log_{10} P_{\rm c}$ (bar)
& $\mathcal{U}(-6, 3)$ &  $-1.25^{+2.57}_{-2.70}$
    & $\mathcal{U}(-6, 3)$ &  $-1.25^{+2.67}_{-3.06}$
        & $\mathcal{U}(-6, 3)$ &   $-1.73^{+3.05}_{-2.85}$
            & $\mathcal{U}(-6, 3)$ & $-1.37^{+2.73}_{-2.70}$\\
$f_{\rm haze}$ (dex) 
& $\mathcal{U}(-3, 6)$ &  $1.39^{+2.71}_{-2.76}$
    & $\mathcal{U}(-3, 6)$ &  $2.53^{+2.15}_{-3.34}$
        & $\mathcal{U}(-3, 6)$ &  $2.34^{+2.38}_{-3.23}$
            & $\mathcal{U}(-3, 6)$  & $1.76^{+2.66}_{-2.97}$\\
$\rm MMR_{Na}$ (dex)
& $\mathcal{U}(-10, 0)$ &   $-4.96^{+3.06}_{-3.00}$
    & $\mathcal{U}(-10, 0)$ &  $-4.86^{+3.13}_{-3.46}$
        & $\mathcal{U}(-10, 0)$ &  $-5.13^{+3.17}_{-3.17}$
            & $\mathcal{U}(-10, 0)$ & $-4.85^{+3.04}_{-3.31}$\\
$\rm MMR_{K}$ (dex)
& $\mathcal{U}(-10, 0)$ & $-5.18^{+3.10}_{-2.93}$
    & $\mathcal{U}(-10, 0)$ &  $-5.32^{+3.00}_{-3.01}$
        & $\mathcal{U}(-10, 0)$ & $-5.42^{+3.28}_{-3.03}$
            & $\mathcal{U}(-10, 0)$ & $-5.45^{+3.24}_{-2.80}$\\
$\rm MMR_{H_2O}$  (dex)
& $\mathcal{U}(-10, 0)$ &  $-4.96^{+3.10}_{-3.10}$
    & $\mathcal{U}(-10, 0)$ &  $-5.31^{+2.99}_{-3.09}$
        & $\mathcal{U}(-10, 0)$ &  $-5.18^{+3.12}_{-3.13}$
            & $\mathcal{U}(-10, 0)$ & $-5.42^{+3.45}_{-2.97}$\\
$\rm MMR_{TiO}$  (dex)
& $\mathcal{U}(-10, 0)$ & $-5.00^{+2.63}_{-2.80}$
    & n/a &  n/a
        & n/a &  n/a
            & $\mathcal{U}(-10, 0)$ & $-4.72^{+3.02}_{-3.23}$\\
$\rm MMR_{VO}$ (dex)
& $\mathcal{U}(-10, 0)$ & $-6.38^{+3.45}_{-2.36}$
    & n/a &  n/a
        & n/a &  n/a
            & $\mathcal{U}(-10, 0)$ & $-4.69^{+2.91}_{-3.26}$\\
    
    \hline\hline

    & \multicolumn{2}{c}{WASP-36b} 
    & \multicolumn{2}{c}{WASP-39b} 
    & \multicolumn{2}{c}{WASP-49b} 
    & \multicolumn{2}{c}{WASP-156b} \\
    
    Parameters & Priors & Posteriors 
                & Priors & Posteriors 
                & Priors & Posteriors 
                & Priors & Posteriors \\
    \hline 
$\log_{10}P_0$ (bar) 
& $\mathcal{U}(-6, 3)$ & $-3.46^{+2.78}_{-1.65}$
    & $\mathcal{U}(-6, 3)$ &  $-1.98^{+2.22}_{-1.70}$
        & $\mathcal{U}(-6, 3)$ &   $-4.24^{+1.63}_{-1.12}$
            & $\mathcal{U}(-6, 3)$ &  $-1.05^{+2.17}_{-2.43}$\\    
$R_p~(R_{\rm J})$ 
& $\mathcal{N}(1.327, 0.021)$ &  $1.308^{+0.006}_{-0.007}$
    & $\mathcal{N}(1.279, 0.040)$ & $1.288^{+0.029}_{-0.033}$
        & $\mathcal{N}(1.198, 0.046)$ & $1.166^{+0.015}_{-0.019}$
            & $\mathcal{N}(0.51, 0.02)$ &  $0.515^{+0.016}_{-0.016}$\\
$M_{\rm p}~(M_{\rm J})$ 
& $\mathcal{N}(2.361, 0.070)$ &  $2.36^{+0.06}_{-0.06}$
    & $\mathcal{N}(0.281, 0.032)$ &  $0.28^{+0.03}_{-0.03}$
        & $\mathcal{N}(0.396, 0.026)$ &  $0.40^{+0.02}_{-0.02}$
            & $\mathcal{N}(0.128, 0.010)$ &  $0.13^{+0.01}_{-0.01}$\\
$T~(\rm K)$ 
& $\mathcal{U}(500, 2500)$ & $1037^{+693}_{-368}$
    & $\mathcal{U}(500, 2000)$ & $867^{+494}_{-247}$
        & $\mathcal{U}(500, 2000)$ & $1001^{+433}_{-332}$
            & $\mathcal{U}(500, 1500)$ & $944^{+348}_{-295}$\\
$\log_{10} P_{\rm c}$ (bar)
& $\mathcal{U}(-6, 3)$ &  $-1.89^{+2.99}_{-2.59}$
    & $\mathcal{U}(-6, 3)$ &  $-3.41^{+3.64}_{-1.73}$
        & $\mathcal{U}(-6, 3)$ &  $-1.32^{+2.54}_{-2.68}$
            & $\mathcal{U}(-6, 3)$ &  $-1.87^{+2.90}_{-2.57}$\\
$f_{\rm haze}$ (dex) 
& $\mathcal{U}(-3, 6)$ &  $1.88^{+2.66}_{-3.04}$
    & $\mathcal{U}(-3, 6)$ &  $2.72^{+2.34}_{-3.47}$
        & $\mathcal{U}(-3, 6)$ &  $3.28^{+1.51}_{-3.15}$
            & $\mathcal{U}(-3, 6)$  & $2.12^{+2.44}_{-3.26}$ \\
$\rm MMR_{Na}$ (dex)
& $\mathcal{U}(-10, 0)$ &  $-4.89^{+3.07}_{-3.17}$
    & $\mathcal{U}(-10, 0)$ &  $-5.52^{+3.06}_{-2.70}$
        & $\mathcal{U}(-10, 0)$ &  $-4.48^{+3.00}_{-3.38}$
            & $\mathcal{U}(-10, 0)$ & $-4.30^{+2.75}_{-3.58}$\\
$\rm MMR_{K}$ (dex)
& $\mathcal{U}(-10, 0)$ &  $-5.69^{+3.07}_{-2.63}$
    & $\mathcal{U}(-10, 0)$ &  $-5.81^{+3.24}_{-2.71}$
        & $\mathcal{U}(-10, 0)$ &  $-5.43^{+3.05}_{-2.94}$
            & $\mathcal{U}(-10, 0)$ & $-6.14^{+3.46}_{-2.54}$\\
$\rm MMR_{H_2O}$  (dex)
& $\mathcal{U}(-10, 0)$ &  $-4.72^{+3.14}_{-3.31}$
    & $\mathcal{U}(-10, 0)$ &  $-5.31^{+3.01}_{-2.85}$
        & $\mathcal{U}(-10, 0)$ &  $-5.07^{+3.09}_{-2.99}$
            & $\mathcal{U}(-10, 0)$ & $-5.02^{+3.34}_{-3.16}$\\
$\rm MMR_{TiO}$  (dex)
& $\mathcal{U}(-10, 0)$ &  $-4.86^{+3.13}_{-3.11}$
    & n/a &  n/a
        & $\mathcal{U}(-10, 0)$ &  $-6.95^{+4.60}_{-1.91}$
            & n/a & n/a \\
$\rm MMR_{VO}$ (dex)
& $\mathcal{U}(-10, 0)$ &  $-4.90^{+3.02}_{-3.11}$
    & n/a &  n/a
        & $\mathcal{U}(-10, 0)$ & $-6.13^{+3.25}_{-2.18}$
            & n/a & n/a \\
    \hline
    \end{tabular}
    
    \tablefoot{$P_0$: reference pressure; $R_{\rm p}$: planet radius at the reference pressure; $M_{\rm p}$: planet mass; $T$: temperature of the isothermal atmosphere; $P_{\rm c}$: cloud-top pressure; $f_{\rm haze}$: Rayleigh-like scattering factor; MMR: mass mixing ratio; n/a: not applicable.}
    
\end{table*}    

\begin{figure*}[htbp]
    \centering
    \includegraphics[width=\linewidth]{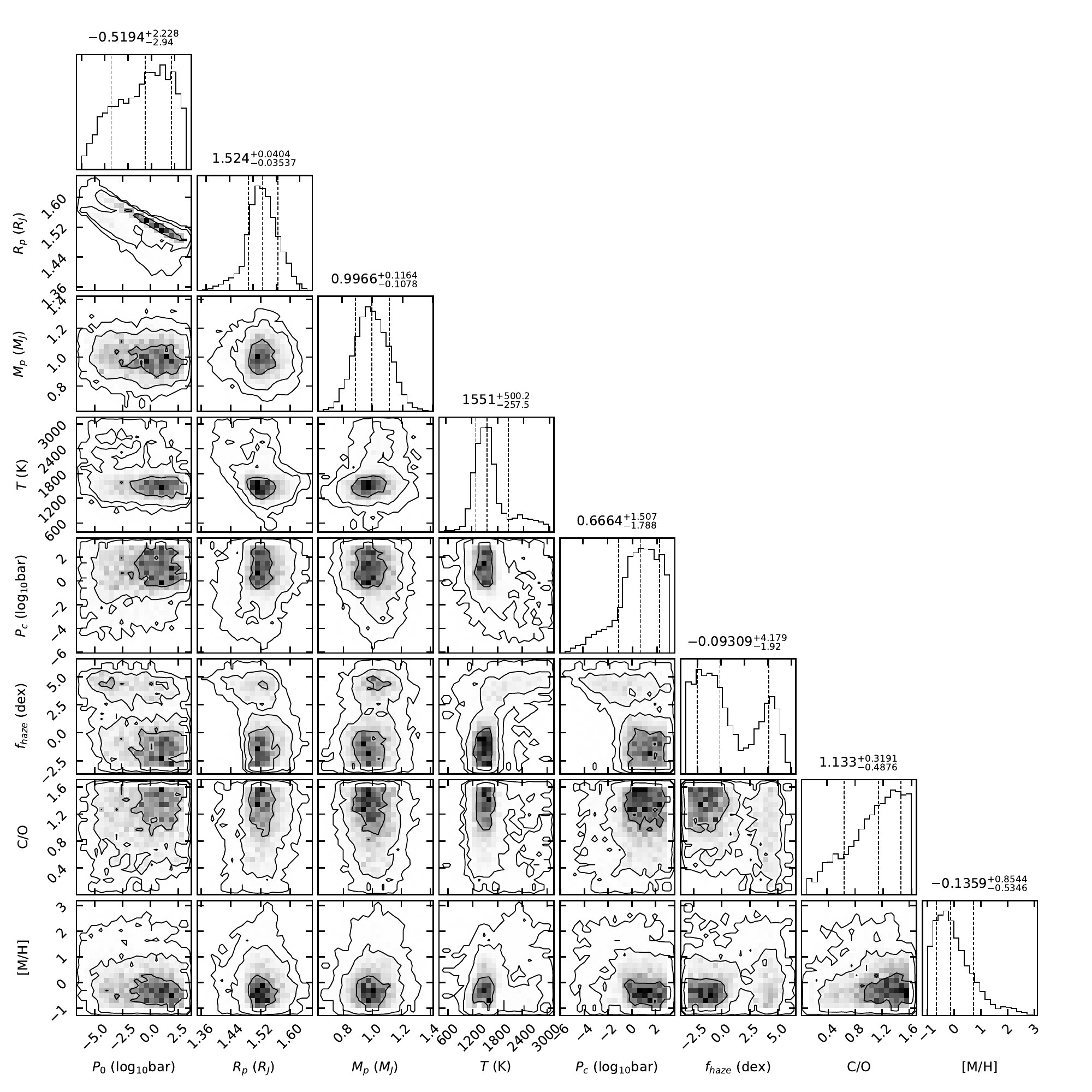}
    \caption{Posterior distribution of the retrieved atmospheric parameters for CoRoT-1b assuming chemical equilibrium based on the OSIRIS transmission spectrum only.}
    \label{fig:corners_retrievals_c1b_ec}
\end{figure*}  

\begin{figure*}[htbp]
    \centering
    \includegraphics[width=\linewidth]{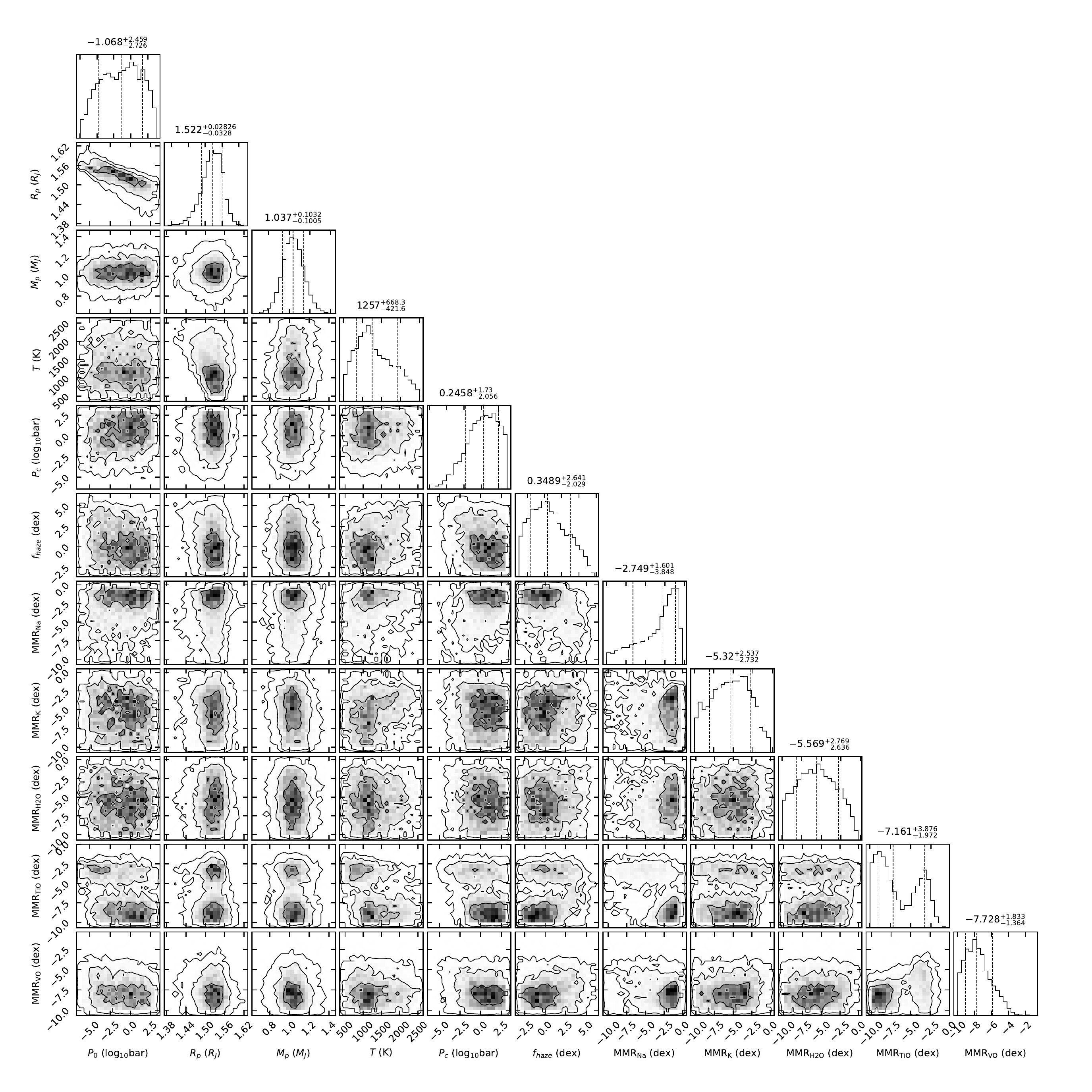}
    \caption{Posterior distribution of the retrieved atmospheric parameters for CoRoT-1b assuming free chemistry based on the OSIRIS transmission spectrum only.}
    \label{fig:corners_retrievals_c1b_fc}
\end{figure*}  

\begin{figure*}[htbp]
    \centering
    \includegraphics[width=\linewidth]{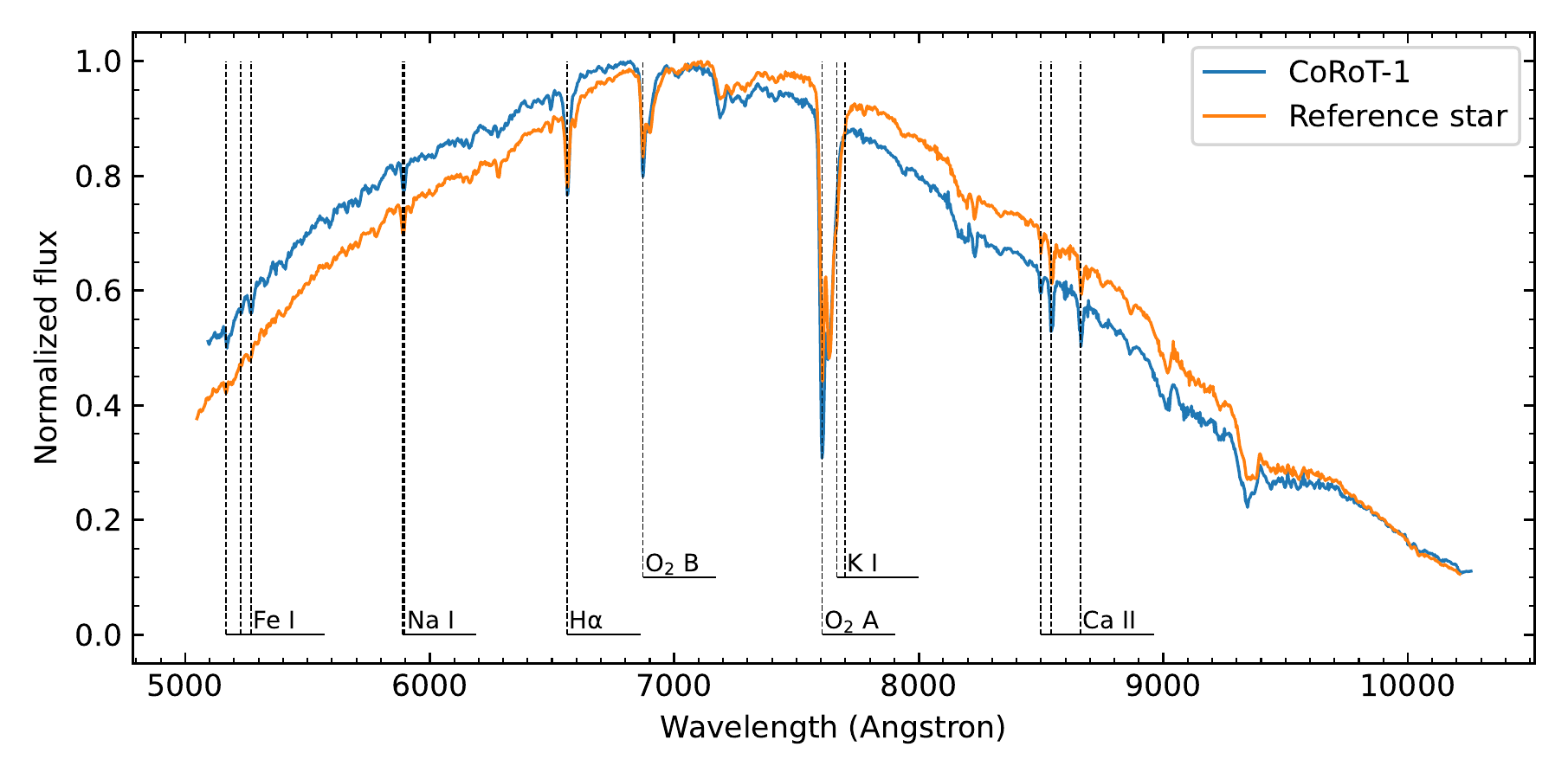}
    \caption{Normalized stellar spectra of CoRoT-1 and its reference star (UCAC4 435-02201). The vertical lines indicate the positions of strong lines.}
    \label{fig:stellar_spectra_c1b}
\end{figure*}

\begin{table*}[htbp]
    
    \caption{Posterior estimates of the atmospheric parameters retrieved from the joint transmission spectra of CoRoT-1b (OSIRIS + WFC3, corresponding to Fig.~\ref{fig:joint_retrieval_22}).}  
    \label{table:atm_posteriors_joint} 
    \centering
    \begin{tabular}{l lc l lc}  
    \hline\hline 
    \multicolumn{3}{c}{Cloudy model} &
    \multicolumn{3}{c}{Equilibrium chemistry} \\ 

    Parameters & Priors & Posteriors &
    Parameters & Priors & Posteriors \\
    \hline 
    
$\log_{10}P_0$ (bar) & $\mathcal{U}(-6, 3)$ & $-1.78^{+2.72}_{-2.30}$ &
$\log_{10}P_0$ (bar) & $\mathcal{U}(-6, 3)$ & $-2.33^{+2.49}_{-2.07}$\\  

$R_p~(R_{\rm J})$ & $\mathcal{N}(1.49, 0.08)$ &  $1.500^{+0.049}_{-0.050}$ &
$R_p~(R_{\rm J})$ & $\mathcal{N}(1.49, 0.08)$ &  $1.500^{+0.038}_{-0.051}$\\

$M_{\rm p}~(M_{\rm J})$ & $\mathcal{N}(1.03, 0.12)$ &  $0.99^{+0.11}_{-0.11}$ &  
$M_{\rm p}~(M_{\rm J})$ & $\mathcal{N}(1.03, 0.12)$ &  $1.01^{+0.10}_{-0.10}$ \\

$T~(\rm K)$ & $\mathcal{U}(500, 2500)$ &  $2132^{+252}_{-398}$ &
$T~(\rm K)$ & $\mathcal{U}(500, 2500)$ &  $2099^{+271}_{-410}$\\

$\log_{10} P_{\rm c}$ (bar) & $\mathcal{U}(-6, 3)$ &  $0.25^{+1.80}_{-2.00}$ &
$\log_{10} P_{\rm c}$ (bar) & $\mathcal{U}(-6, 3)$ &  $-0.36^{2.13}_{-2.25}$ \\

$f_{\rm haze}$ (dex) & $\mathcal{U}(-3, 6)$ &  $3.68^{+1.04}_{-1.22}$ &
$f_{\rm haze}$ (dex) & $\mathcal{U}(-3, 6)$ &  $4.63^{+0.74}_{-0.58}$\\

$\delta$ (ppm) & $\mathcal{U}(-2000, 2000)$ &  $92^{+185}_{-172}$ &
$\delta$ (ppm) & $\mathcal{U}(-2000, 2000)$ &  $10^{+207}_{-164}$\\

- & - & - &
C/O & $\mathcal{U}(0.1, 1.6)$ &  $1.00^{+0.38}_{-0.47}$\\

- & - & - &
[M/H] & $\mathcal{U}(-1, 3)$ &  $-0.21^{+0.83}_{-0.53}$\\

    \hline\hline 
    \multicolumn{3}{c}{Free chemistry w/o TiO\&VO} &
    \multicolumn{3}{c}{Free chemistry with TiO\&VO} \\ 

    Parameters & Priors & Posteriors &
    Parameters & Priors & Posteriors \\
    \hline 
    
$\log_{10}P_0$ (bar) & $\mathcal{U}(-6, 3)$ & $-1.41^{+2.34}_{-2.30}$ & 
$\log_{10}P_0$ (bar) & $\mathcal{U}(-6, 3)$ & $-1.81^{+2.50}_{-2.29}$ \\ 

$R_p~(R_{\rm J})$ & $\mathcal{N}(1.49, 0.08)$ &  $1.507^{+0.045}_{-0.049}$ &
$R_p~(R_{\rm J})$ & $\mathcal{N}(1.49, 0.08)$ &  $1.510^{+0.031}_{-0.044}$ \\

$M_{\rm p}~(M_{\rm J})$ & $\mathcal{N}(1.03, 0.12)$ &  $1.00^{+0.10}_{-0.10}$ &  
$M_{\rm p}~(M_{\rm J})$ & $\mathcal{N}(1.03, 0.12)$ &  $1.01^{+0.10}_{-0.10}$ \\

$T~(\rm K)$ & $\mathcal{U}(500, 2500)$ &  $1997^{+319}_{-631}$ &
$T~(\rm K)$ & $\mathcal{U}(500, 2500)$ &  $1639^{+571}_{-726}$ \\

$\log_{10} P_{\rm c}$ (bar) & $\mathcal{U}(-6, 3)$ & $0.24^{+1.71}_{-1.76}$ &
$\log_{10} P_{\rm c}$ (bar) & $\mathcal{U}(-6, 3)$ & $0.17^{+1.74}_{-3.29}$ \\

$f_{\rm haze}$ (dex) & $\mathcal{U}(-3, 6)$ & $3.10^{+1.30}_{-1.66}$ & 
$f_{\rm haze}$ (dex) & $\mathcal{U}(-3, 6)$ & $1.88^{+2.66}_{-3.04}$ \\

$\delta$ (ppm) & $\mathcal{U}(-2000, 2000)$ & $-54^{+227}_{-165}$ &
$\delta$ (ppm) & $\mathcal{U}(-2000, 2000)$ & $146^{+263}_{-465}$  \\

$\rm MMR_{Na}$ (dex) & $\mathcal{U}(-10, 0)$ &  $-4.02^{+2.41}_{-3.50}$ &
$\rm MMR_{Na}$ (dex) & $\mathcal{U}(-10, 0)$ &  $-5.08^{+2.94}_{-2.92}$ \\ 

$\rm MMR_{K}$ (dex) & $\mathcal{U}(-10, 0)$ &  $-4.49^{+2.12}_{-3.08}$ &
$\rm MMR_{K}$ (dex) & $\mathcal{U}(-10, 0)$ &  $-5.03^{+2.63}_{-2.93}$ \\ 

$\rm MMR_{H_2O}$ (dex) & $\mathcal{U}(-10, 0)$ &  $-6.64^{+2.13}_{-2.01}$ &
$\rm MMR_{H_2O}$ (dex) & $\mathcal{U}(-10, 0)$ &  $-6.23^{+2.29}_{-2.24}$ \\ 

$\rm MMR_{CH_4}$ (dex) & $\mathcal{U}(-10, 0)$ &  $-6.34^{+2.37}_{-2.33}$ &
$\rm MMR_{CH_4}$ (dex) & $\mathcal{U}(-10, 0)$ &  $-5.77^{+2.50}_{-2.56}$ \\ 

$\rm MMR_{CO}$ (dex) & $\mathcal{U}(-10, 0)$ &  $-5.44^{+2.64}_{-2.68}$ &
$\rm MMR_{CO}$ (dex) & $\mathcal{U}(-10, 0)$ &  $-5.30^{+2.68}_{-2.79}$ \\ 

$\rm MMR_{CO_2}$ (dex) & $\mathcal{U}(-10, 0)$ &  $-5.90^{+2.46}_{-2.50}$ &
$\rm MMR_{CO_2}$ (dex) & $\mathcal{U}(-10, 0)$ &  $-5.89^{+2.64}_{-2.44}$ \\ 

- & - & - &
$\rm MMR_{TiO}$ (dex) & $\mathcal{U}(-10, 0)$ &  $-4.84^{+1.98}_{-3.36}$ \\ 

- & - & - &
$\rm MMR_{VO}$ (dex) & $\mathcal{U}(-10, 0)$ &  $-6.78^{+1.98}_{-1.95}$ \\ 

    \hline  
    \end{tabular}

    \tablefoot{$P_0$: reference pressure; $R_{\rm p}$: planet radius at the reference pressure; $M_{\rm p}$: planet mass; $T$: temperature of the isothermal atmosphere; $P_{\rm c}$: cloud-top pressure; $f_{\rm haze}$: Rayleigh-like scattering factor; C/O: atmospheric carbon-to-oxygen ratio; [M/H]: atmospheric metallicity; MMR: mass mixing ratio.}

\end{table*}

\end{appendix}

\end{document}